\documentclass[a4paper,twocolumn,11pt]{quantumarticle}
\pdfoutput=1

\usepackage[utf8]{inputenc}
\usepackage[english]{babel}
\usepackage[T1]{fontenc}
\usepackage{amsmath}
\usepackage{hyperref}
\usepackage{cleveref}
\usepackage{amssymb}
\usepackage{braket}  
\usepackage{euscript}
\usepackage{xcolor}    
\usepackage{tcolorbox}
\tcbuselibrary{raster,skins}

\usepackage{tikz}

\usepackage[acronym]{glossaries}
\glsdisablehyper

\usepackage{booktabs}
\usepackage{colortbl}
\usepackage{graphicx}
\usepackage[numbers,sort&compress]{natbib}
\usepackage{microtype}
\usepackage{stmaryrd}
\usepackage{amsthm}
\usepackage[font=small,format=plain,labelfont=bf,up,textfont=it,up]{caption}
\usepackage{float}
\usetikzlibrary{patterns,positioning,calc,fit,backgrounds,decorations.pathreplacing,decorations.pathmorphing,shapes,arrows.meta,bending}
\usepackage{pgfplots}
\pgfplotsset{compat=1.18}
\usepackage{enumitem}

\usepackage{makecell}

\newtheorem{example}{Example}
\theoremstyle{definition}
\newtheorem{definition}{Definition}

\begin{document}

\title{Methods for Automated Equivalence Checking of Quantum Error-Correction Codes}

\author{Maxie Bichmann}
\affiliation{Chair for Design Automation, Technical University of Munich, Munich, Germany}
\orcid{0009-0006-0944-7716}

\author{Ludwig Schmid}
\affiliation{Chair for Design Automation, Technical University of Munich, Munich, Germany}
\orcid{0000-0002-4246-8125}

\author{Robert Wille}
\affiliation{Chair for Design Automation, Technical University of Munich, Munich, Germany}
\affiliation{Munich Quantum Software Company GmbH, Munich, Germany}
\orcid{0000-0002-4993-7860}
\maketitle

\newcommand{\permutationGroup}[1]{\EuScript{S}_{#1}}
\newcommand{\binField}[1]{\mathbb{F}_2^{#1}}
\newcommand{\linearGroup}[1]{\EuScript{GL}_{#1}}
\newcommand{\hilbertSpace}[1]{\EuScript{H}^{\otimes #1}}
\newcommand{\pauliGroup}[1]{\EuScript{P}_{#1}}
\newcommand{\restrcliffordGroup}[1]{\EuScript{C}_{#1}}
\newcommand{\symplGroup}[1]{\operatorname{Sp}(#1,\binField{})}
\newcommand{\locsymplGroup}[1]{\operatorname{lSp}(#1,\binField{})}
\newcommand{\stabGroup}{\mathcal{S}}
\newcommand{\codeSpace}{\mathcal{C}}

\newcommand{\stab}[1]{\operatorname{S}(#1)}
\newcommand{\powerset}[1]{\mathcal{P}(#1)}

\newcommand{\independence}{\mathcal{I}}
\newcommand{\groundSet}{\text{E}}

\newcommand{\SAT}{\textsc{SAT}}

\definecolor{zxgreen}{HTML}{74D374}
\definecolor{zxgreenmuted}{HTML}{CCFFBF}
\definecolor{zxorangemuted}{HTML}{FFD9B2}
\definecolor{zxgreendark}{HTML}{199900}
\definecolor{zxorangedark}{HTML}{FF7C00}

\definecolor{thesisred}{HTML}{E9341A}

\definecolor{thesisblue}{HTML}{2E61F1}

\definecolor{thesisgreen}{HTML}{98D256}
\definecolor{thesisdarkgreen}{HTML}{57891F}

\definecolor{thesislightlila}{HTML}{AF81DA}
\definecolor{thesislila}{HTML}{53257F}

\definecolor{thesislightcyan}{HTML}{81C9CE}
\definecolor{thesiscyan}{HTML}{3B969C}
\definecolor{thesisdarkcyan}{HTML}{1F6367}

\definecolor{thesislightpink}{HTML}{E48BBF}
\definecolor{thesispink}{HTML}{CF308D}
\definecolor{thesisdarkpink}{HTML}{8C2160}

\colorlet{thesisgrayverylight}{gray!3}
\colorlet{thesisgraylight}{gray!10}
\colorlet{thesisgrayverydark}{gray!40}
\colorlet{thesisgrayveryverydark}{gray}

\colorlet{representationFreedomColor}{black!50}
\colorlet{physicalFreedomColor}{black!80}
\colorlet{representationNodeFill}{representationFreedomColor!10}
\colorlet{physicalNodeFill}{physicalFreedomColor!29}

\colorlet{witnessColor}{thesisred}
\colorlet{refuteColor}{thesisdarkpink}
\colorlet{refineColor}{thesisdarkcyan}
\colorlet{decideColor}{thesislila}
\colorlet{refuteFill}{thesislightpink}
\colorlet{refineFill}{thesislightcyan}
\colorlet{decideFill}{thesislightlila}

\newtcolorbox[auto counter]{equivalenceproblem}[1]{
  enhanced,
  title={#1},
  fonttitle=\small\bfseries,
  fontupper=\small,
  before upper={
    \setlength{\abovedisplayskip}{4pt}
    \setlength{\belowdisplayskip}{4pt}
    \setlength{\abovedisplayshortskip}{2pt}
    \setlength{\belowdisplayshortskip}{2pt}
  },
  coltitle=black,
  colbacktitle=thesisgraylight,
  colback=thesisgrayverylight,
  colframe=thesisgrayveryverydark,
  boxrule=0.4pt,
  titlerule=0.4pt,
  arc=1pt,
  boxsep=0pt,
  left=5pt,
  right=5pt,
  top=3pt,
  bottom=3pt,
  toptitle=2pt,
  bottomtitle=2pt,
  before skip=0.7\baselineskip,
  after skip=0.7\baselineskip
}

\newtcolorbox{methodrolebox}[2][]{
  enhanced,
  title={#2},
  fonttitle=\small\mdseries\upshape,
  halign title=center,
  fontupper=\small,
  coltitle=black,
  colbacktitle=thesisgraylight,
  colback=thesisgrayverylight,
  colframe=thesisgrayveryverydark,
  boxrule=0.4pt,
  titlerule=0.4pt,
  arc=1pt,
  boxsep=0pt,
  left=5pt,
  right=5pt,
  top=3pt,
  bottom=3pt,
  toptitle=3pt,
  bottomtitle=1pt,
  before skip=0pt,
  after skip=0pt,
  #1
}

\newlength{\findingtreeindent}\setlength{\findingtreeindent}{8pt}
\newlength{\findingtreedrop}\setlength{\findingtreedrop}{7.5pt}
\newlength{\findingtreetick}\setlength{\findingtreetick}{4pt}
\newif\iffindingfirst
\NewTColorBox{findingpartbox}{s O{}}{
  enhanced,
  blank,
  fontupper=\small,
  left=\dimexpr\findingtreeindent+\findingtreetick+4pt\relax,
  right=5pt,
  top=3pt,
  bottom=3pt,
  before skip=0pt,
  after skip=0pt,
  overlay={
    \iffindingfirst
      \draw[thesisgrayveryverydark, line width=0.4pt]
        (frame.north west) -- (frame.north east);
    \fi
    \draw[thesisgrayveryverydark, line width=0.5pt]
      ([xshift=\findingtreeindent]frame.north west)
      -- ([xshift=\findingtreeindent, yshift=-\findingtreedrop]frame.north west)
      -- ++(\findingtreetick,0);
    \IfBooleanF{#1}{
      \draw[thesisgrayveryverydark, line width=0.5pt]
        ([xshift=\findingtreeindent]frame.north west)
        -- ([xshift=\findingtreeindent]frame.south west);
    }
  },
  #2
}
\newcommand{\findingpart}[1]{%
  \begin{findingpartbox}#1\end{findingpartbox}%
  \findingfirstfalse
}
\newcommand{\lastfindingpart}[1]{%
  \begin{findingpartbox}*[bottom=4pt]#1\end{findingpartbox}%
  \findingfirstfalse
}
\newtcolorbox{findingbox}[1]{
  enhanced,
  title={\textbf{Finding}\enspace\itshape #1},
  fonttitle=\small,
  fontupper=\small,
  coltitle=black,
  colbacktitle=thesisgraylight,
  colback=thesisgrayverylight,
  colframe=thesisgrayveryverydark,
  boxrule=0.4pt,
  titlerule=0pt,
  arc=1pt,
  boxsep=0pt,
  left=0pt,
  right=0pt,
  top=0pt,
  bottom=0pt,
  lefttitle=5pt,
  righttitle=5pt,
  toptitle=2pt,
  bottomtitle=2pt,
  before skip=0.7\baselineskip,
  after skip=0.7\baselineskip,
  before upper={\findingfirsttrue}
}

\colorlet{bandone}{gray!15}
\colorlet{bandtwo}{gray!30}
\colorlet{bandthree}{gray!45}
\newcommand{\band}[2][bandone]{{\setlength{\fboxsep}{1.2pt}\colorbox{#1}{$\begin{smallmatrix}#2\end{smallmatrix}$}}}
\newcommand{\bandmatrix}[1]{{\renewcommand{\arraystretch}{0.72}\begin{array}{@{}c@{}}#1\end{array}}}
\newcommand{\mapmark}[2][bandone]{{\setlength{\fboxsep}{1pt}\colorbox{#1}{$#2$}}}

\newacronym{qecc}{QECC}{Quantum Error-Correction Code}
\newacronym{qec}{QEC}{Quantum Error Correction}
\newacronym{css}{CSS}{Calderbank-Shor-Steane}
\newacronym{lc}{LC}{local Clifford}
\newacronym{rref}{RREF}{Row Reduced Echelon Form}
\newacronym{kls}{KLS}{Khesin-Lu-Shor}
\newacronym{hk}{HK}{Hu-Khesin}
\newacronym{rq}{RQ}{Research Question}
\newacronym{lse}{LSE}{Linear System of Equations}
\newacronym{oom}{OOM}{Out-of-Memory}
\newacronym{bb}{BB}{Bivariate Bicycle}
\newacronym[longplural={Degrees of Freedom},shortplural={DOF}]{dof}{DOF}{Degrees of Freedom}

\begin{abstract}
  Quantum error-correction code and circuit construction is an active field of research, motivating automated methods for identifying equivalent designs.
  We develop and evaluate automated strategies for deciding whether two descriptions of quantum error-correction codes are equivalent under specified transformations. We formalize permutation equivalence for stabilizer and CSS codes and local-Clifford equivalence for stabilizer codes within a unified framework that separates physical from representation freedom. We develop three decision strategies---group-action search, isomorphism-based methods, and constraint solving---together with invariants and signatures for preprocessing. The empirical evaluation shows that no paradigm dominates across equivalence notions, code families, and parameter ranges. We therefore combine their complementary strengths in input-dependent hybrid checkers following a staged \emph{Refute--Refine--Decide} architecture. 
  Mean hybrid runtimes remain below $10\,\mathrm{s}$ for most evaluated structured-code cases with $n\leq37$; for inequivalent code instances with $n=144$, they are about $310\,\mathrm{s}$ for permutation and $16\,\mathrm{s}$ for local-Clifford equivalence.
\end{abstract}

\newcommand{\steaneDiagram}[3]{%
    \begin{tikzpicture}[
        baseline=(current bounding box.center),
        scale=0.62,
        line join=round,
        tile edge/.style={line width=0.9pt},
        qubit anchor/.style={circle, draw=#1, fill=white, line width=0.8pt,
            minimum size=4.8mm, inner sep=0pt, font=\small},
        gate label/.style={font=\scriptsize, text=thesisred,
            inner sep=0pt, anchor=south west}
    ]
        \coordinate (top) at (0,3.0);
        \coordinate (left) at (-2.6,-1.5);
        \coordinate (right) at (2.6,-1.5);
        \coordinate (leftmid) at (-1.7333,0);
        \coordinate (rightmid) at (0.8667,1.5);
        \coordinate (bottommid) at (0.8667,-1.5);
        \coordinate (center) at (0,0);

        \filldraw[tile edge,fill=zxorangemuted,draw=zxorangedark]
            (top) -- (rightmid) -- (center) -- (leftmid) -- cycle;
        \filldraw[tile edge,fill=zxgreen!55,draw=zxgreendark]
            (left) -- (leftmid) -- (center) -- (bottommid) -- cycle;
        \filldraw[tile edge,fill=thesisblue!50,draw=thesisblue]
            (right) -- (bottommid) -- (center) -- (rightmid) -- cycle;

        \foreach \p/\n in {#2}{
            \node[qubit anchor] at (\p) {\n};
        }
        #3
    \end{tikzpicture}%
}
\newcommand{\steaneExOne}{%
    \steaneDiagram{black}{
        top/4, leftmid/6, rightmid/5, center/7,
        left/2, bottommid/3, right/1
    }{}
}

\newcommand{\iconInvariants}[1][0.12]{%
    \begin{tikzpicture}[
        baseline=(current bounding box.center),
        line cap=round,
        line join=round
    ]
        \fill
            (-0.760, 0.310) -- (-0.675, 0.087) --
            (-0.437, 0.075) -- (-0.622,-0.078) --
            (-0.560,-0.308) -- (-0.760,-0.178) --
            (-0.960,-0.308) -- (-0.898,-0.078) --
            (-1.083, 0.075) -- (-0.845, 0.087) -- cycle;

        \node[inner sep=0pt, outer sep=0pt, font=\huge]
            at (0,#1) {$\neq$};

        \fill
            (0.680,-0.320) rectangle (0.840,0.320)
            (0.440,-0.080) rectangle (1.080,0.080);
    \end{tikzpicture}%
}
\newcommand{\iconSignatures}{%
    \begin{tikzpicture}[
        baseline=(current bounding box.center),
        signature edge/.style={draw=black, line width=0.1pt,
                              line cap=round}
    ]
        \draw[signature edge] (-0.78, 0.38) -- (-0.26,-0.38);
        \draw[signature edge] (-0.78, 0.38) -- ( 0.26,-0.38);
        \draw[signature edge] (-0.26, 0.38) -- (-0.78,-0.38);
        \draw[signature edge] ( 0.26, 0.38) -- (-0.26,-0.38);
        \draw[signature edge] ( 0.26, 0.38) -- ( 0.26,-0.38);
        \draw[signature edge] ( 0.78, 0.38) -- ( 0.78,-0.38);

        \fill (-0.91, 0.25) rectangle (-0.65, 0.51);
        \fill (-0.26, 0.56) -- (-0.42, 0.29) -- (-0.10, 0.29) -- cycle;
        \fill ( 0.13, 0.25) rectangle ( 0.39, 0.51);
        \fill ( 0.78, 0.38) circle[radius=0.14cm];

        \fill (-0.78,-0.20) -- (-0.94,-0.47) -- (-0.62,-0.47) -- cycle;
        \fill (-0.39,-0.51) rectangle (-0.13,-0.25);
        \fill ( 0.13,-0.51) rectangle ( 0.39,-0.25);
        \fill ( 0.78,-0.38) circle[radius=0.14cm];
    \end{tikzpicture}%
}
\newcommand{\iconGroupActionSearch}{%
    \begin{tikzpicture}[
        baseline=(current bounding box.center),
        search edge/.style={draw=black!82, line width=1.1pt,
                           line cap=round, line join=round}
    ]
        \draw[search edge] (0,0.55) -- (-0.48,0.05);
        \draw[search edge] (0,0.55) -- ( 0.48,0.05);
        \draw[search edge] (-0.48,0.05) -- (-0.82,-0.50);
        \draw[search edge] (-0.48,0.05) -- ( 0.00,-0.50);

        \fill[black!30] ( 0.00, 0.55) circle[radius=0.22cm];
        \fill[black!30] (-0.48, 0.05) circle[radius=0.22cm];
        \fill[black!30] ( 0.48, 0.05) circle[radius=0.22cm];
        \fill[black!30] (-0.82,-0.50) circle[radius=0.22cm];
        \fill[black!30] ( 0.00,-0.50) circle[radius=0.22cm];
    \end{tikzpicture}%
}
\newcommand{\iconIsomorphism}{%
    \begin{tikzpicture}[
        baseline=(current bounding box.center),
        graph edge/.style={draw=black!82, line width=1.05pt,
                          line cap=round}
    ]
        \coordinate (leftCenter) at (-0.62,-0.48);
        \coordinate (middleCenter) at (0.00,-0.48);
        \coordinate (rightCenter) at (0.62,-0.48);

        \draw[graph edge] (leftCenter)   -- (-0.80,0.18);
        \draw[graph edge] (leftCenter)   -- (-0.38,0.10);
        \draw[graph edge] (middleCenter) -- (-0.35,0.10);
        \draw[graph edge] (middleCenter) -- ( 0.00,0.15);
        \draw[graph edge] (middleCenter) -- ( 0.38,0.10);
        \draw[graph edge] (rightCenter)  -- ( 0.80,0.18);

        \path[draw=black!30, fill=black!30, line width=0.7pt,
              line cap=round, line join=round]
            (-0.96,0.20)
            .. controls (-1.00,0.36) and (-0.86,0.45) .. (-0.72,0.42)
            .. controls (-0.70,0.60) and (-0.45,0.69) .. (-0.31,0.54)
            .. controls (-0.18,0.78) and ( 0.22,0.78) .. ( 0.34,0.55)
            .. controls ( 0.50,0.69) and ( 0.78,0.60) .. ( 0.77,0.42)
            .. controls ( 0.92,0.45) and ( 1.04,0.34) .. ( 0.98,0.20)
            .. controls ( 0.94,0.08) and ( 0.75,0.04) .. ( 0.64,0.13)
            .. controls ( 0.50,-0.01) and ( 0.27,0.00) .. ( 0.16,0.12)
            .. controls ( 0.02,-0.01) and (-0.24,-0.01) .. (-0.34,0.13)
            .. controls (-0.49,0.01) and (-0.72,0.04) .. (-0.77,0.16)
            .. controls (-0.84,0.10) and (-0.95,0.10) .. cycle;

        \fill[black!82] (leftCenter) circle[radius=0.14cm];
        \fill[black!82] (middleCenter) circle[radius=0.14cm];
        \fill[black!82] (rightCenter) circle[radius=0.14cm];
    \end{tikzpicture}%
}
\newcommand{\iconConstraintSolving}{%
    \begin{tikzpicture}[
        baseline=(current bounding box.center),
        line cap=round,
        line join=round
    ]
        \draw[black!82, line width=0.9pt] (-0.35,0) -- (0,0);
        \draw[black!30, line width=0.9pt] (0,0) -- (0.35,0);
        \fill[black!82] (-0.85,-0.25) rectangle (-0.35,0.25);
        \fill[black!30] (0.35,-0.25) rectangle (0.85,0.25);
    \end{tikzpicture}%
}
\DeclareRobustCommand{\methodTitleIcons}[2]{%
    \texorpdfstring{%
        \nobreak\hspace{0.35em}%
        \raisebox{1mm}{\resizebox*{!}{#1}{#2}}%
    }{}%
}

\definecolor{pmBlue}{HTML}{2E61F1}

\newcommand{\pmTextScale}{1.3}
\newlength{\pmU}\setlength{\pmU}{\pmTextScale pt}
\newcommand{\pmSetTextScale}[1]{\renewcommand{\pmTextScale}{#1}\setlength{\pmU}{#1pt}}
\newcommand{\pmFontLS}[3]{%
  \fontsize{\dimexpr#2\dimexpr#1pt\relax\relax}%
           {\dimexpr#3\dimexpr#2\dimexpr#1pt\relax\relax\relax}\selectfont}
\newcommand{\pmFont}[2]{\pmFontLS{#1}{#2}{1.2}}

\pgfplotsset{
  parameter map/.style 2 args={
    anchor=south west, scale only axis,
    xmin=2.5, xmax=#2+0.5, ymin=0.5, ymax=#2+0.5,
    axis lines*=left, axis on top,
    axis line style={black, line width=0.7pt},
    tick align=outside, major tick length=3.5pt,
    tick style={black, line width=0.8pt},
    tick label style={font=\pmFont{7}{#1}, inner sep=0pt},
    xticklabel shift=3.5pt, yticklabel shift=3.5pt,
    xlabel={Physical qubits $n$}, ylabel={Stabilizer generators $r$},
    label style={font=\pmFont{9}{#1}, inner sep=0pt},
    xlabel shift=6pt, ylabel shift=6.5pt,
    title style={at={(0.5,1)}, anchor=base, inner sep=0pt, font=\pmFont{10}{#1}},
    every axis title shift=6pt,
  },
}

\pgfplotsset{
  colormap={runtime}{
    rgb255=(253,250,200) rgb255=(255,217,178) rgb255=(243,147,133)
    rgb255=(233,52,26) rgb255=(167,37,18)},
}

\pgfdeclareverticalshading{pmruntime}{100bp}{%
  rgb(0.000bp)=(0.99216,0.98039,0.78431); %
  rgb(25.000bp)=(0.99216,0.98039,0.78431); %
  rgb(37.500bp)=(1.00000,0.85098,0.69804); %
  rgb(50.000bp)=(0.95294,0.57647,0.52157); %
  rgb(62.500bp)=(0.91373,0.20392,0.10196); %
  rgb(75.000bp)=(0.65490,0.14510,0.07059); %
  rgb(100.000bp)=(0.65490,0.14510,0.07059)}

\tikzset{
  pm oom/.style={cross out, draw=pmBlue, line width=1.1pt,
                 minimum size=5.83pt, inner sep=0pt, outer sep=0pt},
  pm oom key/.style={pm oom, line width=1pt, minimum size=6pt},
  pm cell/.style={draw=none},
  pm outline/.style={line width=0.5pt},            
}

\newcommand{\pmEmptyTriangle}[2][47]{%
  \pgfmathsetmacro{\pmEdge}{#1+0.5}%
  \fill[#2] (2.5,0.5) -- (\pmEdge,0.5) -- (\pmEdge,\pmEdge)
    \foreach \n [evaluate=\n as \l using \n-0.5,
                 evaluate=\n as \t using \n+0.5] in {#1,...,3} {
      -- (\l,\t) -- (\l,\l)}
    -- cycle;
}

\def\pmSplitRGB#1,#2,#3;{\def\pmR{#1}\def\pmG{#2}\def\pmB{#3}}
\newcommand{\pmPartCorners}[4]{%
  \pgfmathsetmacro{\pmXa}{#1-0.5+#3/#4}%
  \pgfmathsetmacro{\pmXb}{#1-0.5+(#3+1)/#4}%
  \pgfmathsetmacro{\pmYa}{#2-0.5}%
  \pgfmathsetmacro{\pmYb}{#2+0.5}%
}
\newcommand{\pmFillPart}[6]{%
  \pgfplotscolormapaccess[0:1]{#6}{#5}%
  \expandafter\pmSplitRGB\pgfmathresult;%
  \pmPartCorners{#1}{#2}{#3}{#4}%
  \edef\pmDo{\noexpand\filldraw[pm cell, fill={rgb,1:red,\pmR;green,\pmG;blue,\pmB}]
    (\pmXa,\pmYa) rectangle (\pmXb,\pmYb);}%
  \pmDo
}
\newcommand{\pmColorPart}[5]{%
  \pmPartCorners{#1}{#2}{#3}{#4}%
  \edef\pmDo{\noexpand\filldraw[pm cell, fill=#5] (\pmXa,\pmYa) rectangle (\pmXb,\pmYb);}%
  \pmDo
}
\newcommand{\pmOutlinePart}[5]{%
  \pmPartCorners{#1}{#2}{#3}{#4}%
  \edef\pmDo{\noexpand\draw[pm outline, draw=#5] (\pmXa,\pmYa) rectangle (\pmXb,\pmYb);}%
  \pmDo
}

\newcommand{\pmRuntimeCell}[3]{\pmFillPart{#1}{#2}{0}{1}{runtime}{#3}}

\section{Introduction}
Quantum information is highly susceptible to noise and errors. Protecting it for reliable computation has driven the development of \gls{qec}, which faces challenges beyond those of its classical counterpart \cite{qecc_intro, gottesman_og, nielsen_chuang}.

As \gls{qec} remains an active research field, several \gls{qec} code constructions and descriptions coexist, making their comparison and evaluation challenging. In particular, it may be nontrivial to determine whether two representations describe inherently distinct codes or equivalent ones, as their relationship can be obscured.

Determining whether two \gls{qec} codes are equivalent under specified transformations is therefore of practical importance. It can reveal related results in the literature and redundant database entries, support the assessment of new codes, and enable resources developed for one code to be reused for an equivalent one.

Automating such a comparison is nontrivial, as an algorithm must account for freedoms in the code representation and the physical transformations defined by the equivalence notion. We therefore distinguish representation freedom from physical freedom and formalize three equivalence problems: permutation equivalence for general stabilizer and specifically CSS codes, and local-Clifford equivalence for stabilizer codes.
Using an interdisciplinary approach that draws on classical coding theory \cite{sendrier, feulner}, graph theory \cite{prac_gi,gi_quasipolynomial, effic_algo_graphs}, and established decision problems \cite{code_eq_isomorphism,matroid_theory,cook_sat}, we develop and implement four complementary technique families for automated equivalence checking:
\begin{itemize}[label=\raisebox{0.2ex}{\tiny$\bullet$}, itemsep=1pt, topsep=2pt, parsep=0pt, partopsep=0pt]
    \item Invariants and signatures,
    \item Group-action search,
    \item Isomorphism-based methods, and
    \item Constraint solving.
\end{itemize}
We then empirically evaluate these strategies across the problems under consideration to assess their practical applicability, showing that none dominates across all settings. Instead, their complementary behaviors yield guidelines for selecting a suitable strategy for a given problem instance. Based on these guidelines, we develop input-dependent
\begin{itemize}[label=\raisebox{0.2ex}{\tiny$\bullet$}, itemsep=1pt, topsep=2pt, parsep=0pt, partopsep=0pt]
    \item Hybrid Equivalence Checkers
\end{itemize}
that combine the techniques within a \emph{Refute--Refine--Decide} architecture. Consequently, our work provides both a practical tool integrated into the \emph{Munich Quantum Toolkit (MQT)} \cite{mqt} and theoretical guidelines, serving as a foundation for further research and a resource for researchers working with \gls{qec} codes.

Following the background in \Cref{sec:background} and the motivation in \Cref{sec:motivation}, \Cref{sec:problems,sec:related-work} formalize the considered problems and review related work. \Cref{sec:strategies,sec:hybrids} present the strategies and hybrid checkers, which are evaluated in \Cref{sec:evaluation} before the discussion in \Cref{sec:discussion}.

\section{Background}\label{sec:background}
This section provides the necessary background on \emph{\glspl{qecc}}, their defining components, relevant subfamilies, and possible representations.
Throughout this work, a $\llbracket n, k, d\rrbracket$ \gls{qecc}, or simply \emph{code}, is characterized by its \emph{code space} $\codeSpace$, which is a subspace of the $n$-qubit Hilbert space $\hilbertSpace{n}$, and a chosen \emph{encoding isometry} $\hilbertSpace{k} \to \hilbertSpace{n}$ that maps logical information to physical qubits \cite{gottesman_og}.

\subsection{Stabilizer Codes}
This work focuses on stabilizer codes, described using the \emph{stabilizer formalism} \cite{gottesman_stab_og}, which is a framework for representing and manipulating \glspl{qecc}. From the single-qubit Pauli operators and identity $\sigma_0 = \operatorname{I}, \sigma_1 = \operatorname{X}, \sigma_2 = \operatorname{Y}, \sigma_3 = \operatorname{Z}$, the $n$-qubit \emph{Pauli operators} are constructed by
\[
P = i^\phi \bigotimes\limits_{j=0}^{n-1} \sigma_{p_j} \enspace \text{ with } \phi,p_j \in \{0,1,2,3\}.
\]
The \emph{Pauli group} $\pauliGroup{n}$ is the set of all $n$-qubit Pauli operators~\cite{nielsen_chuang}. A \emph{stabilizer group} $\stabGroup$ is a commutative subgroup of $\pauliGroup{n}$ with a nontrivial common $+1$ eigenspace of its elements~\cite{qecc_intro, gottesman_og}. It can be specified by $r$ independent \emph{stabilizer generators}.

A \emph{stabilizer code} is a \gls{qecc} whose code space $\codeSpace$ is defined by the common $+1$ eigenspace of a stabilizer group $\stabGroup$ with $r = n - k$ generators~\cite{qecc_intro}. Its encoding isometry can be specified by Pauli operators satisfying canonical commutation relations \cite{qecc_intro, gottesman_og, gottesman_stab_og}, i.e.,
\[
(\overline{\operatorname{X}}_i, \overline{\operatorname{Z}}_i) \in \pauliGroup{n} \enspace \text{ for } i \in [k]
\] 
representing logical $\operatorname{X}$ and $\operatorname{Z}$ operators on logical qubit $i$~\cite{gottesman_og}. They commute with the stabilizers but do not belong to the stabilizer group themselves. We call such an ordered set of tuples a \emph{logical framework}; it fixes one of the possible encodings of the code space.

\subsubsection{CSS Codes}
\emph{\gls{css} codes} are stabilizer codes whose generators can be chosen to be exclusively $\operatorname{X}$-type or $\operatorname{Z}$-type~\cite{nielsen_chuang}. This generally makes them easier to construct, analyze, and implement than general stabilizer codes \cite{nielsen_chuang}.
Moreover, several widely used \gls{qecc} families, including color codes \cite{coco} and bivariate bicycle codes \cite{bb}, naturally belong to the \gls{css} class.

\subsubsection{Graph States}
\emph{Stabilizer states} $\ket{C}$ are stabilizer codes that encode no logical qubits ($k=0$)~\cite{lc_transform}. Hence, their code spaces are one-dimensional and specify a single quantum state up to global phase. In the stabilizer formalism, they are described by $n$ independent stabilizer generators.
\emph{Graph states} $\ket{G}$ are stabilizer states whose $n$ generators can be described by an undirected graph $G = (V, E)$ with $n$ vertices, i.e.,
\[
\operatorname{X}_v \prod\limits_{\{u,v\} \in E} \operatorname{Z}_u \enspace \text{ with } v \in V.
\]

\subsection{Clifford Operations}
The $n$-qubit \emph{Clifford group} $\restrcliffordGroup{n}$ consists of the unitary operators that preserve the Pauli group under conjugation.
Up to global phases, it is precisely the finite group generated by the three \emph{Clifford operations}, or \emph{Cliffords}, $\operatorname{H}$, $\operatorname{S}$, and $\operatorname{CNOT}$ \cite{gottesman_og}.

\emph{Local}---non-entangling---Cliffords play a central role in this work, and are generated by $\operatorname{H}$ and $\operatorname{S}$ operations. In the phase-free context, their single-qubit actions are represented modulo Pauli conjugations by $\{\operatorname{I},\operatorname{H},\operatorname{S},\operatorname{SH},\operatorname{HS},\operatorname{HSH}\}$.

\subsection{Binary Symplectic Representation}
\subsubsection{Fundamentals} 
The \emph{binary symplectic representation} maps the Pauli group to binary vectors $\pauliGroup{n} \to \binField{2n}$, separating the $X$ and $Z$ components of a Pauli operator while discarding its phase~\cite{gottesman_og}. Hence, this phase-free representation identifies stabilizer codes related by Pauli operations. Pauli multiplication modulo phase corresponds to binary vector addition, while the \emph{symplectic product} of two representations determines whether the Paulis commute.

\subsubsection{Code Representation}
As the stabilizer formalism describes \glspl{qecc} using Pauli operators, the binary symplectic representation can express stabilizer codes as binary matrices. A code space is represented indirectly by its stabilizer group: the binary representations of $r$ independent generators form the rows of an $\binField{r \times 2n}$ matrix called the code's \emph{stabilizer} or \emph{symplectic tableau}. Stabilizer-code subclasses may have distinct tableau structures, as shown in \Cref{tab:stabilizer-tableaus}. A \gls{css} code can be represented compactly by two check matrices, $H_X \in \binField{r_x \times n}$ and $H_Z \in \binField{r_z \times n}$; a graph state requires only the adjacency matrix $\Gamma \in \binField{n \times n}$ of its underlying graph. A code's encoding map is characterized by the binary symplectic forms of its logical operators.

\begin{table}[t]
  \centering
  \caption{Binary symplectic tableaus of the considered code classes; $H_X$ and $H_Z$ are the check matrices of a \gls{css} code; $\Gamma$ is the adjacency matrix of a graph state.}
  \label{tab:stabilizer-tableaus}
  \small
  \renewcommand{\arraystretch}{1.6}
  \begin{tabular}{ccc}
    \toprule
    \makecell{\textbf{General} \\ \textbf{stabilizer codes}} & \textbf{CSS codes} & \textbf{Graph states} \\
    \midrule
    $\left[\,X \,\middle|\, Z\,\right]$
      & $\left[\,\begin{smallmatrix} H_X \\ 0 \end{smallmatrix} \,\middle|\,\begin{smallmatrix} 0 \\ H_Z \end{smallmatrix}\,\right]$
      & $\left[\,I \,\middle|\, \Gamma \,\right]$ \\
    \bottomrule
  \end{tabular}
\end{table}

\subsubsection{Clifford Representation}
The action of a Clifford operation on a code space is reflected by conjugating the stabilizer group by that operation. In the binary symplectic representation, this transformation is represented by a symplectic matrix $Q \in \symplGroup{2n}$, a $2n \times 2n$ binary matrix that acts on a symplectic vector or tableau and preserves its symplectic inner product. A local Clifford operation \cite{lc_transform, symp_iso_stab_codes} is represented by a \emph{local symplectic matrix} $Q \in \locsymplGroup{2n}$, a symplectic matrix with diagonal blocks $A, B, C, D$ in the form
\[
\begin{aligned}
&Q = \left[\,\begin{smallmatrix} A \\ B \end{smallmatrix} \,\begin{smallmatrix} C \\ D \end{smallmatrix}\,\right] \in \symplGroup{2n} \text{ with }\\
&Q_i = \left[\,\begin{smallmatrix} a_{ii} \\ b_{ii} \end{smallmatrix}\, \begin{smallmatrix} c_{ii} \\ d_{ii} \end{smallmatrix}\,\right] \in \symplGroup{2}.
\end{aligned}
\]

\section{Motivation}\label{sec:motivation}

In \gls{qecc} research, distinct published constructions or realizations may define structurally identical codes, even when this is not apparent from their representations.
In particular, codes related by local Clifford operations or a relabeling of physical qubits share the same abstract fault-tolerant properties, although these transformations can obscure their relationship.

\begin{example}[Steane Code]
    \Cref{fig:motivation-steane-ex} shows three illustrative realizations of the Steane code; two of them obtained based on the original version, one by relabeling the vertices and the other by additionally applying local Clifford operations during construction.
    Given only the extracted stabilizer generators---the content within the red boxes---the relationship between the original and transformed codes is not immediately apparent.
\newcommand{\stabilizerHighlight}[2]{%
    \begingroup
    \setlength{\fboxsep}{0.8pt}%
    \colorbox{#1}{\ensuremath{\color{black}#2}}%
    \endgroup
}

\begin{figure}[htbp]
\centering
\resizebox{0.95\columnwidth}{!}{%
    \begin{tikzpicture}[
        >=Latex,
        construction/.style={inner sep=0pt, outer sep=0pt},
        group/.style={inner sep=1pt, outer sep=1pt},
        transformation arrow/.style={
            ->,
            thick,
            rounded corners=1mm,
            shorten >=1.5mm
        },
        transformation label/.style={
            font=\small,
            align=center,
            inner sep=1.5pt
        },
        correspondence node/.style={circle, draw=black, fill=white,
            line width=0.8pt, minimum size=4.8mm, inner sep=0pt, font=\small}
    ]
        \node[construction] (codePerm) at (-2.05,0) {\steaneDiagram{black}{
            top/2, leftmid/4, rightmid/3, center/1,
            left/5, bottommid/6, right/7
        }{}};
        \node[construction] (codeOne) at (3.35,0) {\steaneExOne};
        \node[construction] (codeLC) at (8.75,0) {\steaneDiagram{thesisred}{
            top/4, leftmid/6, rightmid/5, center/7,
            left/2, bottommid/3, right/1
        }{%
            \begin{scope}[overlay]
                \node[gate label] at ([xshift=3.5mm,yshift=3.5mm]top) {$\operatorname{H}$};
                \node[gate label] at ([xshift=4.8mm,yshift=3.5mm]left) {$\operatorname{H}$};
                \node[gate label] at ([xshift=3.5mm,yshift=3.5mm]right) {$\operatorname{H}$};
                \node[gate label] at ([xshift=4.8mm,yshift=3.5mm]leftmid) {$\operatorname{HS}$};
                \node[gate label] at ([xshift=3.5mm,yshift=3.5mm]rightmid) {$\operatorname{HS}$};
                \node[gate label] at ([xshift=3.5mm,yshift=3.5mm]bottommid) {$\operatorname{HS}$};
            \end{scope}
        }};

        \coordinate (leftArrowTopStart) at
            ([xshift=-1mm,yshift=4.75mm]codeOne.north);
        \coordinate (leftArrowTopEnd) at
            ([yshift=4.75mm]codePerm.north);
        \draw[transformation arrow]
            ([xshift=-1mm,yshift=1.75mm]codeOne.north)
            -- (leftArrowTopStart)
            -- node[midway, above, transformation label]
                {Relabeling}
               (leftArrowTopEnd)
            -- ([yshift=-0.5mm]codePerm.north);

        \coordinate (rightArrowTopStart) at
            ([xshift=1mm,yshift=4.75mm]codeOne.north);
        \coordinate (rightArrowTopEnd) at
            ([yshift=4.75mm]codeLC.north);
        \draw[transformation arrow]
            ([xshift=1mm,yshift=1.75mm]codeOne.north)
            -- (rightArrowTopStart)
            -- node[midway, above, transformation label]
                {Local Cliffords}
               (rightArrowTopEnd)
            -- ([yshift=-0.5mm]codeLC.north);

        \coordinate (middle1) at ($(codeOne.center)+(1.612cm,-1.395cm)$);
        \coordinate (middle2) at ($(codeOne.center)+(-1.612cm,-1.395cm)$);
        \coordinate (middle3) at ($(codeOne.center)+(0.5374cm,-1.395cm)$);
        \coordinate (middle4) at ($(codeOne.center)+(0cm,1.395cm)$);
        \coordinate (middle5) at ($(codeOne.center)+(0.5374cm,0.465cm)$);
        \coordinate (middle6) at ($(codeOne.center)+(-1.0746cm,-0.465cm)$);
        \coordinate (middle7) at ($(codeOne.center)+(0cm,-0.465cm)$);

        \coordinate (right1) at ($(codePerm.center)+(0cm,-0.465cm)$);
        \coordinate (right2) at ($(codePerm.center)+(0cm,1.395cm)$);
        \coordinate (right3) at ($(codePerm.center)+(0.5374cm,0.465cm)$);
        \coordinate (right4) at ($(codePerm.center)+(-1.0746cm,-0.465cm)$);
        \coordinate (right5) at ($(codePerm.center)+(-1.612cm,-1.395cm)$);
        \coordinate (right6) at ($(codePerm.center)+(0.5374cm,-1.395cm)$);
        \coordinate (right7) at ($(codePerm.center)+(1.612cm,-1.395cm)$);

        \foreach \side in {middle,right}{
            \foreach \n in {1,...,7}{
                \node[correspondence node] (correspondence-\side-\n)
                    at (\side\n) {\n};
            }
        }

        \foreach \n in {2,4,5,6}{
            \draw[
                thesisred,
                line width=0.9pt
            ] (correspondence-middle-\n) -- (correspondence-right-\n);
        }
        \draw[
            thesisred,
            line width=0.9pt
        ] (correspondence-middle-1)
            to[out=160,in=-10] (correspondence-right-1);
        \draw[
            thesisred,
            line width=0.9pt
        ] (correspondence-middle-3)
            to[bend left=3] (correspondence-right-3);
        \draw[
            thesisred,
            line width=0.9pt
        ] (correspondence-middle-7)
            to[bend left=5] (correspondence-right-7);

        \node[group] (groupLeft) at (-2.05,-4.25) {$\scriptstyle
            \begin{aligned}
                \langle \medspace
                    &\stabilizerHighlight{zxorangemuted!55}{\operatorname{XXXXIII}}, \\[-0.15em]
                    &\stabilizerHighlight{zxgreen!30}{\operatorname{XIIXXXI}}, \\[-0.15em]
                    &\stabilizerHighlight{thesisblue!28}{\operatorname{XIXIIXX}}, \\[-0.15em]
                    &\stabilizerHighlight{zxorangemuted!55}{\operatorname{ZZZZIII}}, \\[-0.15em]
                    &\stabilizerHighlight{zxgreen!30}{\operatorname{ZIIZZZI}}, \\[-0.15em]
                    &\stabilizerHighlight{thesisblue!28}{\operatorname{ZIZIIZZ}} \medspace
                \rangle
            \end{aligned}$};

        \node[group] (groupOne) at (3.35,-4.25) {$\scriptstyle
            \begin{aligned}
                \langle \medspace
                    &\stabilizerHighlight{zxorangemuted!55}{\operatorname{IIIXXXX}}, \\[-0.15em]
                    &\stabilizerHighlight{zxgreen!30}{\operatorname{IXXIIXX}}, \\[-0.15em]
                    &\stabilizerHighlight{thesisblue!28}{\operatorname{XIXIXIX}}, \\[-0.15em]
                    &\stabilizerHighlight{zxorangemuted!55}{\operatorname{IIIZZZZ}}, \\[-0.15em]
                    &\stabilizerHighlight{zxgreen!30}{\operatorname{IZZIIZZ}}, \\[-0.15em]
                    &\stabilizerHighlight{thesisblue!28}{\operatorname{ZIZIZIZ}} \medspace
                \rangle
            \end{aligned}$};

        \node[group] (groupTwo) at (8.75,-4.25) {$\scriptstyle
            \begin{aligned}
                \langle \medspace
                    &\stabilizerHighlight{zxorangemuted!55}{\operatorname{IIIZYYX}}, \\[-0.15em]
                    &\stabilizerHighlight{zxgreen!30}{\operatorname{IZYIIYX}}, \\[-0.15em]
                    &\stabilizerHighlight{thesisblue!28}{\operatorname{ZIYIYIX}}, \\[-0.15em]
                    &\stabilizerHighlight{zxorangemuted!55}{\operatorname{IIIXXXZ}}, \\[-0.15em]
                    &\stabilizerHighlight{zxgreen!30}{\operatorname{IXXIIXZ}}, \\[-0.15em]
                    &\stabilizerHighlight{thesisblue!28}{\operatorname{XIXIXIZ}} \medspace
                \rangle
            \end{aligned}$};

        \draw[->, thick, shorten <=3mm, shorten >=2mm]
            ([yshift=2mm]codePerm.south) -- ([yshift=2mm]groupLeft.north);
        \draw[->, thick, shorten <=3mm, shorten >=2mm]
            ([yshift=2mm]codeOne.south) -- ([yshift=2mm]groupOne.north);
        \draw[->, thick, shorten <=3mm, shorten >=2mm]
            ([yshift=2mm]codeLC.south) -- ([yshift=2mm]groupTwo.north);

        \draw[thesisred, very thick, dashed]
            ([xshift=-1mm,yshift=1mm]groupLeft.north west)
            rectangle
            ([xshift=1mm,yshift=-1mm]groupOne.south east);
        \draw[thesisred, very thick, dashed]
            ([xshift=-2.2mm,yshift=2.2mm]groupOne.north west)
            rectangle
            ([xshift=2.2mm,yshift=-2.2mm]groupTwo.south east);

        \node[font=\Large, text=thesisred] at (0.65,-4.25)
            {$\overset{?}{\thicksim}$};
        \node[font=\Large, text=thesisred] at (6.05,-4.25)
            {$\overset{?}{\thicksim}$};
    \end{tikzpicture}
}
\caption{Three realizations of the Steane code: the original construction (center) and variants with relabeled qubits (left) or additional local Cliffords (right).}
\label{fig:motivation-steane-ex}
\end{figure}
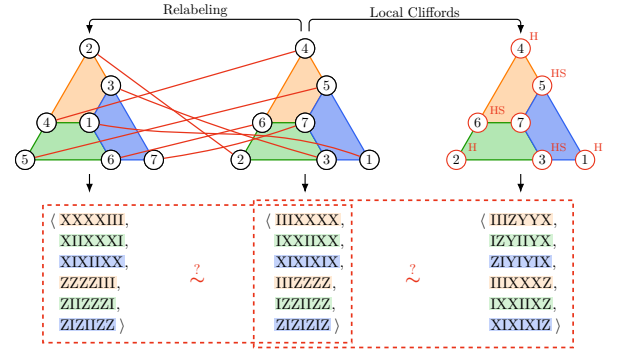
\end{example}

This ambiguity makes it difficult to recognize when results in the literature or entries in code databases are related. Additionally, the obscurity of their relationship can prevent the reuse of costly-to-construct encoding circuits \cite{aut_circ_prep,det_ft_state_prep} for an actually equivalent code.
Addressing this ambiguity in practice gives rise to two distinct needs, which---to our knowledge---prior work does not fully address in this form. First, algorithms should automate equivalence checking rather than require obscured relationships to be resolved manually. Second, systematic method selection is important in practice, as a particular algorithmic technique might be valuable for a given instance despite the problem's theoretical complexity \cite{eq_decidability}, and its usefulness can vary across code families and equivalence notions.

This work addresses these needs by formalizing the considered equivalence problems in a common framework, developing and implementing several algorithmic equivalence-checking strategies, and empirically evaluating their performance. Consequently, in addition to a practical tool, this work provides a theoretical framework that guides the choice of method for a given problem instance.

The following sections define the relevant transformations and examine their implications for practical equivalence-checking methods.

\section{Considered Equivalence Problems}\label{sec:problems}

This section formalizes the concrete equivalence problems considered in this work, so they can be solved algorithmically.

\subsection{Degrees of Freedom}
\begin{figure}[htbp]
\centering
\resizebox{0.85\columnwidth}{!}{
    \begin{tikzpicture}[
        x=1cm,
        y=1cm,
        >=Latex,
        domainbox/.style={
            rounded corners=2.5mm,
            draw,
            thick
        },
        code/.style={
            ellipse,
            draw=physicalFreedomColor,
            fill=white,
            very thick,
            minimum width=2.25cm,
            minimum height=9mm,
            font=\small
        },
        representation/.style={
            rounded corners=1.5mm,
            draw=representationFreedomColor,
            fill=white,
            thick,
            minimum width=2.25cm,
            minimum height=7mm,
            font=\small
        },
        map/.style={
            ->,
            draw=black,
            thick
        },
        freedom/.style={
            very thick,
            decorate,
            decoration={
                snake,
                amplitude=0.4mm,
                segment length=2mm
            }
        },
        freedomtip/.style={
            -{Latex[length=1.8mm, width=1.4mm]},
            very thick
        },
        label/.style={
            align=center,
            font=\footnotesize
        }
    ]
        \node[domainbox, draw=physicalFreedomColor,
              fill=physicalFreedomColor!29,
              minimum width=3.15cm, minimum height=6.35cm]
              (mathdomain) at (0,0) {};
        \node[domainbox, draw=representationFreedomColor,
              fill=representationFreedomColor!10,
              minimum width=4.70cm, minimum height=6.35cm]
              (compdomain) at (4.625,0) {};

        \node[code] (codeC)  at (0, 1.65) {Code $C$};
        \node[code] (codeCp) at (0,-1.65) {Code $C'$};

        \draw[freedom, draw=physicalFreedomColor]
            ($(codeC.south)+(4mm,-1.8mm)$) --
            ($(codeCp.north)+(4mm,1.8mm)$);
        \draw[freedomtip, draw=physicalFreedomColor]
            ($(codeC.south)+(4mm,-1.8mm)$) --
            ($(codeC.south)+(4mm,0)$);
        \draw[freedomtip, draw=physicalFreedomColor]
            ($(codeCp.north)+(4mm,1.8mm)$) --
            ($(codeCp.north)+(4mm,0)$);
        \node[anchor=east, align=right, font=\footnotesize,
              text=physicalFreedomColor]
            at (0.31,0) {Physical\\Freedom};

        \node[representation] (repC1) at (4.35, 2.35)
            {$\operatorname{repr}_{1}(C)$};
        \node[representation] (repC2) at (4.35, 0.55)
            {$\operatorname{repr}_{2}(C)$};
        \node[representation] (repCp1) at (4.35,-0.55)
            {$\operatorname{repr}_{1}(C')$};
        \node[representation] (repCp2) at (4.35,-2.35)
            {$\operatorname{repr}_{2}(C')$};

        \draw[map] (codeC.east)  -- (repC1.west);
        \draw[map] (codeC.east)  -- (repC2.west);
        \draw[map] (codeCp.east) -- (repCp1.west);
        \draw[map] (codeCp.east) -- (repCp2.west);

        \draw[freedom, draw=representationFreedomColor]
            ($(repC1.south)+(5mm,-1.8mm)$) --
            ($(repC2.north)+(5mm,1.8mm)$);
        \draw[freedomtip, draw=representationFreedomColor]
            ($(repC1.south)+(5mm,-1.8mm)$) --
            ($(repC1.south)+(5mm,0)$);
        \draw[freedomtip, draw=representationFreedomColor]
            ($(repC2.north)+(5mm,1.8mm)$) --
            ($(repC2.north)+(5mm,0)$);
        \node[anchor=east, align=right, font=\footnotesize,
              text=representationFreedomColor]
            at (4.80,1.45) {Representation\\Freedom};

        \coordinate (reprstemtop) at ($(repC1.east)+(8mm,0)$);
        \coordinate (reprstembottom) at ($(repCp2.east)+(8mm,0)$);
        \draw[draw=thesisred, very thick, dashed]
            (reprstemtop) -- (reprstembottom);
        \draw[->, draw=thesisred, very thick, dashed, shorten >=0.5pt]
            (reprstemtop) -- (repC1.east);
        \draw[->, draw=thesisred, very thick, dashed, shorten >=0.5pt]
            ($(repC2.east)+(8mm,0)$) -- (repC2.east);
        \draw[->, draw=thesisred, very thick, dashed, shorten >=0.5pt]
            ($(repCp1.east)+(8mm,0)$) -- (repCp1.east);
        \draw[->, draw=thesisred, very thick, dashed, shorten >=0.5pt]
            (reprstembottom) -- (repCp2.east);
        \node[
            anchor=west,
            inner sep=0pt,
            font=\Large,
            text=thesisred
        ] at ($(repC2.east)+(9.8mm,-5.5mm)$)
            {$\overset{!}{\thicksim}$};

        \draw[freedom, draw=representationFreedomColor]
            ($(repCp1.south)+(5mm,-1.8mm)$) --
            ($(repCp2.north)+(5mm,1.8mm)$);
        \draw[freedomtip, draw=representationFreedomColor]
            ($(repCp1.south)+(5mm,-1.8mm)$) --
            ($(repCp1.south)+(5mm,0)$);
        \draw[freedomtip, draw=representationFreedomColor]
            ($(repCp2.north)+(5mm,1.8mm)$) --
            ($(repCp2.north)+(5mm,0)$);
        
        \draw[->, black, very thick]
            ($(mathdomain.north east)+(-5mm,0)$) to[bend left=30]
            node[midway, above, label] {Computational Encoding}
            ($(compdomain.north west)+(5mm,0)$);

        \node[font=\small\bfseries, text=physicalFreedomColor]
            at (0,-3.48) {Mathematical Domain};
        \node[font=\small\bfseries, text=representationFreedomColor]
            at (4.625,-3.48) {Computational Domain};
    \end{tikzpicture}
}
\caption{Representation freedom (light gray) relates encodings of the same code; physical freedom (dark gray) relates equivalent codes. A checker must recognize their combined equivalence relation (red).}
\label{fig:degrees-of-freedom}
\end{figure}
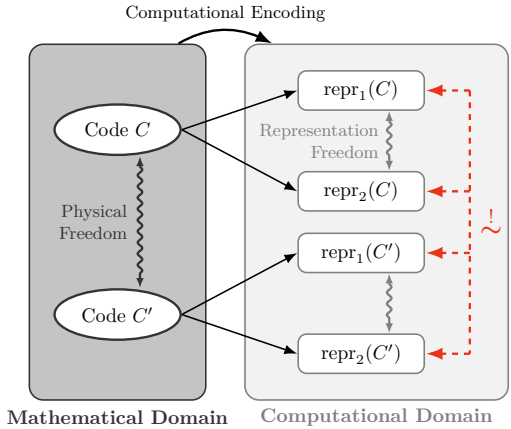

An equivalence-checking algorithm must determine whether two \glspl{qecc} are related by certain allowed transformations. Two independent sources of apparent non-uniqueness complicate this task. First, the same abstract code may have several computational representations. Second, the chosen equivalence notion may identify physically distinct code spaces. As illustrated in \Cref{fig:degrees-of-freedom}, we refer to these sources as
\begin{enumerate}[topsep=3pt, partopsep=0pt]
    \item \emph{Representation Freedom}\\\textcolor{representationFreedomColor}{(equivalent representations of one \gls{qecc})}
    \item \emph{Physical Freedom}\\\textcolor{physicalFreedomColor}{(equivalent \glspl{qecc} under a specific notion)}
\end{enumerate}

These freedoms act at different levels. Representation transformations change only the stored description while leaving the underlying code fixed. Physical transformations instead map one distinct code to another which the chosen equivalence notion declares equivalent. Consequently, equivalent inputs need not be identical representations or describe identical codes. A checker must recognize compatible transformations of both types, as shown in red in \Cref{fig:degrees-of-freedom}. The following subsections formalize these two \glspl{dof} and the transformations considered in this work.

\subsubsection{Representation Freedom}
Automated reasoning requires encoding the relevant information of an abstract \gls{qecc}---here, specifically its code space---from the \emph{mathematical domain} into a representation amenable to algorithmic processing in the \emph{computational domain}, as indicated by the solid black arrows in \Cref{fig:degrees-of-freedom}. When an abstract mathematical object $C$ is encoded into a data representation, arbitrary choices introduce non-uniqueness. As a result, the same object $C$ may yield distinct but equivalent representations $\operatorname{repr}_{1}(C)$ and $\operatorname{repr}_{2}(C)$. This \emph{representation freedom} is independent of the physical notion of code equivalence, but is rooted in data abstraction.

\begin{definition}[Generator-Basis Freedom] \emph{Generator-basis freedom} constitutes the freedom to choose among multiple generator bases that represent the same stabilizer group. 
Concretely, generators may be reordered or multiplied without changing the group they generate.
When the code space $\codeSpace$ of a stabilizer code is represented using the binary symplectic representation, the equivalence problem is conveniently mapped to \emph{linear algebra}, which is particularly well suited for algorithmic treatment. However, as part of the representation process, a \emph{concrete generator basis} of the stabilizer group has to be chosen arbitrarily. Thus, instead of comparing the resulting matrices directly, their row spaces must be compared, as changing the generator basis corresponds to applying invertible \emph{row operations}.
\end{definition}

\begin{example}[Generator Bases of the Steane Code]
\Cref{fig:steane-repr-freedom-ex} illustrates several generator-basis representations of the same Steane code. Although equivalent, their tableaus can differ substantially in apparent complexity.

\newcommand{\matrixZero}{\textcolor{thesisgrayverydark}{0}}
\newcommand{\representationLabel}[1]{%
    \mbox{\Large\ensuremath{#1}}%
}
\newcommand{\continuousMatrixRowHighlight}[2]{%
    \multicolumn{7}{c}{%
        \begingroup
        \setlength{\fboxsep}{0.7pt}%
        \colorbox{#1}{%
            \ensuremath{\begin{array}{@{}ccccccc@{}}#2\end{array}}%
        }%
        \endgroup
    }%
}

\begin{figure}[htbp]
\centering
\resizebox{\columnwidth}{!}{%
    \begin{tikzpicture}[
        >=Latex,
        construction/.style={inner sep=0pt, outer sep=0pt},
        representation/.style={
            inner xsep=3pt,
            inner ysep=2pt,
            font=\large
        },
        representation arrow/.style={
            ->,
            thick,
            shorten <=1.5mm,
            shorten >=1.5mm
        },
        row swap arrow/.style={
            <->,
            >={Latex[length=1.5mm,width=1.5mm]},
            draw=thesisred,
            line width=1.25pt,
            rounded corners=1mm
        },
        row add arrow/.style={
            ->,
            >={Latex[length=1.5mm,width=1.5mm]},
            draw=thesisred,
            line width=1.25pt,
            rounded corners=1mm
        },
        basis squiggle/.style={
            <->,
            >={Latex[length=3mm,width=2mm]},
            draw=thesisred,
            line width=0.95pt,
            shorten <=1mm,
            shorten >=1mm,
            decorate,
            decoration={
                snake,
                amplitude=0.3mm,
                segment length=2mm,
                pre length=2mm,
                post length=2mm
            }
        }
    ]
        \node[construction] (steane) at (0,0) {\scalebox{0.90}{\steaneExOne}};

        \coordinate (representationAxis) at (9.8,0);

        \node[representation, anchor=center] (reprOne)
            at ([yshift=3cm]representationAxis) {$
            \representationLabel{H_X = H_Z =}\;
            \left[
            \begin{array}{ccccccc}
                \matrixZero & \matrixZero & \matrixZero & 1 & 1 & 1 & 1 \\
                \matrixZero & 1 & 1 & \matrixZero & \matrixZero & 1 & 1 \\
                1 & \matrixZero & 1 & \matrixZero & 1 & \matrixZero & 1 \\
            \end{array}
            \right]
        $};

        \node[representation, anchor=east] (reprTwo)
            at ([xshift=-3mm]representationAxis) {$
        \representationLabel{H_X =}\;
            \left[
            \begin{array}{ccccccc}
                \matrixZero & \matrixZero & \matrixZero & 1 & 1 & 1 & 1 \\
                \continuousMatrixRowHighlight{thesisred!5}{%
                    1 & \matrixZero & 1 & \matrixZero & 1 & \matrixZero & 1
                } \\
                \continuousMatrixRowHighlight{thesisred!5}{%
                   \matrixZero & 1 & 1 & \matrixZero & \matrixZero & 1 & 1
                } \\
            \end{array}
            \right]
        $};

        \node[representation, anchor=west] (reprTwoHz)
            at ([xshift=3mm]representationAxis) {$
            \representationLabel{H_Z =}\;
            \left[
            \begin{array}{ccccccc}
                \continuousMatrixRowHighlight{thesisred!5}{%
                    \matrixZero & 1 & 1 & 1 & 1 & \matrixZero & \matrixZero
                } \\
                \matrixZero & 1 & 1 & \matrixZero & \matrixZero & 1 & 1 \\
                1 & \matrixZero & 1 & \matrixZero & 1 & \matrixZero & 1 \\
            \end{array}
            \right]
        $};

        \draw[row swap arrow]
            ([xshift=-1mm]reprTwo.east)
            -- ([xshift=3mm]reprTwo.east)
            -- ([xshift=3mm,yshift=-5mm]reprTwo.east)
            -- ([xshift=-1mm,yshift=-5mm]reprTwo.east);

        \draw[row add arrow]
            ([xshift=-1mm]reprTwoHz.east)
            -- ([xshift=3mm]reprTwoHz.east)
            -- node[midway, right=0.6mm, inner sep=0pt,
                    font=\small, text=thesisred] {$+$}
               ([xshift=3mm,yshift=5mm]reprTwoHz.east)
            -- ([xshift=-1mm,yshift=5mm]reprTwoHz.east);

        \node[representation, anchor=center] (reprThree)
            at ([yshift=-2.7cm]representationAxis) {$
            \representationLabel{\left[\,X \,\middle|\, Z\,\right] =}\;
            \left[
            \begin{array}{ccccccc|ccccccc}
                \matrixZero & 1 & 1 & 1 & 1 & \matrixZero & \matrixZero &
                1 & \matrixZero & 1 & \matrixZero & 1 & \matrixZero & 1 \\
                1 & 1 & \matrixZero & \matrixZero & 1 & 1 & \matrixZero &
                \matrixZero & 1 & 1 & \matrixZero & \matrixZero & 1 & 1 \\
                \matrixZero & \matrixZero & \matrixZero & \matrixZero & \matrixZero & \matrixZero & \matrixZero &
                \matrixZero & \matrixZero & \matrixZero & 1 & 1 & 1 & 1 \\
                \matrixZero & \matrixZero & \matrixZero & \matrixZero & \matrixZero & \matrixZero & \matrixZero &
                \matrixZero & 1 & 1 & \matrixZero & \matrixZero & 1 & 1 \\
                1 & 1 & \matrixZero & \matrixZero & 1 & 1 & \matrixZero &
                1 & 1 & \matrixZero & \matrixZero & 1 & 1 & \matrixZero \\
                1 & \matrixZero & 1 & \matrixZero & 1 & \matrixZero & 1 &
                \matrixZero & \matrixZero & \matrixZero & \matrixZero & \matrixZero & \matrixZero & \matrixZero
            \end{array}
            \right]
        $};

        \draw[basis squiggle]
            (reprOne.south)
            -- node[midway, right=1.2mm, font=\Large, text=thesisred]
                {\Large $R_x, R_z \in \linearGroup{3}$}
               ($(reprOne.south |- reprTwo.north)+(0,-3mm)$);

        \draw[representation arrow]
            ([yshift=6mm]steane.east) -- (reprOne.west);
        \draw[representation arrow]
            (steane.east) -- (reprTwo.west);
        \draw[representation arrow]
            ([yshift=-6mm]steane.east) -- (reprThree.west);
    \end{tikzpicture}%
}
\caption{Generator bases of the Steane code's stabilizer group related by row operations (red).}
\label{fig:steane-repr-freedom-ex}
\end{figure}
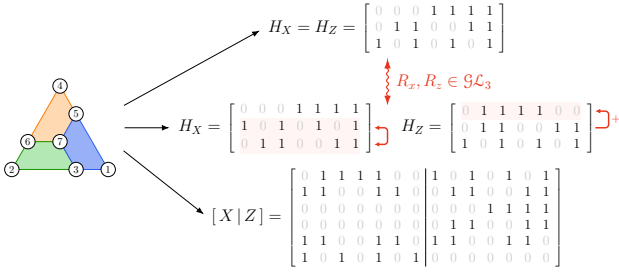
\end{example}

\begin{definition}[Logical-Framework Freedom]
\label{def:logical-framework-freedom}
\emph{Logical-framework freedom} expresses the freedom to choose among multiple logical encodings within the same code space.
When a \gls{qecc} is primarily characterized by its encoding map, its code space $\codeSpace$ is recovered as the image of that map. Based on this encoding map, a stabilizer code can be represented as a graph state, utilizing a \emph{Choi-Jamiołkowski}-based transformation detailed in \Cref{sec:code-to-state}.
Consequently, in this representation, certain \emph{graph algorithms} can be used to check equivalence. However, in the course of the representation process, an arbitrary \emph{logical framework}---a concrete encoding of the code space---is fixed. For equivalence notions that depend solely on the code space, the encoding maps must be compared through their images rather than as maps themselves. Logical frameworks differ precisely by precomposing \emph{arbitrary Clifford operations} with the encoding map, which has to be accounted for in the equivalence-checking process.
\end{definition}

\begin{example}[Logical Frameworks of the Steane Code]
The logical framework 
\[
\big(\bar{\operatorname{X}}_1 = XXXXXXX, \bar{\operatorname{Z}}_1 = ZZZZZZZ\big)
\]
fixes a specific encoding of the Steane code's code space, whereas the framework
\[
\big(\bar{\operatorname{X}}_1 = ZZZZZZZ, \bar{\operatorname{Z}}_1 = XXXXXXX\big)
\]
fixes a different encoding map, related by an $H$ operation on the \emph{logical} qubit. As both distinct maps have the same image, as shown in \Cref{fig:steane-log-freedom-ex}, they describe the same code space. 

\begin{figure}[htbp]
\centering
\resizebox{\columnwidth}{!}{%
    \begin{tikzpicture}[
        >=Latex,
        font=\normalfont,
        space/.style={
            draw=black,
            rounded corners=1.5mm,
            line width=0.7pt
        },
        framework space/.style={
            space,
            minimum width=2.35cm,
            minimum height=1.15cm
        },
        encoding/.style={
            ->,
            thick,
            shorten <=1mm,
            shorten >=1mm
        },
        framework change/.style={
            -{Latex[length=2mm,width=1.5mm]},
            draw=thesisred,
            line width=0.8pt,
            shorten <=0.2mm,
            shorten >=0.2mm
        }
    ]
        \node[framework space] (logical) at (0,-0.2) {};
        \node[font=\normalfont\scriptsize, anchor=north west]
            at ([xshift=1mm,yshift=-0.8mm]logical.north west)
            {logical space};
        \node at ([yshift=-2mm]logical.center) {$\hilbertSpace{k}$};
        \draw[framework change, overlay]
            ([xshift=-1.8mm]logical.north)
            to[out=120,in=60,looseness=4]
            node[midway, above=-0.3mm, font=\normalfont\small, text=thesisred]
                {$U\in\restrcliffordGroup{k}$}
            ([xshift=1.8mm]logical.north);

        \node[space, minimum width=4.2cm, minimum height=2.1cm]
            (physical) at (5.0,0) {};
        \node[font=\normalfont\scriptsize, anchor=north west]
            at ([xshift=1mm,yshift=-0.8mm]physical.north west)
            {physical space $\hilbertSpace{n}$};
        \node[
            framework space,
            fill=representationNodeFill,
        ] (code) at ([yshift=-2.5mm]physical.center)
            {$\codeSpace$};

        \draw[encoding]
            ([yshift=1.8mm]logical.east)
            to[bend left=9]
            node[pos=0.30, above=0.2mm, font=\normalfont\small] {$V$}
            ([yshift=1.8mm]code.west);
        \draw[encoding]
            ([yshift=-1.8mm]logical.east)
            to[bend right=9]
            node[pos=0.30, below=0.2mm, font=\normalfont\small]
                {$V\!\circ\!\textcolor{thesisred}{U}$}
            ([yshift=-1.8mm]code.west);
    \end{tikzpicture}%
}
\caption{Encoding map $V:\hilbertSpace{k}\to\hilbertSpace{n}$ and its precomposition with a logical Clifford $U$ have the same image $\codeSpace$, but yield different logical frameworks.}
\label{fig:steane-log-freedom-ex}
\end{figure}
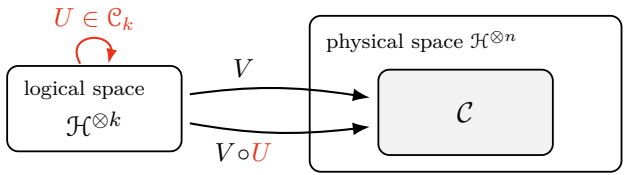
\end{example}

\subsubsection{Physical Freedom}
An equivalence notion specifies which physical transformations between \glspl{qecc} are allowed. We consider notions defined solely on code spaces: two codes are equivalent if and only if an allowed transformation maps one code space onto the other. Unlike a representation transformation, a physical transformation generally changes the code space itself. Because a stabilizer group $\stabGroup$ uniquely determines its code space $\codeSpace$, these transformations can also be characterized directly on the stabilizers.

\begin{definition}[Permutation Equivalence]
Two codes are \emph{permutation-equivalent} ($C \thicksim_{\text{P}} C'$) if and only if they have the same code space up to a permutation of the physical qubits. This corresponds to relabeling the physical qubits, as shown on the left in \Cref{fig:motivation-steane-ex}.
\[
\textcolor{thesisred}{\exists  \pi \in \permutationGroup{n}:} \enspace \codeSpace \medspace = \medspace \{ \medspace \textcolor{thesisred}{\pi(}\ket{\psi}\textcolor{thesisred}{)} \medspace \mid \medspace \ket{\psi} \in \codeSpace' \medspace \}
\]
On the stabilizers, this permutes the single-qubit Pauli factors within each tensor product.
\end{definition}

\begin{definition}[Local-Clifford Equivalence]
Two codes are \emph{local-Clifford-equivalent} ($C \thicksim_{\text{LC}} C'$) if and only if they have the same code space up to a local Clifford operation on the physical qubits, as shown on the right in \Cref{fig:motivation-steane-ex}.
\[
\textcolor{thesisred}{\exists U \in \restrcliffordGroup{1}^{\otimes n}:}\enspace \codeSpace \medspace = \medspace  \{ \medspace \textcolor{thesisred}{U} \ket{\psi} \medspace \mid \medspace \ket{\psi} \in \codeSpace' \medspace\}
\]
On the stabilizers, this corresponds to conjugating the Pauli operators by the Clifford operation.
\end{definition}

\subsection{Resulting Pairwise Equivalence Problems} \label{sec:pairwise-equivalence-problems}
These freedoms yield the pairwise equivalence problem
\begin{center}
\emph{Given \gls{qec} codes $C$ and $C'$, are they equivalent under the chosen notion?}
\end{center}
We instantiate this question for permutation equivalence on stabilizer and \gls{css} codes and for local-Clifford equivalence on stabilizer codes.
Using the phase-free binary symplectic representation, we disregard stabilizer phases and thus identify codes that differ additionally by Pauli conjugation, which broadens the previous definition of permutation equivalence. Under this convention, the problem takes the algebraic form
\[
 S' = R \cdot S \cdot T
\]
where the representations $S$ and $S'$ of the two codes, the physical transformation $T$, and the representation freedom encoded by an element $R$ of the linear group $\linearGroup{}$ over $\binField{}$ are specified separately for each equivalence problem:

\begin{equivalenceproblem}{Permutation Equivalence on Stabilizer Codes}
\[
\resizebox{\columnwidth}{!}{$\displaystyle
S=\left[\,X \,\middle|\, Z\,\right]\in\binField{r\times 2n}, R\in\linearGroup{r}, T=\left[\begin{smallmatrix}P&0\\0&P\end{smallmatrix}\right]\text{ for }P\in\permutationGroup{n}
$}
\]
\tcblower
\textcolor{representationFreedomColor}{\emph{Repr.\ freedom:} invertible row operations}

\textcolor{physicalFreedomColor}{\emph{Phys.\ freedom:} paired permutation of the tableau's $X$- and $Z$-columns}

\end{equivalenceproblem}

\begin{equivalenceproblem}{Permutation Equivalence on CSS Codes}
\[
\resizebox{\columnwidth}{!}{$\displaystyle
\alpha\in\{X,Z\}: \,
S_\alpha=\operatorname{H}_\alpha\in \binField{r_\alpha\times n}, R_\alpha\in\linearGroup{r_\alpha}, T=P\in\permutationGroup{n}
$}
\]
\tcblower
\textcolor{representationFreedomColor}{\emph{Repr.\ freedom:} independent invertible row operations on the separate check matrices}

\textcolor{physicalFreedomColor}{\emph{Phys.\ freedom:} common column permutation of both check matrices}
\end{equivalenceproblem}

\begin{equivalenceproblem}{Local-Clifford Equivalence on Stabilizer Codes}
\[
\resizebox{\columnwidth}{!}{$\displaystyle
  S=\left[\,X \,\middle|\, Z\,\right]\in\binField{r\times 2n}, R\in\linearGroup{r}, T = Q \in\locsymplGroup{2n}
$}
\]

\tcblower
\textcolor{representationFreedomColor}{\emph{Repr.\ freedom:} invertible row operations}

\textcolor{physicalFreedomColor}{\emph{Phys.\ freedom:} independent invertible symplectic transformations of each qubit's $X/Z$-column pair}
\end{equivalenceproblem}

\section{Related Work}\label{sec:related-work} 

Although automated solutions to the equivalence problems defined above have received limited attention, relevant techniques appear across several fields.
Work on classical code equivalence provides algorithms based on invariants, canonical representatives, and permutation-group techniques \cite{sendrier,feulner,leon}.
Complexity-theoretic approaches reduce related equivalence problems to isomorphism problems \cite{code_eq_isomorphism,eq_decidability,complexity_matroid_isomorphism,prac_gi,about_code_equivalence}.
For local-Clifford equivalence, some strategies use strong invariants \cite{fin_set_invariants,loc_inv_qb,local_invariants}, while others combine graph-state representations \cite{schlingemann,choi_zx,zx_enc_resp,elliott,opt_prep_graph_states} with normal forms \cite{eq_classes_qecc,kissinger,hk,kls_long} or graph algorithms \cite{effic_algo_states,lc_transform,small_subsystem,LC_Orbit}.
Related studies also consider equivalence of unitary operations \cite{eq_check_unit,effic_eq_check_cliff} and characterize related code-equivalence notions using binary symplectic representations \cite{symp_iso_stab_codes} or the ZX calculus \cite{eq_classes_qecc}.

In contrast, we develop algorithms tailored to the equivalence problems defined above and empirically evaluate their practical performance.
Together, the aforementioned lines of work motivate the interdisciplinary approach of the following sections, which adapts, combines, and evaluates methods from the different fields.

\section{Strategies}\label{sec:strategies}

This section presents techniques and strategies for solving the equivalence problems defined above.

\subsection{Anatomy of a Check} \label{sec:anatomy-check}
The proposed methods exhibit different correctness guarantees, characterized by soundness and completeness.
\begin{definition}[Soundness] A decision algorithm is \emph{sound} if every \textsc{Equivalent} verdict is correct; it produces no \emph{false positives}.
\end{definition}

\begin{definition}[Completeness] A decision algorithm is \emph{complete} if every \textsc{Not Equivalent} verdict is correct; it produces no \emph{false negatives}.
\end{definition}

\newcommand{\methodRoleIcon}[1]{%
  \raisebox{0.6mm}{\scalebox{0.28}{#1}}%
}
\newcommand{\methodRoleIconSlot}[1]{%
  \makebox[5.8mm][c]{\methodRoleIcon{#1}}%
}

\begin{tcbraster}[
  raster columns=7,
  raster equal height=rows,
  raster column skip=6pt,
  raster before skip=0.7\baselineskip,
  raster after skip=0.7\baselineskip
]
  \begin{methodrolebox}[raster multicolumn=3]{\textbf{Complete}}
    \emph{Roles}\par
    \textcolor{refuteColor}{Refute Equivalence}\\
    \textcolor{refineColor}{Refine Check}
    \tcblower
    \emph{Related section}\par
    \begin{itemize}[leftmargin=0.52cm, labelsep=0.10cm, itemsep=0pt, topsep=0pt, parsep=0pt]
      \item[\textcolor{black}{\scriptsize\ref*{sec:invariants-signatures}}]
        \begin{tabular}[t]{@{}c@{\hspace{0.5mm}}c@{}}
          \raisebox{0.4mm}{\methodRoleIconSlot{\iconInvariants}} &
          \hyperref[sec:invariants-signatures]{\textcolor{black}{\scriptsize Invariants}} \\[-0.5mm]
          & \hyperref[sec:invariants-signatures]{\textcolor{black}{\scriptsize and}} \\[-0.5mm]
          \methodRoleIconSlot{\iconSignatures} &
          \hyperref[sec:invariants-signatures]{\textcolor{black}{\scriptsize Signatures}}
        \end{tabular}
    \end{itemize}
  \end{methodrolebox}
  \begin{methodrolebox}[raster multicolumn=4]{\textbf{Complete + Sound}}
    \emph{Role}\par
    \textcolor{decideColor}{Decide (In)Equivalence}\\
    \phantom{Refine Check}
    \tcblower
    \emph{Related sections}\par
    \begin{itemize}[leftmargin=0.52cm, labelsep=0.10cm, itemsep=0.4mm, topsep=0pt, parsep=0pt]
      \item[\textcolor{black}{\scriptsize\ref*{sec:group-action-search}}]
        \methodRoleIconSlot{\iconGroupActionSearch}\hspace{0.5mm}%
        \hyperref[sec:group-action-search]{\textcolor{black}{\scriptsize Group-Action Search}}
      \item[\textcolor{black}{\scriptsize\ref*{sec:isomorphism-reductions}}]
        \methodRoleIconSlot{\iconIsomorphism}\hspace{0.5mm}%
        \hyperref[sec:isomorphism-reductions]{\textcolor{black}{\scriptsize Isomorphisms}}
      \item[\textcolor{black}{\scriptsize\ref*{sec:constraint-solving}}]
        \methodRoleIconSlot{\iconConstraintSolving}\hspace{0.5mm}%
        \hyperref[sec:constraint-solving]{\textcolor{black}{\scriptsize Constraint Solving}}
    \end{itemize}
  \end{methodrolebox}
\end{tcbraster}

Therefore, they can serve different roles within the equivalence-checking process. \emph{Invariants and signatures} can safely reject inequivalent instances or refine the search, although they cannot confirm equivalence. The sound and complete decision methods are distinguished by their strategies for handling the two \glspl{dof}. We consider three possible strategies: \emph{group-action search} traverses the physical transformation space and resolves representation freedom at each step; \emph{isomorphisms} remove representation freedom through suitable encodings and embed physical freedom within the resulting structures; and \emph{constraint solving} handles both freedoms simultaneously using algebraic techniques.
The following sections state explicitly how each strategy handles the two freedoms. Further details are available in the public code repository \cite{eq_check_code_repo}, where each approach is implemented.

\subsection[Invariants and Signatures]{%
  Invariants and Signatures%
  \methodTitleIcons{4.2mm}{\protect\iconInvariants\hspace{10mm}\protect\iconSignatures}%
}
\label{sec:invariants-signatures}

\paragraph{Global Invariants} An invariant is a property of a code or its representation that is preserved under the considered equivalence relation. While differing invariants prove inequivalence og two inputs, matching invariants cannot confirm their equivalence. Relying solely on invariant agreement can therefore produce false positives. The less \emph{discriminant} an invariant is, the fewer objects it can distinguish and the more false positives it produces. Its usefulness is determined by the tradeoff between computational cost and discriminative power.

\begin{figure}[H]
\centering
\resizebox{0.96\columnwidth}{!}{%
    \ensuremath{%
    \overset{\smash{\raisebox{-1mm}{{\footnotesize$\color{thesisdarkcyan}\begin{array}{cccc}
    a & b & a & c
    \end{array}$}}}}{
    \left[
    \begin{array}{cccc}
    \vdots & \vdots & \vdots & \vdots
    \end{array}
    \right]}
    \enspace
    \thicksim_P
    \enspace
    \overset{\smash{\raisebox{-1mm}{{\footnotesize$\color{thesisdarkcyan}\begin{array}{cccc}
    b & a & a & c
    \end{array}$}}}}{
    \left[
    \begin{array}{cccc}
    \vdots & \vdots & \vdots & \vdots
    \end{array}
    \right]}
    \longrightarrow
    \begin{tikzpicture}[
        baseline={([yshift=-2.5pt]current bounding box.center)},
        node distance=1.5mm and 12mm,
        entry/.style={
            draw,
            line width=0.4pt,
            rounded corners=0.6mm,
            minimum width=17mm,
            minimum height=0pt,
            inner xsep=4pt,
            inner ysep=2pt,
            align=center,
            font=\scriptsize
        },
        connection/.style={
            -{Latex[length=2mm]},
            thick,
            draw=thesisdarkcyan
        },
        faint connection/.style={
            -{Latex[length=2mm]},
            thick,
            draw=thesiscyan!12
        }
    ]
        \node[entry] (a1) {column 0};
        \node[entry, below=of a1] (a2) {column 1};
        \node[entry, below=of a2] (a3) {column 2};
        \node[entry, below=of a3] (a4) {column 3};

        \node[entry, right=of a1] (b1) {column 0};
        \node[entry, below=of b1] (b2) {column 1};
        \node[entry, below=of b2] (b3) {column 2};
        \node[entry, below=of b3] (b4) {column 3};

        \draw[faint connection] (a1.east) -- (b1.west);
        \draw[faint connection] (a1.east) -- (b4.west);
        \draw[faint connection] (a2.east) -- (b2.west);
        \draw[faint connection] (a2.east) -- (b3.west);
        \draw[faint connection] (a2.east) -- (b4.west);
        \draw[faint connection] (a3.east) -- (b1.west);
        \draw[faint connection] (a3.east) -- (b4.west);
        \draw[faint connection] (a4.east) -- (b1.west);
        \draw[faint connection] (a4.east) -- (b2.west);
        \draw[faint connection] (a4.east) -- (b3.west);

        \draw[connection] (a1.east) -- (b2.west);
        \draw[connection] (a1.east) -- (b3.west);
        \draw[connection] (a2.east) -- (b1.west);
        \draw[connection] (a3.east) -- (b2.west);
        \draw[connection] (a3.east) -- (b3.west);
        \draw[connection] (a4.east) -- (b4.west);
    \end{tikzpicture}
    }
}
\caption{Column signatures of a classical linear code: a permutation may only map columns with equal signatures onto each other, refining the search space.}
\label{fig:signatures-prep}
\end{figure}
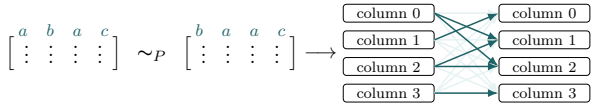

\paragraph{Signatures} Signatures are fingerprints of a coordinate or vertex, utilized in classical coding theory \cite{sendrier} and practical graph-isomorphism algorithms \cite{prac_gi}. Although not unique, they are preserved under permutations of the considered components. They can refine the permutation search space because only components with identical signatures can be mapped to one another. \Cref{fig:signatures-prep} illustrates this concept for a classical linear code, assigning each matrix column---code coordinate---a signature based on its properties. Signatures can also serve as global invariants because a mismatch immediately proves inequivalence.

\par\noindent
Because invariants are not sound, they can only be used as a fast preprocessing step within a longer equivalence-checking process by rejecting inequivalent codes, thus \emph{refuting} equivalence in some cases. Signatures are likewise not sound but can additionally reduce the number of possible permutations, thus \emph{refining} the search space for subsequent methods.

\subsubsection{Representation Freedom}
Properties invariant within representation freedom must be preserved by row operations on the stabilizer tableau or check matrix. Examples include the \emph{rank} of the full tableau or check matrix, the ranks of the $X$- and $Z$-submatrices, the number or positions of \emph{zero columns}, the multiplicities of \emph{duplicate columns}, and \emph{linear dependencies} among a fixed number of columns. All are computable in polynomial time, while the cost of finding linear dependencies can be controlled through the maximum number of columns considered.

\subsubsection{Physical Freedom}
Under both qubit permutations and local Clifford operations, the \emph{code parameters} $n$ and $k$ are preserved, inexpensive to compare, and constrain the tableau dimensions.

Defining the weight of a Pauli operator as its number of nonidentity entries, the simple \emph{weight enumerator} of the code's stabilizer group
\[
    W(\stabGroup)_i = \left|\{s \in \stabGroup \mid \operatorname{weight}(s)=i\}\right|
\]
is also an invariant under both permutations and local Cliffords.

For classical code equivalence, Sendrier \cite{sendrier} proposed as a coordinate signature the \emph{weight enumerator of the hull of the classical linear code punctured at that coordinate}.
This signature can be adapted to quantum stabilizer codes using the quaternary representation of the stabilizer tableau \cite{gf4_qecc, gf4_macaree}. For \gls{css} codes, the check-matrix representation as two classical linear codes simplifies that computation.
Because local Clifford operations preserve the symplectic inner product and Pauli weight \cite{symp_iso_stab_codes}, these qubit signatures can serve as invariants for local-Clifford equivalence as well.
However, they are most effective for permutation equivalence, for which they reduce the search space as described above.

The computational cost of these enumerators cannot be tuned; computing any adapted weight enumerator scales exponentially in the dimension of the stabilizer group or its hull.

\paragraph{Permutation Equivalence}
For permutation equivalence, the standard weight enumerator can be refined into a \emph{Pauli-sensitive weight enumerator}. Rather than grouping by total weight, it counts the occurrences of each single-qubit Pauli within a Pauli operator:
\[
\resizebox{0.9\columnwidth}{!}{$\displaystyle
    W(\stabGroup)_{(w_{\sigma_1},w_{\sigma_2},w_{\sigma_3})} \\
    = \Bigl|\Bigl\{s \in \stabGroup \;\Bigm|\;
    \begin{array}{@{}l@{}}
        \operatorname{weight}_{\sigma_j}(s)=w_{\sigma_j}\\[-2pt]
        \text{for all }j \in \{1,2,3\}
    \end{array}
    \Bigr\}\Bigr|$}.
\]

\paragraph{Local-Clifford Equivalence}
For local-Clifford-invariant properties, \cite{fin_set_invariants} proposes a \emph{local weight distribution} for quantum states that also applies to codes. Moreover, \cite{loc_inv_qb} introduces a family of \emph{local invariants} under local Clifford operations. From both invariant families a member with a low degree---such as $2$---can be chosen and have a bounded maximum subset size, making it computable in polynomial time.

\subsection[Group-Action Search]{%
  Group-Action Search%
  \methodTitleIcons{4.2mm}{\protect\iconGroupActionSearch}%
}
\label{sec:group-action-search}

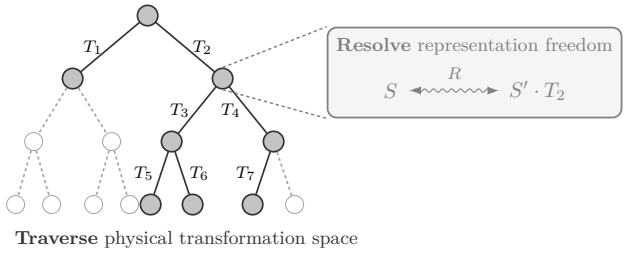
\begin{figure}[htbp]
\centering
\resizebox{\columnwidth}{!}{%
    \begin{tikzpicture}[
        x=1cm, y=1cm, >=Latex,
        visited/.style={
            circle, draw=physicalFreedomColor, fill=physicalFreedomColor!29,
            thick, inner sep=0pt, minimum size=3.4mm
        },
        skipped/.style={
            circle, draw=physicalFreedomColor!45, fill=white,
            thin, inner sep=0pt, minimum size=3mm
        },
        tedge/.style={draw=physicalFreedomColor, thick, line cap=round},
        sedge/.style={draw=physicalFreedomColor!45, thick,
                      dash pattern=on 1.5pt off 1.4pt},
        elabel/.style={font=\scriptsize, inner sep=1.2pt},
        callout/.style={draw=representationFreedomColor,
                        fill=representationFreedomColor!8,
                        rounded corners=1.5mm, line width=0.85pt},
        zoom/.style={draw=representationFreedomColor, line width=0.7pt,
                     dash pattern=on 1.6pt off 1.3pt},
        squig/.style={<->, line width=0.5pt, decorate,
                      decoration={snake, amplitude=0.28mm, segment length=1.5mm,
                                  pre length=1.4pt, post length=1.4pt}},
        lbl/.style={font=\footnotesize, align=center}
    ]
        \node[visited] (r)  at (-2.55, 1.75) {};
        \node[visited] (n1) at (-3.80, 0.70) {};
        \node[visited] (n2) at (-1.30, 0.70) {};

        \node[skipped] (a1) at (-4.45,-0.35) {};
        \node[skipped] (a2) at (-3.15,-0.35) {};
        \node[visited] (b1) at (-2.15,-0.35) {};
        \node[visited] (b2) at (-0.45,-0.35) {};

        \node[skipped] (l1) at (-4.75,-1.40) {};
        \node[skipped] (l2) at (-4.15,-1.40) {};
        \node[skipped] (l3) at (-3.45,-1.40) {};
        \node[skipped] (l4) at (-2.85,-1.40) {};
        \node[visited] (l5) at (-2.50,-1.40) {};
        \node[visited] (l6) at (-1.80,-1.40) {};
        \node[visited] (l7) at (-0.80,-1.40) {};
        \node[skipped] (l8) at (-0.10,-1.40) {};

        \draw[tedge] (r)  -- (n1) node[elabel, midway, left=2pt]  {$T_1$};
        \draw[tedge] (r)  -- (n2) node[elabel, midway, right=2pt] {$T_2$};
        \draw[tedge] (n2) -- (b1) node[elabel, midway, left=2pt]  {$T_3$};
        \draw[tedge] (n2) -- (b2) node[elabel, midway, left=2pt]  {$T_4$};
        \draw[tedge] (b1) -- (l5) node[elabel, midway, left=2pt]  {$T_5$};
        \draw[tedge] (b1) -- (l6) node[elabel, midway, right=2pt] {$T_6$};
        \draw[tedge] (b2) -- (l7) node[elabel, midway, left=2pt]  {$T_7$};

        \draw[sedge] (b2) -- (l8);
        \draw[sedge] (n1) -- (a1);  \draw[sedge] (n1) -- (a2);
        \draw[sedge] (a1) -- (l1);  \draw[sedge] (a1) -- (l2);
        \draw[sedge] (a2) -- (l3);  \draw[sedge] (a2) -- (l4);

        \node[lbl, text=physicalFreedomColor, anchor=north west, align=left]
            at (-4.90,-1.70)
            {\textbf{Traverse} physical transformation space};

        \node[callout, minimum width=4.9cm, minimum height=1.5cm]
            (co) at (2.90,0.8) {};
        \draw[zoom] (n2.north) -- (co.north west);
        \draw[zoom] (n2.south) -- (co.south west);

        \node[lbl, text=representationFreedomColor, anchor=north] at (2.90,1.52)
            {\textbf{Resolve} representation freedom};

        \node[font=\small,text=representationFreedomColor] at (1.50,0.5) {$S$};
        \draw[squig, draw=representationFreedomColor] (1.82,0.5) -- (3.30,0.5);
        \node[anchor=south, font=\scriptsize, align=center,
              text=representationFreedomColor] at (2.56,0.55)
            {$R$};
        \node[font=\small,text=representationFreedomColor] at (3.95,0.5) {$S' \cdot T_2$};
    \end{tikzpicture}%
}
\caption{Traversal of the physical transformation space (dashed branches superfluous, skipped), resolving the representation freedom at each visited transformation.}
\label{fig:group-action-search}
\end{figure}

To handle the two \glspl{dof}, this strategy uses a structured exploration of the physical transformation space as an outer layer, as visualized in \Cref{fig:group-action-search}. While exploring the action of the admissible physical transformation group---represented in dark gray---at each step, the current physical transformation is applied to one code. It is then compared to the other one by resolving only the representation freedom---shown in light gray.
Variants of this strategy differ in how they traverse the physical search space---potentially avoiding redundant paths---and handle the representation freedom.

The primary expected bottleneck is traversing the physical transformation space, as the two \glspl{dof} are separated by applying physical transformations sequentially. Consequently, the runtime is expected to scale poorly with the number of physical qubits because the physical transformation space grows factorially or exponentially with $n$.

\subsubsection{Representation Freedom}

At each search step, it is necessary to determine whether two representations are equivalent under representation transformations alone.

\paragraph{Row-space Check} A straightforward way to determine whether two stabilizer tableaus describe the same stabilizer group is to verify the mutual inclusion of their row spaces $\operatorname{rs}(S)$ and $\operatorname{rs}(S')$. This is tested by stacking the matrices vertically, row-reducing, and comparing ranks:
\[
\operatorname{rk}(S) = \operatorname{rk}(\left[\begin{smallmatrix} S \\ S' \end{smallmatrix}\right]) = \operatorname{rk}(S') \medspace \implies \medspace \operatorname{rs}(S) = \operatorname{rs}(S'),
\]
so this check runs in polynomial time.

\paragraph{Canonicalization} Representation freedom can also be handled by transforming each representation into a canonical form and comparing the resulting forms for exact equality. For stabilizer tableaus or check matrices, one may choose a lexicographic ordering of the entries with a unique minimum and apply canonicalization algorithms similar to \cite{feulner}.
Using the encoding-based graph-state representation, both graph states must be transformed into a unique canonical form under arbitrary Clifford operations on the input nodes. The normal form proposed in \cite{kls_long} satisfies exactly that property, and can be computed in polynomial time: two encoding-respecting graph states describe the same code space if and only if their normal forms agree.

\paragraph{Exhaustive Search} A naive strategy for resolving representation freedom is to enumerate all representations of the same mathematical object. For an adapted graph-state representation \cite{elliott}, this corresponds to checking all possible logical frameworks---equivalently, all $k$-qubit Clifford operations---and requires superexponential time in $k$ \cite{gottesman_og}.

\subsubsection{Physical Freedom}

This strategy focuses on traversing the physical transformation space. The key question is how to traverse it efficiently despite its poor scaling---i.e., how to align physical freedom compactly along the time axis.

\paragraph{Exhaustive Search} 
The straightforward approach is to iterate over all $n!$ permutations or $6^n$ local Clifford operations and apply them to the code's symplectic representation. This baseline brute-force method traverses the entire search tree unless it succeeds earlier.

\paragraph{Upfront Reduction} To avoid traversing superfluous physical transformations, information can be computed beforehand to restrict the search space even before traversing it. For permutations, one example is the \emph{automorphism group} $\operatorname{Aut}(\codeSpace)$ \cite{matrix_aut, aut_hao} of a code space. Because every automorphism $a$ satisfies $a(\codeSpace)=\codeSpace$, all permutations in the same right coset $\pi \circ \operatorname{Aut}(\codeSpace)$ produce the same image of $\codeSpace$. It therefore suffices to examine one representative per coset.
Although this can reduce the search space when the automorphism group is given, computing the group is nontrivial as well.
The signatures described above likewise provide an upfront search-space reduction.

\paragraph{Dynamic Pruning} 
A related method dynamically prunes the search tree when it encounters a redundant or unpromising branch. Feulner \cite{feulner} combined such pruning with representation canonicalization to obtain a practical equivalence-checking algorithm for classical codes.\\
For the graph-state representation, dynamic pruning corresponds to a breadth-first \emph{orbit traversal}, discarding previously visited graph states to avoid redundant paths.
The purpose of the representation of a stabilizer code as a graph state is to enable the use of certain graph algorithms for equivalence checking. This is possible because a local Clifford operation on a graph state corresponds to a \emph{local complementation} of the underlying graph at the relevant node \cite{effic_algo_states,rec_loc_equiv_graphs}. Local complementation is a polynomial-time transformation that complements the subgraph induced by the node's neighborhood, as visualized in \Cref{fig:local-complementation}.
The \emph{local-Clifford orbit} of a graph state is the set of graph states reachable by sequences of local Clifford operations, or equivalently local complementations \cite{LC_Orbit, opt_prep_graph_states}, and its size can be exponential in the number of qubits \cite{lc_orbit_size,LC_Orbit}.

\newcommand{\locComplGraph}[1]{%
    \begin{tikzpicture}[
        line cap=round,
        line join=round,
        thick,
        every node/.style={font=\large},
        vertex/.style={circle, draw, minimum size=7mm, inner sep=0pt},
        selected vertex/.style={vertex, draw=thesislila,
                                fill=thesislila!35, text=thesislila},
        neighbor vertex/.style={vertex, draw=thesislightlila,
                                fill=thesislightlila!15, text=thesislightlila},
        wire/.style={draw, thick}
    ]
        \node[selected vertex] (v0) at (0,1.6) {0};
        \node[neighbor vertex] (v1) at (1.5,0.5) {1};
        \node[neighbor vertex] (v2) at (0.9,-1.2) {2};
        \node[vertex] (v3) at (-0.9,-1.2) {3};
        \node[neighbor vertex] (v4) at (-1.5,0.5) {4};

        \draw[wire, thesislila] (v0) -- (v1);
        \draw[wire, thesislila] (v0) -- (v2);
        \draw[wire, thesislila] (v0) -- (v4);
        #1
        \draw[wire] (v1) -- (v3);
        \draw[wire] (v3) -- (v4);
    \end{tikzpicture}%
}

\begin{figure}[H]
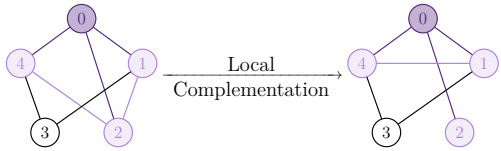

\centering
\resizebox{0.8\columnwidth}{!}{%
    $\displaystyle
    \vcenter{\hbox{\locComplGraph{%
        \draw[wire, thesislightlila] (v2) -- (v4);
        \draw[wire, thesislightlila] (v1) -- (v2);
    }}}
    \xrightarrow[\enspace\text{\Large Complementation}\enspace]{\text{\Large Local}}
    \vcenter{\hbox{\locComplGraph{%
        \draw[wire, thesislightlila] (v1) -- (v4);
    }}}
    $%
}
\caption{Local complementation at node $0$: complementing the subgraph induced by its neighborhood (light lilac).}
\label{fig:local-complementation}
\end{figure}

\subsection[Isomorphisms]{%
  Isomorphisms%
  \methodTitleIcons{4.2mm}{\protect\iconIsomorphism}%
}
\label{sec:isomorphism-reductions}

This approach removes representation freedom by design, yielding a representation independent of arbitrary choices but at greater cost.
Physical freedom is encoded within the resulting structure, converting both physical and representation transformations into a unified paradigm, reducing code equivalence to a graph-isomorphism problem \cite{prac_gi,gi_quasipolynomial, code_eq_isomorphism}. Specifically, a colored graph $G$ is constructed from each code such that
\[
C \thicksim_{\; \text{P},\text{LC}} C' \iff G_{\text{P},\text{LC}}(C) \cong G_{\text{P},\text{LC}}(C'),
\]
and the equivalence check is delegated to a practical graph-isomorphism algorithm.
The overall cost also includes constructing the graph and depends strongly on the size of the resulting representation.
Thus, explicitly encoding representation freedom is expected to make both runtime and size scale poorly with the representation cost.

\begin{example}[Graph Constructions]
    \Cref{fig:full-stabilizer-perm-example-graph,fig:full-stabilizer-lc-example-graph,fig:matroid-example-graph} illustrate the graph constructions for different equivalence relations and code families. They use the \gls{css} code with the stabilizer group $\stabGroup = \langle \operatorname{IXXX},\ \operatorname{ZZIZ} \rangle = \{\operatorname{IIII}, \operatorname{IXXX}, \operatorname{ZZIZ}, -\operatorname{ZYXY}\}$, disregarding global phases, as running example.
\end{example}

\subsubsection{Representation Freedom}
\paragraph{Full-Group Encoding} A naive encoding removes the generator-basis freedom by representing \emph{every} stabilizer-group element as a node rather than selecting generators. Assigning these nodes a common color permits arbitrary orderings, as shown by the light-gray nodes in \Cref{fig:full-stabilizer-perm-example-graph,fig:full-stabilizer-lc-example-graph}. This direct encoding closely reflects the abstract stabilizer group, but its cost is exponential in the group rank $r$, making it infeasible for large codes but potentially useful for small ones.

\paragraph{Matroid-Circuit Encoding}
A \emph{matroid} $M = (\groundSet, \independence)$ consists of a finite \emph{ground set} $\groundSet$ and a collection $\independence \subseteq \powerset{\groundSet}$ of \emph{independent subsets} satisfying specific axioms, thereby generalizing the notion of independence \cite{matroid_theory}. A binary linear code induces a matroid whose ground set comprises the code coordinates and whose independent sets index linearly independent columns of a representation matrix. Its \emph{circuits}---not to be confused with quantum or hardware circuits---are the unique minimal dependent sets and are preserved under matroid isomorphisms \cite{code_eq_isomorphism, matroid_theory, complexity_matroid_isomorphism, qc_symp_matroid}.
A colored graph listing every circuit therefore uniquely represents the matroid---and thus a binary check matrix of a \gls{css} code---independently of its basis. The lighter nodes in \Cref{fig:matroid-example-graph} represent circuits of $H_X$ and $H_Z$, distinguished by color.
This shifts the exponential dependency of computation and representation costs from the stabilizer rank $r$ to the number of physical qubits $n$.

\subsubsection{Physical Freedom}

Given a basis-independent representation, certain vertex structures and colors encode the allowed physical transformations.

\begin{figure}[htbp]
\centering
\resizebox{0.6\columnwidth}{!}{
    \begin{tikzpicture}[
        >=Stealth,
        topnode/.style={
            ellipse, draw=representationFreedomColor, fill=representationNodeFill,
            thick, minimum width=11mm, minimum height=7mm, inner sep=0pt,
            font=\normalsize\bfseries
        },
        bottomnode/.style={
            circle, draw=physicalFreedomColor, fill=physicalNodeFill,
            thick, minimum size=7mm, inner sep=0pt,
            font=\normalsize\bfseries
        },
        orangeedge/.style={draw=zxorangedark, very thick},
        greenedge/.style={draw=zxgreendark, very thick}
    ]
    \node[topnode] (t0) at (0,2.3) {$\operatorname{IIII}$};
    \node[topnode] (t1) at (1.8,2.3) {$\operatorname{IXXX}$};
    \node[topnode] (t2) at (3.6,2.3) {$\operatorname{ZZIZ}$};
    \node[topnode] (t3) at (5.4,2.3) {$-\operatorname{ZYXY}$};

    \node[bottomnode] (b0) at (0,0) {$0$};
    \node[bottomnode] (b1) at (1.8,0) {$1$};
    \node[bottomnode] (b2) at (3.6,0) {$2$};
    \node[bottomnode] (b3) at (5.4,0) {$3$};

    \draw[orangeedge] (t1) -- (b1);
    \draw[orangeedge] (t1) -- (b2);
    \draw[orangeedge] (t1) -- (b3);
    \draw[orangeedge] (t3) to[bend left=4] (b1);
    \draw[orangeedge] (t3) -- (b2);
    \draw[orangeedge] (t3) to[bend left=4] (b3);

    \draw[greenedge] (t2) -- (b0);
    \draw[greenedge] (t2) -- (b1);
    \draw[greenedge] (t2) -- (b3);
    \draw[greenedge] (t3) -- (b0);
    \draw[greenedge] (t3) to[bend right=4] (b1);
    \draw[greenedge] (t3) to[bend right=4] (b3);
\end{tikzpicture}
}
\caption{Full-group construction of the running example for permutation equivalence: stabilizer elements as interchangeable representation nodes (light gray), qubits as interchangeable physical nodes (dark gray), edges for $X$ (orange) and $Z$ (green) support.}
\label{fig:full-stabilizer-perm-example-graph}
\end{figure}
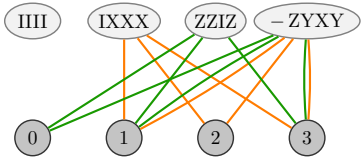

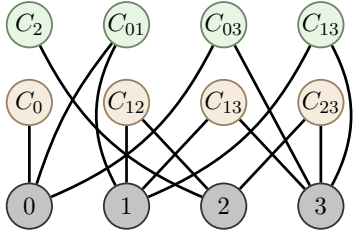
\begin{figure}[htbp]
\centering
\resizebox{0.6\columnwidth}{!}{
    \begin{tikzpicture}[
        >=Stealth,
        midnode/.style={
            circle, draw=representationFreedomColor!80!zxorangedark,
            fill=representationNodeFill!70!zxorangemuted,
            thick, minimum size=7mm, inner sep=0pt,
            font=\normalsize\bfseries
        },
        bottomnode/.style={
            circle, draw=physicalFreedomColor, fill=physicalNodeFill,
            thick, minimum size=7mm, inner sep=0pt,
            font=\normalsize\bfseries
        },
        topnode/.style={
            circle, draw=representationFreedomColor!80!zxgreendark,
            fill=representationNodeFill!70!zxgreenmuted,
            thick, minimum size=7mm, inner sep=0pt,
            font=\normalsize\bfseries
        },
        graphedge/.style={draw, very thick}
    ]
    \node[topnode] (t3) at (0,2.8) {$C_{2}$};
    \node[topnode] (t12) at (1.5,2.8) {$C_{01}$};
    \node[topnode] (t14) at (3.0,2.8) {$C_{03}$};
    \node[topnode] (t24) at (4.5,2.8) {$C_{13}$};

    \node[midnode] (m1) at (0,1.6) {$C_{0}$};
    \node[midnode] (m23) at (1.5,1.6) {$C_{12}$};
    \node[midnode] (m24) at (3.0,1.6) {$C_{13}$};
    \node[midnode] (m34) at (4.5,1.6) {$C_{23}$};

    \node[bottomnode] (b1) at (0,0) {$0$};
    \node[bottomnode] (b2) at (1.5,0) {$1$};
    \node[bottomnode] (b3) at (3.0,0) {$2$};
    \node[bottomnode] (b4) at (4.5,0) {$3$};

    \draw[graphedge] (m1) -- (b1);
    \draw[graphedge] (m23) -- (b2);
    \draw[graphedge] (m23) -- (b3);
    \draw[graphedge] (m24) -- (b2);
    \draw[graphedge] (m24) -- (b4);
    \draw[graphedge] (m34) -- (b3);
    \draw[graphedge] (m34) -- (b4);

    \draw[graphedge] (t3) to[bend right=20] (b3);
    \draw[graphedge] (t12) to[bend right=10] (b1);
    \draw[graphedge] (t12) to[bend right=28] (b2);
    \draw[graphedge] (t14) to[bend left=20] (b1);
    \draw[graphedge] (t14) -- (b4);
    \draw[graphedge] (t24) to[bend left=20] (b2);
    \draw[graphedge] (t24) to[bend left=28] (b4);
\end{tikzpicture}
}
\caption{Matroid-circuit construction of the running example: circuits of $H_X$ (orange) and $H_Z$ (green) as interchangeable representation nodes (lighter), qubits as interchangeable physical nodes (dark gray), edges for circuit membership.}
\label{fig:matroid-example-graph}
\end{figure}

\paragraph{Permutation Equivalence} Uniformly colored qubit nodes permit arbitrary qubit permutations within the graph isomorphism, as shown by the dark-gray nodes in \Cref{fig:full-stabilizer-perm-example-graph,fig:matroid-example-graph}. Edges connecting representation and qubit nodes encode incidence, such as a stabilizer element's support or a qubit's membership in a circuit.

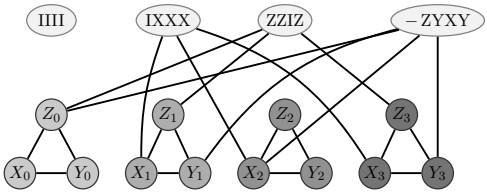
\begin{figure}[htbp]
\centering
\resizebox{0.8\columnwidth}{!}{
    \begin{tikzpicture}[
            topnode/.style={
                ellipse, draw=representationFreedomColor, fill=representationNodeFill,
                thick, minimum width=11mm, minimum height=7mm, inner sep=0pt,
                font=\normalsize\bfseries
            },
            qubitnode/.style={
                circle, draw=physicalFreedomColor,
                thick, minimum size=7mm, inner sep=0pt,
                font=\normalsize\bfseries
            },
            nodeA/.style={qubitnode, fill=physicalFreedomColor!26},
            nodeB/.style={qubitnode, fill=physicalFreedomColor!40},
            nodeC/.style={qubitnode, fill=physicalFreedomColor!54},
            nodeD/.style={qubitnode, fill=physicalFreedomColor!68},
            graphedge/.style={draw, very thick},
        ]
        \node[topnode] (t0) at (0.9,3.55) {$\operatorname{IIII}$};
        \node[topnode] (t1) at (3.5,3.55) {$\operatorname{IXXX}$};
        \node[topnode] (t2) at (6.1,3.55) {$\operatorname{ZZIZ}$};
        \node[topnode] (t3) at (9.5,3.55) {$-\operatorname{ZYXY}$};

        \node[nodeA] (x0) at (0.2,0.15) {$X_0$};
        \node[nodeA] (y0) at (1.6,0.15) {$Y_0$};
        \node[nodeB] (x1) at (2.9,0.15) {$X_1$};
        \node[nodeB] (y1) at (4.1,0.15) {$Y_1$};
        \node[nodeC] (x2) at (5.4,0.15) {$X_2$};
        \node[nodeC] (y2) at (6.8,0.15) {$Y_2$};
        \node[nodeD] (x3) at (8.1,0.15) {$X_3$};
        \node[nodeD] (y3) at (9.5,0.15) {$Y_3$};

        \node[nodeA] (z0) at (0.9,1.45) {$Z_0$};
        \node[nodeB] (z1) at (3.5,1.45) {$Z_1$};
        \node[nodeC] (z2) at (6.1,1.45) {$Z_2$};
        \node[nodeD] (z3) at (8.7,1.45) {$Z_3$};

        \draw[graphedge] (t1) to[bend right=10] (x1);
        \draw[graphedge] (t1) -- (x2);
        \draw[graphedge] (t1) to[bend left=20] (x3);
        \draw[graphedge] (t2) -- (z0);
        \draw[graphedge] (t2) -- (z1);
        \draw[graphedge] (t2) -- (z3);
        \draw[graphedge] (t3) -- (z0);
        \draw[graphedge] (t3) to[bend right=20] (y1);
        \draw[graphedge] (t3) -- (x2);
        \draw[graphedge] (t3) -- (y3);

        \draw[graphedge] (x0) -- (y0);
        \draw[graphedge] (x0) -- (z0);
        \draw[graphedge] (z0) -- (y0);
        \draw[graphedge] (x1) -- (y1);
        \draw[graphedge] (x1) -- (z1);
        \draw[graphedge] (z1) -- (y1);
        \draw[graphedge] (x2) -- (y2);
        \draw[graphedge] (x2) -- (z2);
        \draw[graphedge] (z2) -- (y2);
        \draw[graphedge] (x3) -- (y3);
        \draw[graphedge] (x3) -- (z3);
        \draw[graphedge] (z3) -- (y3);

    \end{tikzpicture}
}
\caption{Full-group construction of the running example for local-Clifford equivalence: stabilizer elements as interchangeable representation nodes (light gray), each physical qubit as a triangle of permutable Pauli nodes (distinctly colored dark grays), edges for the Pauli acting on each qubit.}
\label{fig:full-stabilizer-lc-example-graph}
\end{figure}

\paragraph{Local-Clifford Equivalence} A construction inspired by \cite{small_subsystem} represents each qubit by a triangle of its three nonidentity Pauli operators. As shown in \Cref{fig:full-stabilizer-lc-example-graph}, the triangle symmetry permits Pauli permutations, matching the action of local Clifford operations on a nonidentity Pauli, while a distinct color for each triangle fixes the qubit order. Edges encode which Pauli each stabilizer element applies to each qubit.

\subsection[Constraint Solving]{%
  Constraint Solving%
  \methodTitleIcons{2.4mm}{\protect\iconConstraintSolving}%
}
\label{sec:constraint-solving}

The third strategy expresses the two \glspl{dof} as a constraint-feasibility problem. Equivalence requires a choice of representation and physical transformations that act compatibly on the inputs. The search for a witness transformation can therefore be translated into a constraint satisfaction problem that characterizes compatible choices algebraically.

\begin{figure}[H]
\centering
\resizebox{0.96\columnwidth}{!}{%
    \begin{tikzpicture}[
        >=Latex,
        node distance=2.6cm,
        lbl/.style={font=\small, inner sep=1.5pt, above},
        phys/.style={draw=physicalFreedomColor, text=physicalFreedomColor, thick},
        repr/.style={draw=representationFreedomColor, text=representationFreedomColor, thick},
        physbox/.style={fill=physicalFreedomColor!29, rounded corners=1.5mm, inner sep=4pt},
        reprbox/.style={fill=representationFreedomColor!10, rounded corners=1.5mm, inner sep=4pt}
    ]
        \node (S)   [text=physicalFreedomColor] {$S$};
        \node (aux) [right=of S, inner xsep=8pt] {$A$};
        \node (Sp)  [right=of aux, text=representationFreedomColor] {$S'$};
        \draw[->, phys] (S)   -- node[lbl] (lP) {$P$ or $Q$} (aux);
        \draw[<-, repr] (aux) -- node[lbl] (lR) {$R$} (Sp);
        \coordinate (auxL) at ($(aux.west)+(1pt,0)$);
        \coordinate (auxR) at ($(aux.east)-(1pt,0)$);
        \begin{scope}[on background layer]
            \node[physbox, fit=(S) (lP) (auxL)] {};
            \node[reprbox, fit=(Sp) (lR) (auxR)] {};
        \end{scope}
    \end{tikzpicture}
}
\caption{\SAT{} encoding: auxiliary tableau $A$, from $S$ by physical transformations ($P$ or $Q$) and from $S'$ by representation transformations $R$.}
\label{fig:sat-concept}
\end{figure}
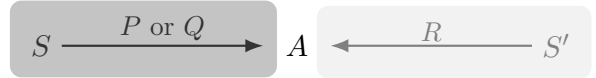

\paragraph{\SAT{} Encoding} A common approach encodes such constraints as an instance of the Boolean Satisfiability Problem (\SAT{}) \cite{cook_sat}. Using the symplectic representation and an auxiliary tableau $A$, an equivalence instance is translated into a Boolean formula that is satisfiable if and only if compatible representation and physical transformations map one code to the other.
As illustrated in \Cref{fig:sat-concept}, $A$ is obtained from $S$ by a physical transformation and from $S'$ by a representation transformation.
The resulting formula is then passed to a practical \SAT{} solver \cite{z3}.

\paragraph{Graph-State LSE} Van den Nest, Dehaene, and De Moor proposed an efficient algorithm for deciding local-Clifford equivalence of graph states~\cite{effic_algo_states}, building on graph-theoretic results about local complementation \cite{effic_algo_graphs, rec_loc_equiv_graphs}. This polynomial-time algorithm uses the two adjacency matrices to formulate a \gls{lse} over $\mathbb{F}_2$, subject to constraints that ensure a valid witness transformation. Through the aforementioned graph-state representations, it can also address local-Clifford equivalence of stabilizer codes.

\subsubsection{Representation Freedom}
\paragraph{\SAT{} Encoding} Following the formal equivalence problems in \Cref{sec:pairwise-equivalence-problems}, the encoding depicts representation freedom as a row-transformation matrix $R$ with one Boolean variable $r_{ij}$ per entry. Its action on the full stabilizer tableau or check matrices is expressed as a linear combination of rows whose coefficients are the entries of $R$:
\[
\operatorname{row}_i(A) \leftrightarrow \bigoplus_{j=0}^{r-1} \bigl( r_{ij} \cdot \operatorname{row}_j(S')\bigr).
\]

\paragraph{Graph-State LSE} The generator-basis freedom is fully captured by the \gls{lse}, but due to utilizing the graph-state representation of a stabilizer code, the logical-framework freedom of the representation would need to be additionally accounted for, as established in \Cref{def:logical-framework-freedom}. However, the rigid algorithm cannot incorporate arbitrary logical Clifford transformations and is therefore applicable only to codes with $k<2$. For $k=0$, the code already specifies a stabilizer state and has no logical-framework freedom that needs to be considered. For $k=1$, this freedom comprises a single-qubit Clifford, which is among the transformations considered by the algorithm.

\subsubsection{Physical Freedom}
\paragraph{\SAT{} Encoding} Similar to representation freedom, physical freedom is encoded by a permutation matrix $P$ with Boolean variables $p_{ij}$ or by local-Clifford indicators $c_{ij}$. In addition to constraints that ensure the physical transformation is valid, its effect on the matrix representation is expressed using implication clauses:
\[
\begin{aligned}
p_{ij} &\implies \bigl(\operatorname{qubit}_i(S) \leftrightarrow \operatorname{qubit}_j(A)\bigr) \\
c_{ij} &\implies \bigl(\operatorname{qubit}_j(S) \leftrightarrow \operatorname{qubit}_j(A)\cdot Q_i\bigr)
\end{aligned}.
\]

\paragraph{Graph-State LSE}
The algorithm in \cite{effic_algo_states} constructs a \gls{lse} over $\mathbb{F}_2$ with $4n$ unknowns to find a witness transformation $Q$, using both adjacency matrices as coefficients. Additional constraints ensure that the solution represents a valid local Clifford operation. By exploiting a simplification for large solution spaces, a constraint-satisfying solution can be found in polynomial time.

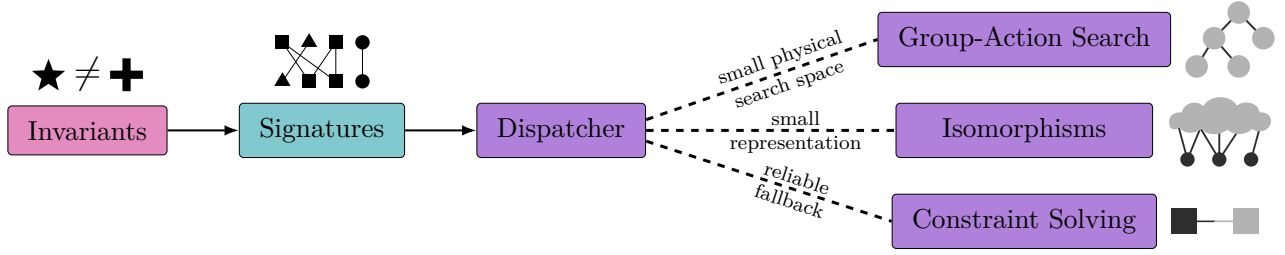
\begin{figure*}[htbp]
\centering
\resizebox{\textwidth}{!}{
    \begin{tikzpicture}[
        qecc state/.style={
            draw,
            line width=0.4pt,
            rounded corners=0.6mm,
            inner xsep=8pt,
            inner ysep=6pt,
            align=center,
            font=\normalsize
        },
        decision state/.style={
            qecc state,
            minimum width=36mm
        },
        qecc arrow/.style={-{Latex[length=2mm]}, thick},
        graphedge/.style={draw, very thick, dashed},
        edge label/.style={
            font=\scriptsize,
            inner sep=1.5pt,
            sloped
        },
        method icon/.style={
            inner sep=0pt,
            outer sep=0pt
        },
        node distance=1cm
        ]
        \node[qecc state, fill=refuteFill] (inv) {Invariants};
        \node[qecc state, right=of inv, fill=refineFill] (sig) {Signatures};
        \node[qecc state, right=of sig, fill=decideFill] (disp) {Dispatcher};
        \node[decision state, right=35mm of disp, fill=decideFill] (gi) {Isomorphisms};
        \node[decision state, above=5mm of gi, fill=decideFill] (bf) {Group-Action Search};
        \node[decision state, below=5mm of gi, fill=decideFill] (sat) {Constraint Solving};

        \node[method icon, anchor=south]
            at ([yshift=2mm]inv.north) {\scalebox{0.72}{\iconInvariants[-0.30]}};
        \node[method icon, anchor=south]
            at ([yshift=2mm]sig.north) {\scalebox{0.72}{\iconSignatures}};

        \node[method icon, anchor=west]
            at ([xshift=2mm]bf.east) {\scalebox{0.72}{\iconGroupActionSearch}};
        \node[method icon, anchor=west]
            at ([xshift=2mm]gi.east) {\scalebox{0.72}{\iconIsomorphism}};
        \node[method icon, anchor=west]
            at ([xshift=2mm]sat.east) {\scalebox{0.72}{\iconConstraintSolving}};

        \draw[qecc arrow] (inv) -- (sig);
        \draw[qecc arrow] (sig) -- (disp);
        \draw[graphedge] ($(disp.east)+(0,1.5mm)$) -- (bf.west)
            node[pos=0.6, above, edge label] {small physical}
            node[pos=0.6, below, edge label] {search space};
        \draw[graphedge] ($(disp.east)+(0,-1.5mm)$) -- (sat.west)
            node[pos=0.6, above, edge label] {reliable}
            node[pos=0.6, below, edge label] {fallback};
        \draw[graphedge] (disp) -- (gi)
            node[pos=0.6, above, edge label] {small}
            node[pos=0.6, below, edge label] {representation};

    \end{tikzpicture}
}
\caption{Refute--Refine--Decide architecture of the hybrid checkers: invariants refute (pink), signatures refine (cyan), and an input-dependent dispatcher selects the decision method (lilac).}
\label{fig:hybrid-architecture-flow}
\end{figure*}

\section{Hybrid Equivalence Checkers}\label{sec:hybrids}

The method roles established in \Cref{sec:anatomy-check} allow the individual techniques to be composed as successive stages of a hybrid equivalence checker.

As the proposed decision methods handle the two \glspl{dof} differently, their expected performance and applicability vary across problem instances, as will be validated in \Cref{sec:evaluation}. Consequently, they are \emph{no universally optimal methods for equivalence checking}. However, an input-dependent dispatcher can exploit their complementary characteristics in the decision stage.

This section proposes a staged hybrid architecture that combines these methods to improve practical performance across diverse scenarios.

\subsection{Architecture} The proposed hybrid checkers follow a \emph{\textcolor{refuteColor}{Refute}--\textcolor{refineColor}{Refine}--\textcolor{decideColor}{Decide}} architecture, assigning methods to stages according to their soundness and completeness guarantees. The flowchart in \Cref{fig:hybrid-architecture-flow} illustrates the overall process. In the decision stage, a dispatcher selects a suitable strategy based on input properties such as code dimensions, code family, and equivalence notion.
  
\subsection{Design}

The hybrid approach should be tailored to each specific equivalence problem in \Cref{sec:pairwise-equivalence-problems}, because the code family and equivalence notion determine which specialized methods apply. Nevertheless, the overall design of the hybrid checker remains similar.

\paragraph{Refute}
Invariants are used when their expected benefit justifies their cost. Cheap constant- or polynomial-time invariants are applied to all inputs, so that they reject easy inequivalent cases and verify preconditions for later stages, whereas more expensive---potentially exponentially scaling---invariants are restricted to empirically tractable instances.

\paragraph{Refine}
For permutation equivalence, signatures can further refine the search space, but their exponential cost limits them to suitable problem sizes and code families.

\paragraph{Decide}
As indicated in \Cref{fig:hybrid-architecture-flow}, group-action search is suitable while the physical search space remains small; beyond this regime, traversal becomes too time-consuming.
Isomorphism-based methods remain appropriate while their encoding size is manageable, and should be used conservatively until their representation cost becomes too high. More compact encodings of representation freedom should be preferred when supported by the code family.
For larger codes, constraint-solving methods serve as a fallback because they typically scale better in practice than traversal on large physical search spaces and incur less representation overhead than isomorphism encodings.

In the next section, we empirically validate this approach and determine the exact usability thresholds.

\section{Evaluation}\label{sec:evaluation}

This section empirically evaluates the proposed methods and hybrid approaches in terms of runtime and memory use, testing our theoretical expectations and their practical applicability.

\subsection{Experimental Methodology}

\subsubsection{Input Instances}
\paragraph{Generation} 
In addition to evaluating specific known codes, we generate seeded randomized instances of the equivalence problems, as described in \Cref{sec:instance-generation}, to cover a broad and dense range of input sizes. This allows us to assess the practical performance of each method across diverse scenarios and extract precise code parameters as usability thresholds.
An \emph{input instance} of an equivalence problem is a pair of codes $(C,C')$ from the same family that is guaranteed to be equivalent or inequivalent. Each code is represented by its stabilizer group in the binary symplectic representation appropriate to its code family.

\paragraph{Measurements} 
For each \emph{parameter setting} $(n,k)$, we generate 10 randomized instances with different seeds, measure the relevant quantities in the environment described in \Cref{sec:reproducibility}, and aggregate the results. While this provides a more robust characterization of each algorithm's behavior, it also reveals variation caused by input characteristics other than size.

\paragraph{Selection} To characterize scaling without exhaustively measuring every possible parameter setting $(n,k)$, we select a subset of possible $k$ for $n$ ranging from $3$ to $47$.

\subsubsection{Resource Limits}
A method is considered \emph{practical} for an instance if it completes within the runtime and memory budgets of $1.5\,\mathrm{h}$ and $13\,\mathrm{GiB}$, respectively. Precise runtimes are recorded only for runs that finish before the timeout.

\subsection[Invariants and Signatures]{%
  Invariants and Signatures%
  \methodTitleIcons{4.2mm}{\protect\iconInvariants\hspace{10mm}\protect\iconSignatures}%
}
\definecolor{relEmpty}{HTML}{FFFFFF} 
\definecolor{relGray}{HTML}{808080}  
\definecolor{relTimeout}{HTML}{202020}
\definecolor{relInvTimeout}{HTML}{E9341A}
\definecolor{relBlue}{HTML}{1F45B2}
\definecolor{relRed}{HTML}{A72512}
\colorlet{relPink}{thesispink}

\pgfplotsset{
  colormap={relative}{
    rgb255(0cm)=(31,69,178) rgb255(0.20209cm)=(46,97,241)
    rgb255(0.35641cm)=(148,171,234) rgb255(0.45877cm)=(200,216,248)
    rgb255(0.5cm)=(253,250,200)
    rgb255(0.54123cm)=(255,217,178) rgb255(0.64359cm)=(243,147,133)
    rgb255(0.79791cm)=(233,52,26) rgb255(1cm)=(167,37,18)},
  relative panel/.style={parameter map={\pmTextScale}{47},
                         xtick={3,5,10,15,20,25,30,35,40,47}, ytick={3,5,10,15,20,25,30,35,40,47},
                         width=3.1653in, height=3.3060in,
                         set layers},
}

\newcommand{\relCell}[5]{\pmFillPart{#1}{#2}{#3}{#4}{relative}{#5}}
\newcommand{\relBackendTimeout}[4]{\pmOutlinePart{#1}{#2}{#3}{#4}{relTimeout}}
\tikzset{rel threshold/.style={draw=relPink, line width=1.8pt}}
\newcommand{\relInvariantTimeout}[4]{\pmOutlinePart{#1}{#2}{#3}{#4}{relInvTimeout}}

\pgfdeclareverticalshading{relbar}{100bp}{%
  rgb(0.000bp)=(0.12157,0.27059,0.69804); %
  rgb(25.000bp)=(0.12157,0.27059,0.69804); %
  rgb(35.105bp)=(0.18039,0.38039,0.94510); %
  rgb(42.820bp)=(0.58039,0.67059,0.91765); %
  rgb(47.939bp)=(0.78431,0.84706,0.97255); %
  rgb(50.000bp)=(0.99216,0.98039,0.78431); %
  rgb(52.061bp)=(1.00000,0.85098,0.69804); %
  rgb(57.179bp)=(0.95294,0.57647,0.52157); %
  rgb(64.895bp)=(0.91373,0.20392,0.10196); %
  rgb(75.000bp)=(0.65490,0.14510,0.07059); %
  rgb(100.000bp)=(0.65490,0.14510,0.07059)}

\newcommand{\relKeyPatch}[1]{\tikz[baseline=0pt]{%
  \useasboundingbox (0,0) rectangle (8\pmU,8\pmU);
  \path[#1] (0,0) rectangle (8\pmU,8\pmU);}}

\newcommand{\relKeyHalf}[2]{\tikz[baseline=0pt]{%
  \useasboundingbox (0,0) rectangle (8\pmU,8\pmU);
  \fill[relGray] (#1,0) rectangle (#2,8\pmU);
  \draw[relGray, line width=1pt] (0,0) rectangle (8\pmU,8\pmU);}}

\begin{figure*}[htbp]
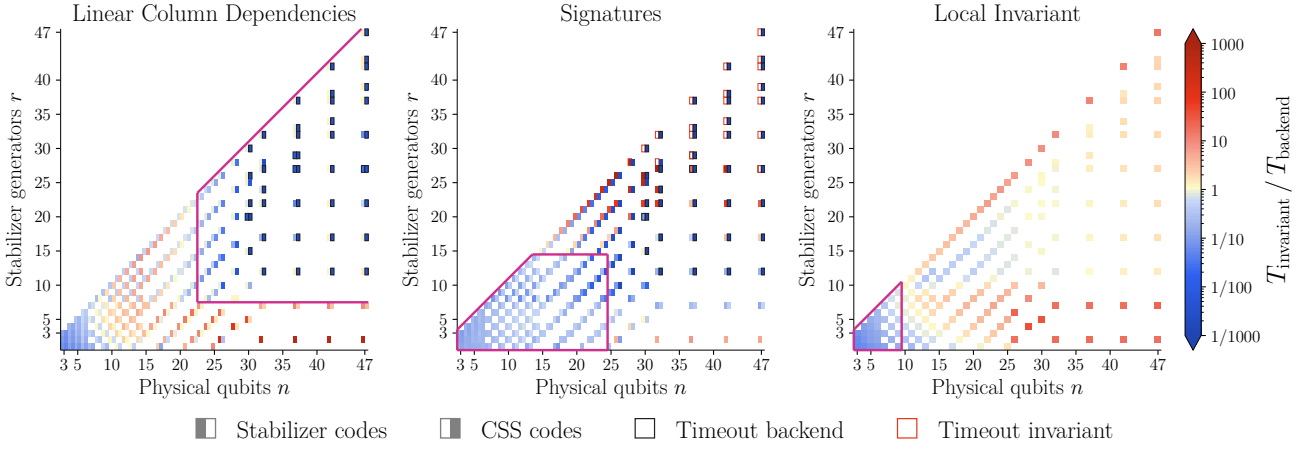

\centering
\resizebox{\textwidth}{!}{%
\pmSetTextScale{1.8}%
\begin{tikzpicture}
  \begin{axis}[relative panel, at={(1.0851in,1.2540in)},
               title={Linear Column Dependencies}]
    \pmEmptyTriangle{relEmpty}
    \input{material/graphics/data/a3_linear_dependency.tex}
    \begin{pgfonlayer}{axis foreground}
      \draw[rel threshold] (22.5,7.5) -- (22.5,23.5);
      \draw[rel threshold] (22.5,7.5) -- (47.5,7.5);
      \draw[rel threshold] (22.5,23.5) -- (46.5,47.5);
    \end{pgfonlayer}
  \end{axis}

  \begin{axis}[relative panel, at={(5.1661in,1.2540in)}, title={Signatures}]
    \pmEmptyTriangle{relEmpty}
    \input{material/graphics/data/a3_signatures.tex}
    \begin{pgfonlayer}{axis foreground}
      \draw[rel threshold] (13.5,14.5) -- (24.5,14.5);
      \draw[rel threshold] (24.5,0.5) -- (24.5,14.5);
      \draw[rel threshold] (2.5,3.5) -- (13.5,14.5);
      \draw[rel threshold] (2.5,0.5) -- (2.5,3.5);
      \draw[rel threshold] (2.5,0.5) -- (24.5,0.5);
    \end{pgfonlayer}
  \end{axis}

  \begin{axis}[relative panel, at={(9.2471in,1.2540in)}, title={Local Invariant}]
    \pmEmptyTriangle{relEmpty}
    \input{material/graphics/data/a3_local_invariant.tex}
    \begin{pgfonlayer}{axis foreground}
      \draw[rel threshold] (9.5,0.5) -- (9.5,10.5);
      \draw[rel threshold] (2.5,3.5) -- (9.5,10.5);
      \draw[rel threshold] (2.5,0.5) -- (2.5,3.5);
      \draw[rel threshold] (2.5,0.5) -- (9.5,0.5);
    \end{pgfonlayer}
  \end{axis}

  \shade[shading=relbar] (12.6558in,1.4043in) rectangle (12.8211in,4.4097in);
  \fill[relBlue] (12.6558in,1.4043in) -- (12.7384in,1.2540in) -- (12.8211in,1.4043in) -- cycle;
  \fill[relRed]  (12.6558in,4.4097in) -- (12.7384in,4.5600in) -- (12.8211in,4.4097in) -- cycle;
  \draw[black, line width=0.7pt]
    (12.6558in,1.4043in) -- (12.7384in,1.2540in) -- (12.8211in,1.4043in) --
    (12.8211in,4.4097in) -- (12.7384in,4.5600in) -- (12.6558in,4.4097in) -- cycle;
  \foreach \k/\lbl in {0/{1/1000},1/{1/100},2/{1/10},3/1,4/10,5/100,6/1000} {
    \draw[black, line width=0.8pt]
      (12.8211in, {1.4043in+\k*0.5009in}) -- ++(3.5pt,0);
    \node[anchor=base west, inner sep=0pt, font=\pmFont{7}{\pmTextScale}, name=relTick\k]
      at (12.9183in, {1.4043in+\k*0.5009in-2.7\pmU}) {\lbl};
  }
  \foreach \k in {0,...,5} \foreach \m in {2,...,9} {
    \draw[black, line width=0.6pt]
      (12.8211in, {1.4043in+(\k+log10(\m))*0.5009in}) -- ++(2pt,0);
  }

  \node[rotate=90, anchor=north, inner sep=0pt, font=\pmFont{11}{\pmTextScale}]
    at ([xshift=3pt]relTick0.east |- 0,2.9070in) {$T_{\mathrm{invariant}}\,/\,T_{\mathrm{backend}}$};

  \node[anchor=base, inner sep=0pt, font=\pmFont{10}{\pmTextScale}]
    at (7.2in, 0.35in) {%
      \relKeyHalf{0}{4\pmU}\hspace{8.8\pmU}Stabilizer codes\hspace{22\pmU}%
      \relKeyHalf{4\pmU}{8\pmU}\hspace{8.8\pmU}CSS codes\hspace{22\pmU}%
      \relKeyPatch{draw=relTimeout, line width=0.9pt}\hspace{8.8\pmU}Timeout backend\hspace{22\pmU}%
      \relKeyPatch{draw=relInvTimeout, line width=0.9pt}\hspace{8.8\pmU}Timeout invariant};
\end{tikzpicture}%
}
\caption{Mean runtime of an invariant relative to the fastest decision method (backend) on randomized instances; blue favors the invariant, red favors the decision method, pink lines mark the usefulness bounds.}
\label{fig:usefulness-inv}
\end{figure*}

\paragraph{Invariants}
Among the stated properties preserved by the transformations in \Cref{sec:invariants-signatures}, we evaluate only potentially expensive invariants, as expectedly cheap invariants are always used in the hybrid checkers. Preliminary experiments identified the best-performing candidates, namely \textbf{signatures} and \textbf{linear column dependencies} for permutation equivalence, and a low-degree \textbf{local invariant} for local-Clifford equivalence.

The primary purpose of invariants is to \emph{reject} inequivalent instances. To assess their discriminative power, we measure the percentage of inequivalent instances rejected by each invariant across several parameter settings.
The results in \Cref{tab:rejection-rate} show that the permutation invariants reject a high fraction of instances, particularly inequivalent \gls{css} codes. These rejection rates empirically establish their potential as preprocessing in a hybrid checker.

\begin{table}[htbp]
  \centering
  \caption{Mean percentage of randomized inequivalent instances rejected by each invariant across all parameter settings.}
  \label{tab:rejection-rate}
  \footnotesize
  \renewcommand{\arraystretch}{1.15}%
  \setlength{\tabcolsep}{4pt}%
  \begin{tabular}{@{}cc|cc|c@{}}
    \toprule
    \multicolumn{2}{c|}{\makecell{Linear column\\dependencies}} &
    \multicolumn{2}{c|}{Signatures} &
    \makecell{Local\\invariant} \\
    \midrule
    \makecell{Non-CSS} & CSS &
    \makecell{Non-CSS} & CSS &
    \makecell{General} \\
    \midrule
    $43.8\%$ & $90.5\%$ & $46.0\%$ & $99.0\%$ & $29.4\%$ \\
    \bottomrule
  \end{tabular}%
\end{table}

However, practical usefulness also depends on their computational cost---an invariant should cost less than the subsequent decision work it aims to avoid.
\Cref{fig:usefulness-inv} compares the runtime of each invariant with that of the fastest decision method for the same parameter setting, providing a conservative criterion for usefulness. Linear column dependencies become worthwhile for sufficiently large codes, whereas signatures on general stabilizer codes are fast enough only for small instances and scale poorly primarily with the number of generators $r$. The modified signature implementation for \gls{css} codes is beneficial across all evaluated settings. The local invariant is useful only for small $n$, where its discriminative power is greatest.
High rejection rates therefore do not guarantee a runtime benefit within the overall hybrid procedure, and the application of an invariant must depend on the input parameters.

\paragraph{Signatures}
The secondary purpose of signatures is to \emph{refine} the permutation search space by partitioning the qubits into signature classes.
The measurements detailed in \Cref{sec:signature-refinement} show that signatures can distinguish two randomly chosen distinct physical qubits with a mean probability of $86\%$ for general stabilizer codes and $69\%$ for \gls{css} codes.
Signatures therefore provide substantial refinement beyond their use as invariants, particularly for non-\gls{css} stabilizer codes. However, as shown in \Cref{fig:usefulness-inv}, their computation becomes too expensive above a certain threshold.

\definecolor{bpEmpty}{HTML}{FFFFFF}      
\definecolor{bpGray}{HTML}{808080}   

\colorlet{bpSat}{thesislightlila}
\colorlet{bpLse}{thesislila}
\colorlet{bpGraphIso}{thesisgreen}
\colorlet{bpMatroid}{thesisdarkgreen}

\definecolor{bpBruteforce}{HTML}{9AB2F4}
\definecolor{bpKls}{HTML}{648AF2}
\colorlet{bpClassical}{thesisblue}

\pgfplotsset{
  bp panel/.style={parameter map={\pmTextScale}{47},
                   xtick={3,5,10,15,20,25,30,35,40,47}, ytick={3,5,10,15,20,25,30,35,40,47},
                   width=3.0016in, height=3.1350in,
                   title style={align=center}},
}

\newcommand{\bpCell}[3]{\pmColorPart{#1}{#2}{0}{1}{#3}}
\newcommand{\bpRunnerUp}[3]{%
  \pmPartCorners{#1}{#2}{0}{1}%
  \edef\pmDo{\noexpand\fill[#3] (\pmXa,\pmYa) -- (\pmXb,\pmYa) -- (\pmXb,\pmYb) -- cycle;}%
  \pmDo
}

\newcommand{\bpKey}[1]{\tikz[baseline=0pt]{%
  \useasboundingbox (0,0) rectangle (8\pmU,8\pmU);
  \fill[#1] (0,0) rectangle (8\pmU,8\pmU);}}
\newcommand{\bpEntry}[2]{#1\hspace{8.8\pmU}\rule[-2.6\pmU]{0pt}{11\pmU}#2}

\newcommand{\bpRowSep}{\\[6.5\pmU]}

\newcommand{\bpEntryTwo}[3]{#1\hspace{8.8\pmU}%
  \raisebox{\dimexpr4\pmU-0.5\height+0.5\depth\relax}{%
    \renewcommand{\arraystretch}{0.85}\pmFontLS{9}{\pmTextScale}{1.0}%
    \begin{tabular}[b]{@{}l@{}}#2\\#3\end{tabular}}}

\newcommand{\bpTitleIcon}[1]{\raisebox{0.3em}{\scalebox{0.45}{#1}}\hspace{3\pmU}}
\newcommand{\bpLink}[2]{\hyperref[#1]{\textcolor{bpGray}{#2}}}
\newcommand{\bpLegendTitle}[3]{\bpTitleIcon{#1}\bpLink{#2}{#3}}
\newsavebox{\bpBody}\newsavebox{\bpHead}\newlength{\bpW}

\newcommand{\bpLegend}[3]{%
  \pmFont{11}{\pmTextScale}\renewcommand{\arraystretch}{0}%
  \sbox{\bpBody}{#3}%
  \sbox{\bpHead}{\pmFont{9}{\pmTextScale}\bfseries\rule[-2.1\pmU]{0pt}{9\pmU}#2}%
  \setlength{\bpW}{#1}%
  \setlength{\fboxsep}{3.6\pmU}\setlength{\fboxrule}{1pt}%
  \fcolorbox{bpGray}{white}{\vbox{\offinterlineskip
    \hbox to\bpW{\hss\box\bpHead\hss}%
    \vskip 3.5\pmU
    \hbox to\bpW{\box\bpBody\hss}}}}
\newsavebox{\bpLegA}\newsavebox{\bpLegB}\newsavebox{\bpLegC}

\begin{figure*}[!t]
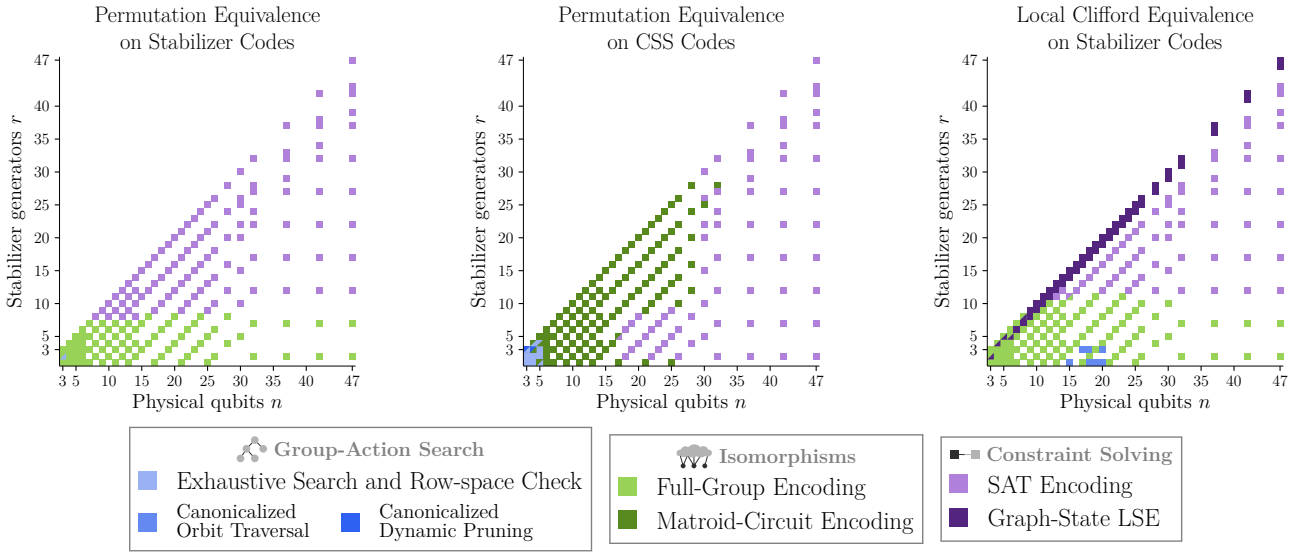

\centering
\resizebox{\textwidth}{!}{%
\pmSetTextScale{1.65}
\sbox{\bpLegA}{\bpLegend{325.45709pt}{\bpLegendTitle{\iconGroupActionSearch}{sec:group-action-search}{Group-Action Search}}{%
  \begin{tabular}{@{}l@{\hspace{14\pmU}}l@{}}
    \multicolumn{2}{@{}l@{}}{\bpEntry{\bpKey{bpBruteforce}}{Exhaustive Search and Row-space Check}} \bpRowSep
    \bpEntryTwo{\bpKey{bpKls}}{Canonicalized}{Orbit Traversal} &
    \bpEntryTwo{\bpKey{bpClassical}}{Canonicalized}{Dynamic Pruning}
  \end{tabular}}}%
\sbox{\bpLegB}{\bpLegend{216.34222pt}{\bpLegendTitle{\iconIsomorphism}{sec:isomorphism-reductions}{Isomorphisms}}{%
  \begin{tabular}{@{}l@{}}
    \bpEntry{\bpKey{bpGraphIso}}{Full-Group Encoding} \bpRowSep
    \bpEntry{\bpKey{bpMatroid}}{Matroid-Circuit Encoding}
  \end{tabular}}}%
\sbox{\bpLegC}{\bpLegend{163.33751pt}{\bpLegendTitle{\iconConstraintSolving}{sec:constraint-solving}{Constraint Solving}}{%
  \begin{tabular}{@{}l@{}}
    \bpEntry{\bpKey{bpSat}}{SAT Encoding} \bpRowSep
    \bpEntry{\bpKey{bpLse}}{Graph-State LSE}
  \end{tabular}}}%
\begin{tikzpicture}
  \begin{axis}[bp panel, at={(1.2841in,1.4820in)},
               title={Permutation Equivalence\\on Stabilizer Codes}]
    \pmEmptyTriangle{bpEmpty}
    \input{material/graphics/data/a5_pm_stb.tex}
  \end{axis}

  \begin{axis}[bp panel, at={(5.9872in,1.4820in)},
               title={Permutation Equivalence\\on CSS Codes}]
    \pmEmptyTriangle{bpEmpty}
    \input{material/graphics/data/a5_pm_css.tex}
  \end{axis}

  \begin{axis}[bp panel, at={(10.6903in,1.4820in)},
               title={Local Clifford Equivalence\\on Stabilizer Codes}]
    \pmEmptyTriangle{bpEmpty}
    \input{material/graphics/data/a5_lc_stb.tex}
  \end{axis}

  \node[anchor=south west, inner sep=0pt] at (143.54169pt,-0.40in) {\usebox{\bpLegA}};
  \node[anchor=south west, inner sep=0pt] at (495.36742pt,-23.46849pt) {\usebox{\bpLegB}};
  \node[anchor=south west, inner sep=0pt] at (738.07828pt,-21.85222pt) {\usebox{\bpLegC}};
\end{tikzpicture}%
}
\caption{Fastest method per parameter setting on randomized instances; split cells: runner-up within 5\% in the lower-right half.}
\label{fig:best-approach}
\end{figure*}

\begin{findingbox}{Invariants and signatures are beneficial only in parts of the parameter space.}
    \findingpart{They \emph{reject} most inequivalent instances, and signatures additionally \emph{refine} the permutation search space substantially.}
    \lastfindingpart{Their computation can cost more than the decision procedure it is intended to accelerate, so their use must be \emph{parameter-dependent}.}
\end{findingbox}

\subsection[Group-Action Search]{%
  Group-Action Search%
  \methodTitleIcons{4.2mm}{\protect\iconGroupActionSearch}%
}
\Cref{fig:best-approach} identifies the empirically best-performing algorithms based on runtime and memory use. Canonicalized orbit traversal and dynamic pruning of the permutation search tree are fastest for isolated parameter settings, but show no consistent trend. Tableau-based exhaustive search, traversing the physical permutation space and applying the rank-based row-space check, is effective for \gls{css} codes with few physical qubits, as expected. However, the same advantage is not observed for other code families or equivalence notions.

\begin{findingbox}{Group-action search is competitive only for instances with a small physical search space.}
    \findingpart{It is competitive for \gls{css} codes with few \emph{physical qubits}, where the search space is small.}
    \lastfindingpart{It becomes impractical for larger $n$, as the search space grows factorially or exponentially in the number of qubits.}
\end{findingbox}

\subsection[Isomorphism]{%
  Isomorphism%
  \methodTitleIcons{4.2mm}{\protect\iconIsomorphism}%
}
Eliminating representation freedom through an explicit encoding is effective when that encoding remains small, as shown clearly in \Cref{fig:best-approach}. The full-group reduction is the fastest method for codes with few stabilizer generators, whereas the matroid-circuit encoding performs the best at low $n$, reflecting their different size parameters. As detailed in \Cref{sec:performance-isomorphisms}, both runtime and---more severely---memory use rise sharply with the exponentially growing graph size.

\begin{findingbox}{Isomorphism-based methods are most effective when the basis-independent encoding remains small.}
    \findingpart{Graph constructions are fastest when the \emph{encoded object} is small: few stabilizer elements (low $r$) or few matroid circuits (low $n$).}
    \lastfindingpart{As the encoded object grows, \emph{graph size} causes prohibitive runtime and, especially, memory costs.}
\end{findingbox}

\subsection[Constraint Solving]{%
  Constraint Solving%
  \methodTitleIcons{2.4mm}{\protect\iconConstraintSolving}%
}

\subsubsection{\SAT}
\Cref{fig:best-approach} makes clear that the \SAT{}-based methods are the strongest general baselines for all equivalence problems. They encounter no memory issues for the larger codes evaluated, which is important for a fallback method that should not exhaust the available memory. The \SAT{} methods also have the lowest runtimes across the widest parameter ranges for local-Clifford and permutation equivalence on stabilizer codes.
These properties make \SAT{} a suitable fallback when more specialized methods cease to apply.

\def\satLogVmin{-1.786746}
\def\satLogSpan{5.519139}

\definecolor{satcEmpty}{HTML}{FFFFFF} 

\pgfplotsset{
  satc panel/.style={parameter map={\pmTextScale}{25},
                   xtick={3,5,10,15,20,25}, ytick={3,5,10,15,20,25},
                   width=3.5300in, height=3.8370in,
                   title style={align=center, font=\pmFont{12}{\pmTextScale}}, ylabel shift=2pt,
                   label style={font=\pmFont{11}{\pmTextScale}}},
}

\newcommand{\satcBarY}[1]{0.5155in+(log10(#1)-(\satLogVmin))/\satLogSpan*4.0290in}

\begin{figure}[H]
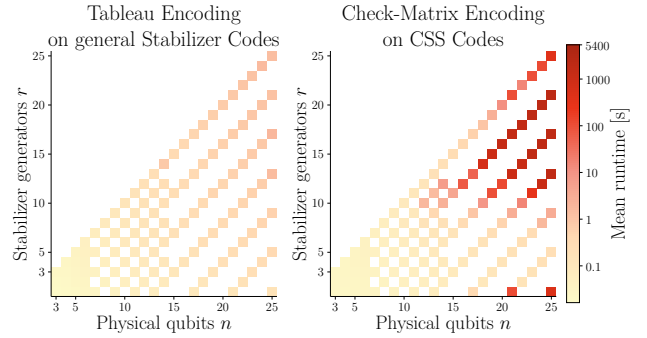

\centering
\resizebox{\columnwidth}{!}{%
\pmSetTextScale{2.0}%
\begin{tikzpicture}
  \begin{axis}[satc panel, at={(0.7140in,0.6115in)},
               title={Tableau Encoding\\on general Stabilizer Codes}]
    \pmEmptyTriangle[25]{satcEmpty}
    \input{material/graphics/data/a6_pm_stb_sat_on_stabilizer.tex}
  \end{axis}

  \begin{scope}[xshift=0.2in]
  \begin{axis}[satc panel, at={(4.8794in,0.6115in)},
               title={Check-Matrix Encoding\\on CSS Codes}]
    \pmEmptyTriangle[25]{satcEmpty}
    \input{material/graphics/data/a6_pm_css_sat_on_css.tex}
  \end{axis}

  \shade[shading=pmruntime] (8.5705in,0.5155in) rectangle (8.7720in,4.5445in);
  \draw[black, line width=0.7pt] (8.5705in,0.5155in) rectangle (8.7720in,4.5445in);
  \foreach \v in {0.1,1,10,100,1000} {
    \draw[black, line width=0.8pt] (8.7720in,{\satcBarY{\v}}) -- ++(3.5pt,0);
    \node[anchor=base west, inner sep=0pt, font=\pmFont{7}{\pmTextScale}]
      at (8.8692in,{\satcBarY{\v}-2.7\pmU}) {\v};
  }
  \draw[black, line width=0.6pt] (8.7720in,4.5445in) -- ++(3pt,0);
  \node[anchor=base west, inner sep=0pt, font=\pmFont{7}{\pmTextScale}]
    at (8.8609in,{4.5445in-2.7\pmU}) {5400};
  \node[rotate=90, anchor=north, inner sep=0pt, font=\pmFont{11}{\pmTextScale}]
    at (9.2514in,2.5300in) {Mean runtime [s]};
  \end{scope}
\end{tikzpicture}%
}
\caption{Mean \SAT{} runtime for permutation equivalence on randomized instances.}
\label{fig:sat-css}
\end{figure}

For permutation equivalence on \gls{css} codes, however, \Cref{fig:sat-css} shows runtimes several orders of magnitude above those for general stabilizer codes of the same size. The effect persists when \gls{css} codes are represented by full tableaus rather than check matrices, so the concrete representation is not its cause. Instead, the binary-symplectic representation employed in the encoding interacts poorly with the separated $X$/$Z$ structure---a property that usually benefits code-processing methods, e.g., as it does for signature computation. \Cref{sec:sat-limitations} analyzes this behavior for unbalanced \gls{css} codes and presents a potential remedy. Thus, despite avoiding memory limitations, this \SAT{} formulation is not always practical for \gls{css}-code permutation equivalence.

\begin{findingbox}{\SAT{} encodings provide the broadest general baseline, but are weak for permutation equivalence on CSS codes.}
    \findingpart{They achieve the lowest runtimes over the \emph{widest parameter range} for general stabilizer codes without exhausting memory.}
    \lastfindingpart{For \gls{css} codes, the matrix encoding can be orders of magnitude \emph{slower}.}
\end{findingbox}

\subsubsection{Graph-State LSE}
Converting a stabilizer code to a graph state incurs non-negligible overhead for small codes. Nevertheless, within the restricted domain $k<2$, where logical freedom need not be handled explicitly, the polynomial-time graph-state \gls{lse} algorithm is the fastest local-Clifford-equivalence method in \Cref{fig:best-approach}.

\begin{findingbox}{Despite graph-state conversion \emph{overhead}, the polynomial-time \gls{lse} algorithm is fastest within its \emph{restricted} domain.}
\end{findingbox}

\subsection{Hybrid Equivalence Checkers}
We evaluate the complete hybrid checkers from \Cref{sec:hybrids} using conservative selection thresholds derived from the preceding experiments in \Cref{fig:usefulness-inv} and \Cref{fig:best-approach}. The inputs are randomized variants of practical structured codes, including color codes \cite{coco} and bivariate bicycle codes \cite{bb}.

\Cref{tab:hybrid-comp} shows that the selected decision method changes with the instance. Exhaustive search, graph or matroid isomorphism, and the graph-state \gls{lse} dominate suitable small inputs; beyond their respective regimes, the dispatcher shifts primarily to \SAT{}.

The clearest observable effect is the asymmetry between equivalent and inequivalent inputs. Equivalent instances cannot be rejected during preprocessing, whereas cheap invariants terminate many inequivalent permutation instances, and signatures become decisive for larger \gls{css} codes. Mean permutation-equivalence runtimes remain below $10$ minutes through $n=37$, while inequivalent \gls{css} instances remain decidable at $n=144$ in about $310$ seconds. For local-Clifford equivalence, all inequivalent $n=144$ instances complete with a mean below $16$ seconds. At $n=72$, all equivalent instances complete with a mean of $49$ seconds, while larger local-Clifford-equivalent instances increasingly time out. Early refutation therefore extends the practical range significantly for inequivalent inputs.

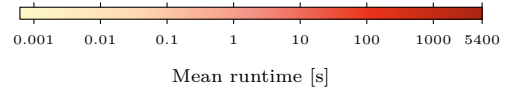
\def\hybLogVmin{-3.209677}
\def\hybLogVmax{3.732395}

\newcommand{\hybRows}{%
  \hybCode{Bell pair} \hybCell{0.0786}{hybText}{GI} \hybCell{0.0035}{hybText}{CI} \hybCell{0.0798}{hybText}{BF} \hybCell{0.0000}{hybText}{CI} \hybCell{0.0648}{hybText}{LSE} \hybCell{0.0210}{hybText}{EI} \\ \hline
  \hybCode{3-bit repetition} \hybCell{0.0843}{hybText}{GI} \hybCell{0.0262}{hybText}{CI} \hybCell{0.0926}{hybText}{BF} \hybCell{0.0265}{hybText}{CI} \hybCell{0.0933}{hybText}{LSE} \hybCell{0.0331}{hybText}{EI} \\ \hline
  \hybCode{5-qubit perfect} \hybCell{0.1145}{hybText}{GI} \hybCellTwo{0.0760}{hybText}{CI}{60\%}{GI}{40\%} \hybNA \hybNA \hybCell{0.1214}{hybText}{LSE} \hybCellTwo{0.1062}{hybText}{LSE}{70\%}{EI}{30\%} \\ \hline
  \hybCode{Steane} \hybCell{0.2089}{hybText}{GI} \hybCell{0.0404}{hybText}{CI} \hybCell{0.1559}{hybText}{MI} \hybCell{0.0352}{hybText}{CI} \hybCell{0.1703}{hybText}{LSE} \hybCell{0.1760}{hybText}{LSE} \\ \hline
  \hybCode{Shor} \hybCell{0.3117}{hybText}{SAT} \hybCell{0.0360}{hybText}{CI} \hybCell{0.1706}{hybText}{MI} \hybCell{0.0237}{hybText}{CI} \hybCell{0.2155}{hybText}{LSE} \hybCellTwo{0.1780}{hybText}{EI}{50\%}{LSE}{50\%} \\ \hline
  \hybCode{Carbon} \hybCell{0.3549}{hybText}{SAT} \hybCellTwo{0.1769}{hybText}{S}{60\%}{CI}{40\%} \hybCell{0.2102}{hybText}{MI} \hybCellTwo{0.1600}{hybText}{S}{60\%}{CI}{40\%} \hybCell{0.2721}{hybText}{GI} \hybCell{0.2726}{hybText}{GI} \\ \hline
  \hybCode{Tetrahedral} \hybCell{0.4192}{hybText}{SAT} \hybCell{0.0739}{hybText}{CI} \hybCell{0.2586}{hybText}{MI} \hybCell{0.0570}{hybText}{CI} \hybCell{0.2153}{hybText}{LSE} \hybCell{0.2324}{hybText}{LSE} \\ \hline
  \hybCode{15-qubit optimal} \hybCell{0.4648}{hybText}{SAT} \hybCellTwo{0.3961}{hybText}{S}{60\%}{SAT}{40\%} \hybNA \hybNA \hybCell{0.3523}{hybText}{SAT} \hybCell{0.3485}{hybText}{SAT} \\ \hline
  \hybCode{Hamming-15} \hybCell{0.3621}{hybText}{SAT} \hybCell{0.0601}{hybText}{CI} \hybCell{0.2861}{hybText}{MI} \hybCell{0.0566}{hybText}{CI} \hybCell{0.1989}{hybText}{GI} \hybCell{0.1954}{hybText}{GI} \\ \hline
  \hybCode{Golay} \hybCell{0.4810}{hybText}{SAT} \hybCell{0.7207}{white}{SAT} \hybCell{0.6475}{hybText}{MI} \hybCell{0.4616}{hybText}{S} \hybCell{0.2629}{hybText}{LSE} \hybCell{0.2821}{hybText}{LSE} \\ \hline
  \hybCode{Rotated surface} \hybCell{0.5168}{hybText}{SAT} \hybCellTwo{0.3945}{hybText}{CI}{70\%}{EI}{20\%} \hybCell{0.4605}{hybText}{MI} \hybCellTwo{0.3643}{hybText}{CI}{70\%}{EI}{20\%} \hybCell{0.2890}{hybText}{LSE} \hybCell{0.2892}{hybText}{LSE} \\ \hline
  \hybCode{CoCo-488} \hybCell{0.7652}{white}{SAT} \hybCellTwo{0.5568}{hybText}{SAT}{60\%}{EI}{40\%} \hybCell{0.5995}{hybText}{SAT} \hybCellTwo{0.5416}{hybText}{S}{60\%}{EI}{40\%} \hybCell{0.3094}{hybText}{LSE} \hybCell{0.3151}{hybText}{LSE} \\ \hline
  \hybCode{Hamming-31} \hybCell{0.5063}{hybText}{SAT} \hybCell{0.1481}{hybText}{CI} \hybCell{0.4987}{hybText}{SAT} \hybCell{0.0764}{hybText}{CI} \hybCell{0.3233}{hybText}{GI} \hybCell{0.3163}{hybText}{GI} \\ \hline
  \hybCode{CoCo-666} \hybCell{0.8601}{white}{SAT} \hybCellTwo{0.6261}{hybText}{SAT}{90\%}{EI}{10\%} \hybCell{0.8161}{white}{SAT} \hybCellTwo{0.5567}{hybText}{S}{90\%}{EI}{10\%} \hybCell{0.3258}{hybText}{LSE} \hybCell{0.3348}{hybText}{LSE} \\ \hline
  \hybCode{BB-72} \hybCell{1.0000}{white}{SAT} \hybCell{0.9832}{white}{SAT} \hybCell{1.0000}{white}{SAT} \hybCell{0.6760}{white}{S} \hybCell{0.7059}{white}{SAT} \hybCell{0.5347}{hybText}{SAT} \\ \hline
  \hybCode{BB-90} \hybCell{1.0000}{white}{SAT} \hybCell{0.1763}{hybText}{CI} \hybCell{1.0000}{white}{SAT} \hybCell{0.1504}{hybText}{CI} \hybCell{0.9784}{white}{SAT} \hybCell{0.5675}{hybText}{SAT} \\ \hline
  \hybCode{BB-108} \hybCell{1.0000}{white}{SAT} \hybCell{1.0000}{white}{SAT} \hybCell{1.0000}{white}{SAT} \hybCell{0.7579}{white}{S} \hybCell{0.9892}{white}{SAT} \hybCell{0.5968}{hybText}{SAT} \\ \hline
  \hybCode{BB-144} \hybCell{1.0000}{white}{SAT} \hybCell{1.0000}{white}{SAT} \hybCell{1.0000}{white}{SAT} \hybCell{0.8213}{white}{S} \hybCell{0.9727}{white}{SAT} \hybCell{0.6345}{hybText}{SAT} \\
}

\definecolor{hybText}{HTML}{202020}
\definecolor{hybFrame}{HTML}{808080} 

\newlength{\hybRowH}\setlength{\hybRowH}{11.5pt}
\newlength{\hybHeadH}\setlength{\hybHeadH}{13pt}
\newlength{\hybSubH}\setlength{\hybSubH}{7pt}
\newlength{\hybCodeW}
\newlength{\hybCellW}
\newlength{\hybPad}\setlength{\hybPad}{1.5pt}
\newlength{\hybAir}\setlength{\hybAir}{1pt}

\newcommand{\hybFontSingle}{\fontsize{7}{7.5}\selectfont\bfseries}
\newcommand{\hybFontStage}{\fontsize{5.5}{6}\selectfont\bfseries}
\newcommand{\hybFontShare}{\fontsize{4.1}{4.3}\selectfont}
\newcommand{\hybFontSecond}{\fontsize{4.7}{5}\selectfont}
\newcommand{\hybFontSecondShare}{\fontsize{3.7}{3.8}\selectfont}
\newcommand{\hybFontCode}{\fontsize{4.6}{5.5}\selectfont}
\newcommand{\hybFontHead}{\fontsize{4.6}{5.5}\selectfont\bfseries}
\newcommand{\hybFontSub}{\fontsize{5}{6}\selectfont}

\newcommand{\hybBox}[3]{\parbox[c][#1][c]{#2}{\fontsize{4}{5}\selectfont\centering #3}}
\newcommand{\hybCellColor}[1]{%
  \pgfplotscolormapaccess[0:1]{#1}{runtime}\expandafter\pmSplitRGB\pgfmathresult;%
  \edef\hybTmp{\noexpand\cellcolor[rgb]{\pmR,\pmG,\pmB}}\hybTmp}
\newcommand{\hybCode}[1]{\parbox[c][\hybRowH][c]{\hybCodeW}{\hybFontCode\raggedright\hspace*{\hybPad}#1}}
\newcommand{\hybNA}{& \hybBox{\hybRowH}{\hybCellW}{\hybFontSecond\bfseries N/A}}
\newcommand{\hybCell}[3]{%
  & \hybCellColor{#1}\hybBox{\hybRowH}{\hybCellW}{\color{#2}%
    {\hybFontSingle\texttt{#3}}}}
\newcommand{\hybCellTwo}[6]{%
  & \hybCellColor{#1}\hybBox{\hybRowH}{\hybCellW}{\color{#2}%
    {\hybFontStage\texttt{#3}}\hspace{0.25em}{\hybFontShare #4}\\[0.3pt]
    {\hybFontSecond\texttt{#5}}\hspace{0.25em}{\hybFontSecondShare #6}}}

\newcommand{\hybSep}{\hspace{\hybAir}{\color{black}\vline}\hspace{\hybAir}}
\newcommand{\hybHead}[1]{\multicolumn{2}{c!{\hybSep}}{\hybHeadBox{#1}}}
\newcommand{\hybHeadLast}[1]{\multicolumn{2}{c}{\hybHeadBox{#1}}}
\newcommand{\hybHeadBox}[1]{\parbox[c][\hybHeadH][c]{\dimexpr2\hybCellW+\arrayrulewidth\relax}{\hybFontHead\centering #1}}
\newcommand{\hybSub}[1]{\hybBox{\hybSubH}{\hybCellW}{\hybFontSub #1}}

\begin{table}[!t]
\centering
\caption{Hybrid performance on equivalent ($+$) and inequivalent ($-$) practical-code instances. Cell color encodes mean runtime; labels identify the stage terminating most runs by decision or timeout, with shares and the runner-up where outcomes split: \texttt{CI}/\texttt{EI} cheap/expensive invariants, \texttt{S} signatures, \texttt{BF} exhaustive search, \texttt{MI}/\texttt{GI} matroid/graph isomorphism, \texttt{SAT} solver, and \texttt{LSE} graph-state LSE.}
\label{tab:hybrid-comp}
\setlength{\tabcolsep}{0pt}%
\setlength{\arrayrulewidth}{0.5pt}%
\renewcommand{\arraystretch}{1}%
\setlength{\hybCodeW}{0.21\linewidth}%

\setlength{\hybCellW}{\dimexpr(\linewidth-\hybCodeW-6\arrayrulewidth-6\hybAir)/6\relax}%
\arrayrulecolor{white}%
\begin{tabular}{@{}l!{\hybSep}c|c!{\hybSep}c|c!{\hybSep}c|c@{}}
\arrayrulecolor{black}\specialrule{\heavyrulewidth}{0pt}{0pt}\arrayrulecolor{white}

\raisebox{-0.5\hybSubH}[0pt][0pt]{%
  \parbox[c][\hybHeadH][c]{\hybCodeW}{\hybFontHead\raggedright\hspace*{\hybPad}Code}}
& \hybHead{Permutation eq.\\on stabilizer codes} & \hybHead{Permutation eq.\\on CSS codes} & \hybHeadLast{LC eq.\\on stabilizer codes} \\
& \hybSub{$+$} & \hybSub{$-$} & \hybSub{$+$} & \hybSub{$-$} & \hybSub{$+$} & \hybSub{$-$} \\
\arrayrulecolor{black}\specialrule{\lightrulewidth}{0pt}{0pt}\arrayrulecolor{white}
\hybRows
\arrayrulecolor{black}\specialrule{\heavyrulewidth}{0pt}{0pt}
\end{tabular}%
\arrayrulecolor{black}%
\\[2pt]

\begin{tikzpicture}[baseline]
  \pgfplotscolorbardrawstandalone[
    colormap name=runtime,
    colorbar horizontal,
    point meta min=\hybLogVmin, point meta max=\hybLogVmax,
    colorbar style={
      width=0.75\linewidth, height=4.5pt,
      xtick={-3,-2,-1,0,1,2,3}, xticklabels={0.001,0.01,0.1,1,10,100,1000},
      extra x ticks={\hybLogVmax}, extra x tick labels={5400},
      extra x tick style={tick style={draw=hybFrame}},
      xtick style={draw=black, line width=0.4pt}, tick align=outside, major tick length=1.8pt,
      xticklabel style={font=\fontsize{5}{6}\selectfont, yshift=-0.5pt},
      axis line style={black, line width=0.4pt},
      xlabel={Mean runtime [s]}, xlabel style={font=\fontsize{7}{8}\selectfont, yshift=2pt},
    }]
\end{tikzpicture}%
\end{table}

\begin{findingbox}{The Refute--Refine--Decide architecture yields \emph{asymmetric} hybrid checkers that decide inequivalent instances faster than equivalent ones.}
\end{findingbox}

\section{Discussion}\label{sec:discussion}

\subsection{Methodological Limitations}
The principal limitation of the empirical evaluation is the sampling procedure for randomized input instances, which primarily generates a partner code guaranteed to be either equivalent or inequivalent. Although the generated instances are intended to reflect practical cases, it may introduce bias, particularly among inequivalent instances. This could affect the measured invariant rejection rates and the hybrid evaluation.

\subsection{Broader Implications}
The results show that code representation is not a neutral implementation choice---it determines which algorithms apply and can dominate runtime and memory requirements. Methods invariant under generator-basis changes are robust but may require exponentially larger encodings; representation-dependent methods avoid this cost but must handle the corresponding freedom explicitly.

Techniques from classical coding theory translate effectively to quantum-code equivalence with appropriate adaptations, notably signature-based preprocessing and matroid-circuit encodings.

Although \gls{css} codes form a particularly well-behaved subclass of stabilizer codes, not every method benefits from their additional structure. In particular, their decoupled check-matrix structure degrades the performance of the proposed \SAT{} encoding unless handled explicitly.

No algorithmic paradigm dominates across equivalence notions, code classes, and parameter ranges. This supports the Refute--Refine--Decide architecture, which rejects or simplifies instances before selecting a decision procedure. In a resource-constrained tool, the asymmetric behavior of the hybrids could be exploited: every returned verdict is conclusive, whereas timeouts are unresolved but may indicate likely equivalence.

\subsection{Summary and Outlook}
This work provides a unified framework for automated equivalence checking of \glspl{qecc}. We separate representation freedom from physical freedom and instantiate three problems: permutation equivalence for stabilizer and \gls{css} codes, and local-Clifford equivalence for stabilizer codes. We implement and evaluate three decision paradigms---group-action search, isomorphism-based methods, and constraint solving---together with invariants and signatures for preprocessing. Their complementary behavior during evaluation motivates the Refute--Refine--Decide hybrid architecture and its problem-dependent method selection, which are implemented as open-source hybrid approaches in \emph{MQT QECC}.

Future work should tailor equivalence notions and algorithms to \emph{more structured code families}. For topological codes, unrestricted qubit permutations may disregard relevant geometry or locality, motivating structure-preserving transformation groups. Further directions include the joint equivalence relation generated by qubit permutations and local Clifford operations, explicit treatment of logical-framework freedom in code-to-state transformations, and parallelization across compatible stages and within individual methods.

\section{Author Contributions}
M.B. designed and implemented the proposed methods, conducted the experimental evaluation, and analyzed and interpreted the results. L.S. supervised the research and helped refine its direction and presentation. R.W. provided senior supervision and strategic guidance.

LLMs were used as a grammar checker and proofreader to improve the clarity and conciseness of the text and to generate and refine auxiliary code.

\section{Acknowledgments}
We thank Tom Peham for conceiving the original idea and supervising the research in its early stages.
This work received funding from the European Research Council (ERC) under the European Union's Horizon 2020 research and innovation programme (grant agreement No.~101001318), and from the Deutsche Forschungsgemeinschaft (DFG, German Research Foundation) through the project ``Towards Software for Fault Tolerant Quantum Computing'' (project No.~563402549). It was also conducted as part of the Munich Quantum Valley, which is supported by the Bavarian state government with funds from the Hightech Agenda Bayern Plus.

\clearpage

\bibliographystyle{quantum}
\bibliography{main}

\clearpage

\appendix
\crefalias{section}{appendix}

\section{Code-to-Graph Transformation} \label{sec:code-to-state}
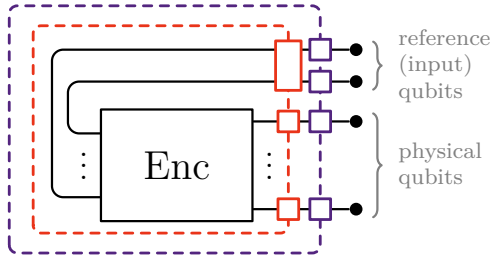
\begin{figure}[H]
\centering
\resizebox{0.8\columnwidth}{!}{%
    \begin{tikzpicture}[
        line cap=round,
        line join=round,
        thick,
        every node/.style={font=\large},
        io/.style={circle, fill=black, inner sep=0pt, minimum size=1.6mm},
        wire/.style={draw, thick},
        boundary gate/.style={
            draw=witnessColor,
            fill=white,
            line width=1pt,
            inner sep=0pt
        },
        outer gate/.style={
            draw=decideColor,
            fill=white,
            line width=1pt,
            inner sep=0pt
        }
    ]
    \node[
        draw,
        minimum width=1.9cm,
        minimum height=1.4cm,
        line width=0.8pt
    ] (enc) at (0,0.15) {\Large Enc};

    \coordinate (bendTop) at ($(enc.west)+(0,0.4)$);
    \coordinate (bendBottom) at ($(enc.west)+(0,-0.4)$);
    \coordinate (outTop) at ($(enc.east)+(0, 0.55)$);
    \coordinate (outBottom) at ($(enc.east)+(0,-0.55)$);
    \coordinate (ioColumn) at (2.25,0.15);

    \node[io] (out1) at (ioColumn |- outTop) {};
    \node[io] (out2) at (ioColumn |- outBottom) {};
    \node[io] (in1) at ($(ioColumn)+(0,1.45)$) {};
    \node[io] (in2) at ($(ioColumn)+(0,1.05)$) {};

    \draw[wire, rounded corners=4pt]
        (bendTop) -- ++(-0.4,0) |- (in2);
    \draw[wire, rounded corners=4pt]
        (bendBottom) -- ++(-0.6,0) |- (in1);
    \draw[wire] (outTop) -- (out1);
    \draw[wire] (outBottom) -- (out2);

    \coordinate (frameRight) at (1.4,0);
    \node at ($(enc.west)+(-0.2,0.1)$) {$\vdots$};
    \node at ($(enc.east)+(0.2,0.1)$) {$\vdots$};

    \draw[witnessColor, dashed, line width=1pt, rounded corners=1mm]
        (-1.8,-0.7) rectangle (1.4,1.9);

    \node[boundary gate, minimum width=2.6mm, minimum height=2.6mm]
        at (frameRight |- outTop) {};
    \node[boundary gate, minimum width=2.6mm, minimum height=2.6mm]
        at (frameRight |- outBottom) {};
    \node[boundary gate, minimum width=3.2mm, minimum height=6.2mm]
        at ($(frameRight |- in1)!0.5!(frameRight |- in2)$) {};

    \coordinate (outerFrameRight) at (1.8,0);
    \draw[decideColor, dashed, line width=1pt, rounded corners=1mm]
        (-2.1,-0.95) rectangle (1.8,2.15);

    \node[outer gate, minimum width=2.6mm, minimum height=2.6mm]
        at (outerFrameRight |- in1) {};
    \node[outer gate, minimum width=2.6mm, minimum height=2.6mm]
        at (outerFrameRight |- in2) {};
    \node[outer gate, minimum width=2.6mm, minimum height=2.6mm]
        at (outerFrameRight |- outTop) {};
    \node[outer gate, minimum width=2.6mm, minimum height=2.6mm]
        at (outerFrameRight |- outBottom) {};

    \draw[draw=thesisgrayveryverydark, text=thesisgrayveryverydark,
          decorate, decoration={brace, amplitude=3pt}]
        ($(in1.east)+(0.12,0.10)$) -- ($(in2.east)+(0.12,-0.10)$)
        node[midway, right=5pt, align=left, font=\scriptsize]
            {reference\\(input)\\qubits};
    \draw[draw=thesisgrayveryverydark, text=thesisgrayveryverydark,
          decorate, decoration={brace, amplitude=3pt}]
        ($(out1.east)+(0.12,0.10)$) -- ($(out2.east)+(0.12,-0.10)$)
        node[midway, right=5pt, align=left, font=\scriptsize]
            {physical\\qubits};

    \end{tikzpicture}%
}
\caption{Code-to-state transformation via the encoding map (Enc): a stabilizer state on $k$ reference and $n$ physical qubits. Code equivalence permits arbitrary $k$-qubit Cliffords on the reference subsystem and local Cliffords on the physical qubits (red); standard local-Clifford equivalence permits only local Cliffords on every qubit (lilac).}
\label{fig:choi}
\end{figure}

A valid code-to-state transformation for our application must preserve and reflect the code-space information so that equivalence of the resulting states holds if and only if the original codes are equivalent. We therefore apply the \emph{Choi-Jamiołkowski isomorphism} to the code's encoding map $\hilbertSpace{k} \to \hilbertSpace{n}$. The resulting state can be interpreted as a purification of the maximally mixed state on the code space \cite{mixed_state_proj}.

This transformation can be implemented in several ways, for example, using the symplectic stabilizer tableau and explicit logical operators $\bar{X}$ and $\bar{Z}$ \cite{mixed_state_proj}, or from a synthesized encoder circuit by bending its input wires into outputs according to the ZX calculus \cite{kls_long, choi_zx, zx_enc_resp}. In either case, the resulting stabilizer state fixes an arbitrary logical framework.
This state has $n+k$ qubits: $n$ physical qubits corresponding to the previous outputs of the encoding map and $k$ reference qubits corresponding to its previous inputs. The ZX-calculus construction is shown in black in \Cref{fig:choi}.

For the subsequent step, we exploit the fact that every stabilizer state, including the resulting state, is local-Clifford equivalent to some graph state \cite{effic_algo_states}. Consequently, in the state-equivalence problem, this step is subsumed by the allowed local-Clifford operations between the two codes.

However, because the first step fixes a logical framework, code equivalence---which depends solely on the code space---requires restoring the freedom to apply arbitrary Clifford operations across the reference subsystem. This required freedom is shown in red in \Cref{fig:choi}.

Standard local-Clifford-equivalence algorithms, such as the polynomial-time algorithm in \Cref{sec:constraint-solving}, permit only single-qubit Clifford operations on all $n+k$ qubits. For $k>1$, this freedom---shown in lilac in \Cref{fig:choi}---is strictly weaker than the required reference-subsystem freedom, so the application of a standard local-Clifford algorithm on the resulting state can produce false negatives for code equivalence. 

\section{Reproducibility} \label{sec:reproducibility}
All empirical results were collected within the hardware and software environment summarized in \Cref{tab:reproducibility-environment}. The exact methodology, including the code for generating input instances, running the experiment, and visualizing the results, is available in the replication package \cite{eq_check_code_repo}.

\begin{table}[htbp]
    \centering
    \caption{Hardware and software environment of the empirical evaluation.}
    \label{tab:reproducibility-environment}
    \footnotesize
    \renewcommand{\arraystretch}{1.12}
    \setlength{\tabcolsep}{3pt}
    \begin{tabular}{ll}
        \toprule
        \multicolumn{2}{@{}l}{\textbf{Hardware}} \\
        \midrule
        Architecture & \texttt{x86\_64} \\
        Processor & AMD Ryzen 7 3700X (8 cores) \\
        Memory & \makecell[l]{32 GB DDR4-3200\\($2 \times 16$ GB, CL16)} \\
        \midrule
        \multicolumn{2}{@{}l}{\textbf{Software}} \\
        \midrule
        Operating system & Debian GNU/Linux 13.5 \\
        Kernel & Linux 6.12.90+deb13.1-amd64 \\
        Python & 3.13.5 \\
        NumPy & 2.4.6 \\
        LDPC & 2.4.1 \\
        Matplotlib & 3.11.1 \\
        PyNauty & 2.8.8.1 \\
        Z3 Solver & 4.16.0.0 \\
        PyZX & 0.10.6 \\
        GAP + Guava & 4.15.1 + 3.21 \\
        \bottomrule
    \end{tabular}
\end{table}

\section{Input Instance Generation} \label{sec:instance-generation}
A randomized input instance $(C,C')$ for a given seed and parameter setting $(n,k)$ or named code is created in two stages.

$C$ is either loaded from a set of known structured codes, such as ``Steane'' or ``Shor,'' or generated by applying random Clifford gates and row operations to a tableau initially formed from independent single-qubit $\operatorname{Z}$ checks.
Its counterpart $C'$ is constructed more carefully to limit selection bias. Equivalent instances are generated by applying a random permutation or local Clifford operation followed by additional row operations. Constructing guaranteed inequivalent instances is more difficult, as they should resemble practical cases that are nontrivial for invariants and decision methods.
For sufficiently small codes, $C'$ is generated by applying arbitrary random Cliffords to $C$, preserving a structural relation between the codes. If $C$ is a \gls{css} code, only $\operatorname{CNOT}$ operations are applied to preserve the \gls{css} property without trivializing the problem.
The pair $(C,C')$ is then checked by a relatively inexpensive equivalence algorithm to guarantee inequivalence. As this verification becomes costly for larger codes, a cheap invariant is instead used to witness inequivalence. We choose an invariant that is neither too weak nor closely related to those evaluated, but this procedure may still introduce selection bias.

\section{Signature Refinement} \label{sec:signature-refinement}

We quantify search-space refinement by the fraction of distinct qubit pairs distinguished by their signatures.
\Cref{fig:search-space} shows this normalized measure, including its variation across randomized instances, for a code with signature partition $J_1,\ldots,J_m$:
\[
     \bar{q} = \frac{n(1-q)}{n-1} \in [0, 1] \text{ with } q = \sum\limits_{i=1}^m \frac{|J_i|^2}{n^2}.
\]
Signatures substantially refine the search space, particularly for non-\gls{css} codes, where a larger number of stabilizer generators $r$ provides more discriminative information.
Although the refinement saturates as $r$ increases, the timed-out instances show that computation time continues to grow as well. Signature computation is therefore unsuitable as a general-purpose preprocessing step for general stabilizer codes.

\definecolor{sigEmpty}{HTML}{FFFFFF}    
\definecolor{sigOutline}{HTML}{808080}

\pgfplotsset{
  colormap={signature}{
    rgb255=(231,245,246) rgb255=(192,229,231) rgb255=(129,203,207)
    rgb255=(64,164,170) rgb255=(39,95,98)},
  signature panel/.style={parameter map={\pmTextScale}{47},
                          xtick={3,5,10,15,20,25,30,35,40,47}, ytick={3,5,10,15,20,25,30,35,40,47},
                          width=3.5282in, height=3.6850in,
                          ylabel shift=2pt,
                          label style={font=\pmFont{11}{\pmTextScale}},
                          title style={font=\pmFont{12}{\pmTextScale}}},
  signature colorbar/.style={
    colorbar, colorbar shift/.style={xshift=0.1611in},
    colorbar style={
      at={(parent axis.north east)}, anchor=north west,
      scale only axis, width=0.1843in, height=3.6850in,
      axis line style={black, line width=0.7pt},
      tick align=outside, major tick length=3.5pt,
      tick style={black, line width=0.8pt},
      ytick pos=right, ytick={0,0.2,...,1},
      yticklabel={\pgfmathprintnumber[fixed, precision=1, zerofill]{\tick}},
      tick label style={font=\pmFont{7}{\pmTextScale}, inner sep=0pt},
      yticklabel shift=3.5pt,
      ylabel={$\bar{q}$},
      ylabel style={font=\pmFont{11}{\pmTextScale}, inner sep=0pt, align=center},
      ylabel shift=4.5pt,
    },
    colormap name=signature, point meta min=0, point meta max=1,
  },
}
\newcommand{\sigCell}[3]{\pmFillPart{#1}{#2}{0}{1}{signature}{#3}}
\newcommand{\sigCensored}[2]{\pmOutlinePart{#1}{#2}{0}{1}{sigOutline}}

\begin{figure}[htbp]
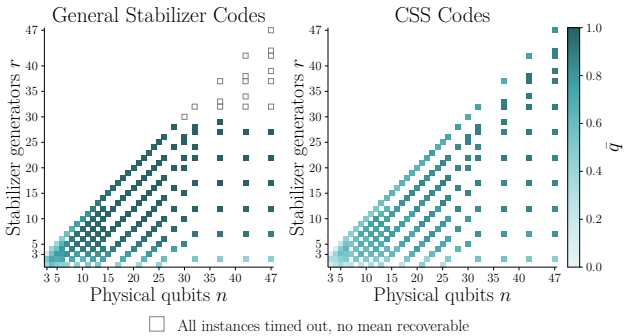

\centering
\resizebox{\columnwidth}{!}{%
\pmSetTextScale{2.0}%
\begin{tikzpicture}
  \begin{axis}[signature panel, at={(0.7158in,0.9350in)},
               title={General Stabilizer Codes}]
    \pmEmptyTriangle{sigEmpty}
    \input{material/graphics/data/a2_pm_stb.tex}
  \end{axis}

  \begin{scope}[xshift=0.2in]
  \begin{axis}[signature panel, at={(4.8812in,0.9350in)},
               title={CSS Codes}, signature colorbar]
    \pmEmptyTriangle{sigEmpty}
    \input{material/graphics/data/a2_pm_css.tex}
  \end{axis}
  \end{scope}

  \node[anchor=base, inner sep=0pt, font=\pmFont{9}{\pmTextScale}] at (4.794in, -0.05in) {%
    \tikz[baseline=0pt]{\draw[fill=sigEmpty, draw=sigOutline, line width=0.9pt]
      (0,0) rectangle (8\pmU,8\pmU);}%
    \hspace{7.2\pmU}All instances timed out, no mean recoverable};
\end{tikzpicture}%
}
\caption{Fraction $\bar{q}$ of qubit pairs distinguished by permutation signatures (0: no refinement, 1: all qubits distinguished).}
\label{fig:search-space}
\end{figure}

\section{Performance of Isomorphisms} \label{sec:performance-isomorphisms}

\Cref{fig:graph-iso} isolates the runtime and memory behavior of the isomorphism-based methods. Both costs increase steeply with graph size, with memory becoming the primary limitation. The heatmaps also expose the dominant scaling parameter---full-group encodings scale with stabilizer rank $r$, whereas the matroid-circuit encoding scales with code length $n$.

Despite the additional Pauli-triangle nodes in the local-Clifford construction, the permutation construction consumes more memory. This inversion is an implementation artifact: because the graph-isomorphism backend does not support colored edges, the permutation encoding represents edge colors by auxiliary nodes.

\def\reprLogVmin{-2.136313}
\def\reprLogSpan{5.868707}

\definecolor{reprEmpty}{HTML}{FFFFFF} 
\colorlet{reprOOMColor}{black!30}
\tikzset{
  repr oom/.style={pm oom, draw=reprOOMColor, line width=1.5pt},
  repr oom box/.style={draw=reprOOMColor, line width=0.8pt},
  repr oom key/.style={pm oom key, draw=reprOOMColor, line width=1.3pt},
}

\pgfplotsset{
  repr panel/.style={parameter map={\pmTextScale}{25},
                     xtick={3,5,10,15,20,25}, ytick={3,5,10,15,20,25},
                     width=3.5019in, height=3.8064in},
  repr panel wide/.style={parameter map={\pmTextScale}{35},
                          xtick={3,5,10,15,20,25,30,35}, ytick={3,5,10,15,20,25,30,35},
                          width=3.5019in, height=3.7142in},
  repr title/.style={title style={align=center, font=\pmFont{15}{\pmTextScale}},
                     label style={font=\pmFont{14}{\pmTextScale}},
                     ylabel shift=2pt},
}
\newcommand{\reprCell}[3]{\pmFillPart{#1}{#2}{0}{1}{runtime}{#3}}
\newcommand{\reprOOM}[2]{%
  \draw[repr oom box] (#1-0.46,#2-0.46) rectangle (#1+0.46,#2+0.46);
  \node[repr oom] at (#1,#2) {};}
\newcommand{\reprBarY}[1]{0.4400in+(log10(#1)-(\reprLogVmin))/\reprLogSpan*3.9050in}

\begin{figure}[htbp]
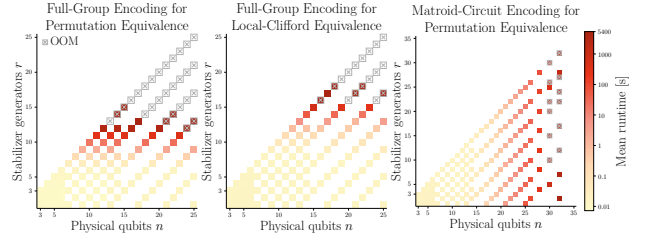

\centering
\resizebox{\columnwidth}{!}{%
\pmSetTextScale{1.5}%
\begin{tikzpicture}
  \begin{axis}[repr panel, repr title, at={(0.7920in,0.4893in)},
               title={Full-Group Encoding for\\Permutation Equivalence}]
    \pmEmptyTriangle[25]{reprEmpty}
    \input{material/graphics/data/a4_pm_stb_graph_iso.tex}
  \end{axis}

  \begin{axis}[repr panel, repr title, at={(4.9243in,0.4893in)},
               title={Full-Group Encoding for\\Local-Clifford Equivalence}]
    \pmEmptyTriangle[25]{reprEmpty}
    \input{material/graphics/data/a4_lc_stb_graph_iso.tex}
  \end{axis}

  \begin{axis}[repr panel wide, repr title, at={(9.0565in,0.5354in)},
               title={Matroid-Circuit Encoding for\\Permutation Equivalence}]
    \pmEmptyTriangle[35]{reprEmpty}
    \input{material/graphics/data/a4_pm_css_matroid.tex}
  \end{axis}

  \draw[repr oom box] ([shift={(-3.75pt,-3.75pt)}]0.9726in,4.1126in)
    rectangle ([shift={(3.75pt,3.75pt)}]0.9726in,4.1126in);
  \node[repr oom key] at (0.9726in,4.1126in) {};
  \node[anchor=base west, inner sep=0pt, font=\pmFont{13}{\pmTextScale}]
    at (1.0837in,4.0635in) {OOM};

  \shade[shading=pmruntime] (12.7410in,0.4400in) rectangle (12.9362in,4.3450in);
  \draw[black, line width=0.7pt] (12.7410in,0.4400in) rectangle (12.9362in,4.3450in);
  \foreach \v in {0.01,0.1,1,10,100,1000} {
    \draw[black, line width=0.8pt] (12.9362in,{\reprBarY{\v}}) -- ++(3.5pt,0);
    \node[anchor=base west, inner sep=0pt, font=\pmFont{7}{\pmTextScale}]
      at (13.0334in,{\reprBarY{\v}-2.7\pmU}) {\v};
  }
  \draw[black, line width=0.6pt] (12.9362in,4.3450in) -- ++(3pt,0);
  \node[anchor=base west, inner sep=0pt, font=\pmFont{7}{\pmTextScale}]
    at (13.0265in,{4.3450in-2.7\pmU}) {5400};
  \node[rotate=90, anchor=north, inner sep=0pt, font=\pmFont{14}{\pmTextScale}]
    at (13.4090in,2.3925in) {Mean runtime [s]};
\end{tikzpicture}%
}
\caption{Mean runtime of the isomorphism-based methods on randomized instances; crosses mark runs out of memory.}
\label{fig:graph-iso}
\end{figure}

\section{\SAT{} Encoding and Its Limitations} \label{sec:sat-limitations}

\Cref{fig:sat-css} shows that this \SAT{} encoding performs poorly on some \gls{css} codes compared with general stabilizer codes. We first rule out the concrete matrix representation and generator basis as causes, then relate the slowdown to how quickly a wrong partial permutation makes the encoded linear systems inconsistent.

\paragraph{Representation Independence} The check-matrix and tableau encodings exhibit almost identical slow regions in \Cref{fig:sat-enc-comp}; using check matrices reduces the mean normalized log-runtime by only $1.46$ percentage points. This observation is not surprising, as for a \gls{css} tableau $\left[\begin{smallmatrix} H_X & 0 \\ 0 & H_Z\end{smallmatrix}\right]$, the zero blocks eliminate half of the terms in the row-operation constraints, and the formula separates into the same two blocks as the check-matrix encoding. The slowdown is therefore tied to the block structure, not to the concrete matrix representation.

Nor can it be removed by hiding that separated structure. Random row operations that mix the $\operatorname{X}$- and $\operatorname{Z}$-type generators leave the code unchanged but obscure its \gls{css} form. However, as \Cref{tab:sat-row-mixing} shows, this increases rather than decreases the number of solver decisions, which is a more direct solver-level measure than runtime.

\def\satLogVmin{-1.786746}
\def\satLogSpan{5.519139}

\definecolor{sateEmpty}{HTML}{FFFFFF} 

\pgfplotsset{
  sate panel/.style={parameter map={\pmTextScale}{25},
                   xtick={3,5,10,15,20,25}, ytick={3,5,10,15,20,25},
                   width=3.5300in, height=3.8370in,
                   title style={align=center, font=\pmFont{12}{\pmTextScale}}, ylabel shift=2pt,
                   label style={font=\pmFont{11}{\pmTextScale}}},
}

\newcommand{\sateBarY}[1]{0.4400in+(log10(#1)-(\satLogVmin))/\satLogSpan*3.9050in}

\begin{figure}[ht]
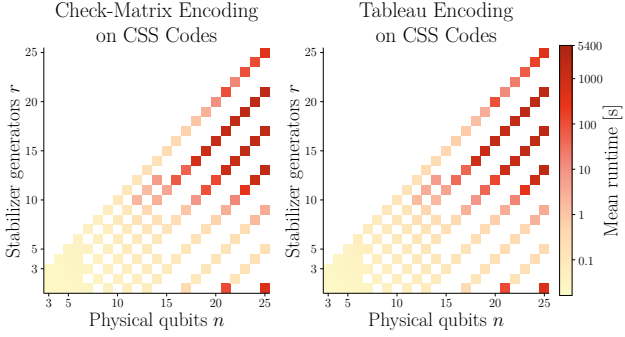

\centering
\resizebox{\columnwidth}{!}{%
\pmSetTextScale{2.0}%
\begin{tikzpicture}
  \begin{axis}[sate panel, at={(0.7140in,0.4740in)},
               title={Check-Matrix Encoding\\on CSS Codes}]
    \pmEmptyTriangle[25]{sateEmpty}
    \input{material/graphics/data/a6_pm_css_sat_on_css.tex}
  \end{axis}

  \begin{scope}[xshift=0.2in]
  \begin{axis}[sate panel, at={(4.8794in,0.4740in)},
               title={Tableau Encoding\\on CSS Codes}]
    \pmEmptyTriangle[25]{sateEmpty}
    \input{material/graphics/data/a6_pm_stb_sat_on_css.tex}
  \end{axis}

  \shade[shading=pmruntime] (8.5705in,0.4400in) rectangle (8.7658in,4.3450in);
  \draw[black, line width=0.7pt] (8.5705in,0.4400in) rectangle (8.7658in,4.3450in);
  \foreach \v in {0.1,1,10,100,1000} {
    \draw[black, line width=0.8pt] (8.7658in,{\sateBarY{\v}}) -- ++(3.5pt,0);
    \node[anchor=base west, inner sep=0pt, font=\pmFont{7}{\pmTextScale}]
      at (8.8630in,{\sateBarY{\v}-2.7\pmU}) {\v};
  }
  \draw[black, line width=0.6pt] (8.7658in,4.3450in) -- ++(3pt,0);
  \node[anchor=base west, inner sep=0pt, font=\pmFont{7}{\pmTextScale}]
    at (8.8547in,{4.3450in-2.7\pmU}) {5400};
  \node[rotate=90, anchor=north, inner sep=0pt, font=\pmFont{11}{\pmTextScale}]
    at (9.2452in,2.3925in) {Mean runtime [s]};
  \end{scope}
\end{tikzpicture}%
}
\caption{Mean \SAT{} runtime of the check-matrix and the tableau encoding on the same randomized \gls{css} instances.}
\label{fig:sat-enc-comp}
\end{figure}

\newcommand{\satMixingRows}{%
      14 & 12 & 686.5 & 2,713.5 \\
      16 & 14 & 1,746 & 8,228 \\
      18 & 16 & 8,320 & 711,414 \\
}

\begin{table}[htbp]
  \centering
  \caption{Median \SAT{} solver decisions on equivalent balanced \gls{css} instances (tableau encoding), with the block-diagonal generator basis (clean) and after random row operations mixing $\operatorname{X}$ and $\operatorname{Z}$ generators (mixed); timed-out runs excluded.}
  \label{tab:sat-row-mixing}
  \footnotesize
  \renewcommand{\arraystretch}{1.15}%
  \setlength{\tabcolsep}{3.5pt}%
  \begin{tabular}{@{}cc|cc@{}}
    \toprule
    $n$ & $r$ & clean & mixed \\
    \midrule
    \satMixingRows
    \bottomrule
  \end{tabular}%
\end{table}

\paragraph{Late Refutation}
Fix one row $h$ of a row-transformation block $R$ in the \SAT{} encoding, containing $\rho\in\{r,r_x,r_z\}$ unknowns. Under a candidate permutation, assigning qubit $i\mapsto j$ fixes column $j$ in the auxiliary matrix to column $i$ of $S$. The corresponding row constraints now become linear equations in the unknowns of row $h$, with coefficients given by the constant column $j$ of $S'$.

In the check-matrix encoding, $H_X$ and $H_Z$ have separate row transformations $R_X$ and $R_Z$. Each assigned qubit therefore yields one equation in $r_x$ unknowns and one independent equation in $r_z$ unknowns:
    \[
    \begin{aligned}
        (H_X)_{hi} &= \bigoplus_{l=0}^{r_x-1} {\textcolor{zxorangedark}{R_{X}}}_{hl}\,(H'_X)_{lj} \, ,
        \\
        (H_Z)_{hi} &= \bigoplus_{l=0}^{r_z-1} {\textcolor{zxgreendark}{R_{Z}}}_{hl} \, (H'_Z)_{lj}
    \end{aligned}
    \]
For a general stabilizer tableau, the $X$ and $Z$ columns of qubit $j$ instead share the same transformation $R$. An assignment then yields two equations in the same $r$ unknowns:
    \[
        X_{hi} = \bigoplus_{l=0}^{r-1} \textcolor{red}{R}_{hl} \, X'_{lj} \, , 
        \quad
        Z_{hi} = \bigoplus_{l=0}^{r-1} \textcolor{red}{R}_{hl} \, Z'_{lj}.
    \]
After $m$ qubit assignments, each system therefore contains $\mu\in\{m,2m\}$ equations in $\rho$ unknowns. The selected columns $c'_{j_0},\ldots,c'_{j_{\mu-1}}$ of $S'$ form the rows of the coefficient matrix:
\[
    \left[\begin{smallmatrix} S'_{0, j_0} & \cdots & S'_{\rho-1, j_0} \\ \vdots & & \vdots \\ S'_{0, j_{\mu-1}} & \cdots & S'_{\rho-1, j_{\mu-1}} \end{smallmatrix}\right] 
    \cdot
    \left[\begin{smallmatrix} R_{h,0} \\ \vdots \\ R_{h,\rho-1} \end{smallmatrix}\right]
    =
    \left[\begin{smallmatrix} S_{h, i_0} \\ \vdots \\ S_{h, i_{\mu-1}} \end{smallmatrix}\right] 
    \]
If these rows are linearly independent, the coefficient matrix has rank $\mu$, and the system has a solution for every possible right-hand side in $\mathbb{F}_2^\mu$. An incorrect mapping merely selects a different right-hand side---i.e., column $i$ in $S$---and cannot yet cause a contradiction. Linear independence is not guaranteed, but typical while only a few columns have been assigned. Once $\mu>\rho$, independence becomes impossible, and further assignments can make the system inconsistent, if the candidate mapping is invalid.

This motivates a heuristic \emph{refutation threshold} near $\mu=\rho$: under the typical full-row-rank assumption above, contradictions cannot arise before this point but become possible once additional equations are added. The threshold is near $m=r/2$ for a general tableau, but near $m=r_x$ and $m=r_z$ for the two \gls{css} blocks. A balanced split of the \gls{css} code, $r_x=r_z=r/2$, therefore resembles the general stabilizer case. With an unbalanced split, the larger block can remain underdetermined until $m=\max(r_x,r_z)$. The smaller block becomes overdetermined earlier, but its lower rank also distinguishes fewer mappings. For instance, a one-row block can reject a mapping from a $1$-column to a $0$-column, but not one between two $1$-columns.

The U-shaped curve in \Cref{fig:sat-rank-sweep} supports this explanation: unbalanced splits require substantially more solver decisions, whereas balanced splits approach the general-stabilizer baseline. Because $r_x$ is sampled uniformly from $0$ to $r$ in \Cref{fig:sat-css}, increasingly costly unbalanced instances raise the mean runtime. For this argument, a general tableau is thus not equivalent to a single $r$-row check block: coupling its $X$ and $Z$ columns through the same $R$ provides two constraints per assigned qubit and approximately halves the threshold. From the solver's perspective, this mirrors group-action search: the large block permits many partial assignments and prunes an invalid branch only at greater depth.

\begin{example}[Refutation threshold]
Take the single check matrix
\[
    H_Z =
    \begin{array}{@{}c@{}cccc@{}c@{}}
        & {\color{gray}\scriptstyle c_0} & {\color{gray}\scriptstyle c_1} & {\color{gray}\scriptstyle c_2} & {\color{gray}\scriptstyle c_3} & \\[-2pt]
        \bigg[ & \begin{array}{@{}c@{}} 1 \\ 0 \end{array}
               & \begin{array}{@{}c@{}} 0 \\ 1 \end{array}
               & \begin{array}{@{}c@{}} 1 \\ 1 \end{array}
               & \begin{array}{@{}c@{}} 1 \\ 0 \end{array} & \bigg]
    \end{array}
\]
This is the only check block of a \gls{css} code with $r_x=0$, $\rho=r_z=2$, and $n=4$. We compare the code with itself, so $S=S'=H_Z$ and the identity is a witness. Its valid permutations include $0\leftrightarrow3$ and $1\leftrightarrow2$, whereas $0\mapsto1$ is invalid. Consider the invalid candidate permutation $\{0\mapsto1,1\mapsto0,3\mapsto3,2\mapsto2\}$, which is checked step by step:
\begin{itemize}
    \item $m = 1$, $\{\mapmark[bandone]{0 \mapsto 1}\}$:\\
    Row 1:
    $\left[\bandmatrix{\band[bandone]{0 & 1}}\right] \left[\begin{smallmatrix} r_{00} \\ r_{01} \end{smallmatrix}\right] = \left[\bandmatrix{\band[bandone]{1}}\right] \rightarrow r_{01} = 1$\\
    Row 2:
    $\left[\bandmatrix{\band[bandone]{0 & 1}}\right] \left[\begin{smallmatrix} r_{10} \\ r_{11} \end{smallmatrix}\right] = \left[\bandmatrix{\band[bandone]{0}}\right] \rightarrow r_{11} = 0$\\
    These equations fix the second column of $R=\left[\begin{smallmatrix} r_{00} & 1 \\ r_{10} & 0 \end{smallmatrix}\right]$ but leave the first column free. All four assignments to that column remain feasible, nothing is refuted.
    \item $m = 2 = \rho$, $\{\mapmark[bandone]{0 \mapsto 1},\, \mapmark[bandtwo]{1 \mapsto 0}\}$:\\
    Row 1:
    $\left[\bandmatrix{\band[bandone]{0 & 1} \\ \band[bandtwo]{1 & 0}}\right] \left[\begin{smallmatrix} r_{00} \\ r_{01} \end{smallmatrix}\right] = \left[\bandmatrix{\band[bandone]{1} \\ \band[bandtwo]{0}}\right] \rightarrow r_{00} = 0, r_{01} = 1$\\
    Row 2:
    $\left[\bandmatrix{\band[bandone]{0 & 1} \\ \band[bandtwo]{1 & 0}}\right] \left[\begin{smallmatrix} r_{10} \\ r_{11} \end{smallmatrix}\right] = \left[\bandmatrix{\band[bandone]{0} \\ \band[bandtwo]{1}}\right] \rightarrow r_{10} = 1, r_{11} = 0$\\
    The system now uniquely determines $R=\left[\begin{smallmatrix} 0 & 1 \\ 1 & 0 \end{smallmatrix}\right]$ and remains consistent, although the partial permutation is invalid.
    \item $m = 3 > \rho$, $\{\mapmark[bandone]{0 \mapsto 1},\, \mapmark[bandtwo]{1 \mapsto 0}, \mapmark[bandthree]{3 \mapsto 3}\}$:\\
    Row 1:
    $\left[\bandmatrix{\band[bandone]{0 & 1} \\ \band[bandtwo]{1 & 0} \\ \band[bandthree]{1 & 0}}\right] \left[\begin{smallmatrix} r_{00} \\ r_{01} \end{smallmatrix}\right] = \left[\bandmatrix{\band[bandone]{1} \\ \band[bandtwo]{0} \\ \band[bandthree]{1}}\right] \times$\\
    Row 2:
    $\left[\bandmatrix{\band[bandone]{0 & 1} \\ \band[bandtwo]{1 & 0} \\ \band[bandthree]{1 & 0}}\right] \left[\begin{smallmatrix} r_{10} \\ r_{11} \end{smallmatrix}\right] = \left[\bandmatrix{\band[bandone]{0} \\ \band[bandtwo]{1} \\ \band[bandthree]{0}}\right] \times$\\
    The third assignment adds equations to an already determined system. Both rows become inconsistent, refuting the candidate permutation.
\end{itemize}
\end{example}

\definecolor{satLilaLight}{HTML}{C3A0E3}
\definecolor{satLila}{HTML}{8E4DCB}
\colorlet{satLilaDark}{thesislila}
\definecolor{satGray}{HTML}{808080}
\definecolor{satFrame}{HTML}{CCCCCC}
\definecolor{satGrid}{HTML}{D8D8D8}

\pgfplotsset{
  sat sweep/.style={
    width=6.45in, height=3.15in, scale only axis,
    xmin=-0.35, xmax=6.35, ymin=50, ymax=3e7,
    xtick={0,1,2,3,4,5,6},
    xticklabels={$0$, $1$, $2$, $\lfloor r/2 \rfloor$, $r-2$, $r-1$, $r$},
    ytick={100,1000,10000,100000,1000000,10000000},
    axis lines*=left, axis line style={black, line width=0.7pt},
    tick align=outside, major tick length=3.5pt,
    tick style={black, line width=0.8pt},
    tick label style={font=\pmFont{8}{\pmTextScale}},
    xticklabel shift=3pt, yticklabel shift=3pt,
    xlabel={X-check rank $r_x$},
    ylabel={Median solver decisions},
    label style={font=\pmFont{11}{\pmTextScale}},
    ymajorgrids, grid style={satGrid, line width=0.4pt},
    clip=false,
    legend style={at={(0.5,0.97)}, anchor=north, draw=satFrame, fill=white,
                  line width=0.5pt, inner xsep=6pt, inner ysep=4pt,
                  font=\pmFont{9}{\pmTextScale}, legend columns=2,
                  /tikz/every even column/.append style={column sep=1.4em}},
    legend cell align=left,
    legend image post style={mark size=2.2pt},
  },
  sat css/.style={dashed, line width=1.1pt, mark=square*, mark size=2.3pt, mark options={solid}},
  sat general/.style={solid, line width=1pt, mark=none},
}

\begin{figure}[htbp]
\centering
\resizebox{\columnwidth}{!}{%
\pmSetTextScale{1.6}%
\begin{tikzpicture}
  \begin{semilogyaxis}[sat sweep]%
    \addlegendimage{satLilaLight, line width=1.1pt}\addlegendentry{$n=14$}
    \addlegendimage{satGray, sat css}\addlegendentry{CSS codes}
    \addlegendimage{satLila, line width=1.1pt}\addlegendentry{$n=16$}
    \addlegendimage{satGray, sat general}\addlegendentry{Non-CSS codes}
    \addlegendimage{satLilaDark, line width=1.1pt}\addlegendentry{$n=18$}
    \addlegendimage{empty legend}\addlegendentry{}
    \input{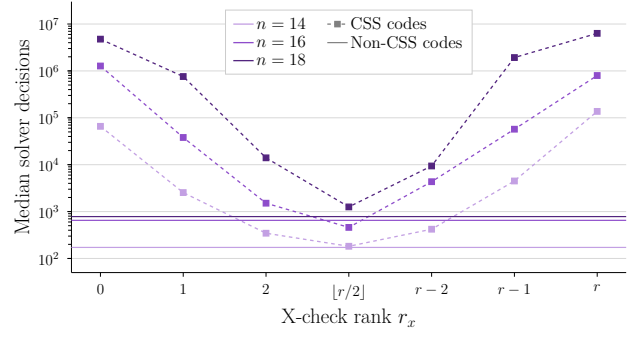}
  \end{semilogyaxis}
\end{tikzpicture}%
}
\caption{Median \SAT{} solver decisions for equivalent \gls{css} instances ($k=4$) as a function of $X$-check rank $r_x$, compared with non-\gls{css} stabilizer codes of the same $n$ and $r$; timed-out runs excluded; lines serve only as visual guides.}
\label{fig:sat-rank-sweep}
\end{figure}

\paragraph{Remedy}
Permutation equivalence of row spaces is equivalent to permutation equivalence of their orthogonal complements. Since 
\[
    \dim\big(\operatorname{rs}(H)\big)+\dim\big(\ker(H)\big)=n
\]
and permutations preserve the inner product, we have
\[
    \resizebox{\columnwidth}{!}{$\displaystyle
    \textcolor{thesisred}{\pi(}\operatorname{rs}(H_1)\textcolor{thesisred}{)} = \operatorname{rs}(H_2) \;\Leftrightarrow\; \textcolor{thesisred}{\pi(}\ker(H_1)\textcolor{thesisred}{)} = \ker(H_2).
    $}
\]
A check block with $\rho>n/2$ rows can therefore be replaced, in both codes, by a basis of its kernel with $n-\rho$ rows. This preserves the permutation-equivalence problem while reducing the refutation threshold to $\min(\rho,n-\rho)\leq n/2$, thus, reducing the number of assignments a wrong candidate permutation usually had to make before being refuted.

\begin{findingbox}{The matrix-based \SAT{} encoding is particularly costly for CSS codes with an unbalanced $X$/$Z$ check split.}
    \findingpart{Within this encoding, the slowdown stems from the uncoupled CSS check structure, rather than the chosen representation or generator basis.}
    \lastfindingpart{An \emph{unbalanced} CSS code needs more decisions before an invalid candidate permutation is refuted, because its $X$ and $Z$ check matrices are uncoupled.}
\end{findingbox}

\end{document}